\documentclass[draft,aps,pre,preprint,showpacs,showkeys,superscriptaddress,preprintnumbers,amsmath,amssymb,floatfix,nofootinbib]{revtex4-1}

\usepackage{setspace}
\usepackage{amsmath}
\usepackage[colorlinks]{hyperref}
\usepackage[final,dvips]{graphicx}
\usepackage{color}
\usepackage[symbol]{footmisc}
\usepackage{bm}
\usepackage{wrapfig}

\usepackage[percent]{overpic}
\usepackage{xcolor}
\usepackage{pict2e}
\usepackage{caption,subcaption}

\newcommand{\beq}{\begin{equation}}
\newcommand{\eeq}{\end{equation}}

\newcommand{\beqa}{\begin{eqnarray}}
\newcommand{\eeqa}{\end{eqnarray}}

\newcommand \rt {\right}
\newcommand \lt {\left}
\newcommand \nline {\nonumber \\}

\newcommand \freem {{\cal F}}

\newcommand \pxpy[2] {\frac{\partial #1}{\partial #2}}

\newcommand \fxfy[2] {\frac{\delta #1}{\delta #2}}

\newcommand{\mbfk}{{\mathbf k}}

\newcommand{\mbfr}{{\mathbf r}}

\usepackage[capitalise]{cleveref}
 
\begin{document}

\title{Self-consistent modeling of anisotropic interfaces and missing orientations: Derivation from phase-field crystal}

\author{N. Ofori-Opoku}
\altaffiliation{Guest Researcher: Material Measurement Laboratory, National Institute of Standards and Technology}
\affiliation{Center for Hierarchical Materials Design, Northwestern University, Evanston IL 60208}
\affiliation{Material Measurement Laboratory, National Institute of Standards and Technology, Gaithersburg MD 20899}

\author{J.A. Warren}
\affiliation{Material Measurement Laboratory, National Institute of Standards and Technology, Gaithersburg MD 20899}

\author{P.W. Voorhees}
\affiliation{Center for Hierarchical Materials Design, Northwestern University, Evanston IL 60208}
\affiliation{Department of Materials Science and Engineering, Northwestern University, Evanston IL 60208}
\affiliation{Department of Engineering Sciences and Applied Mathematics, Northwestern University, Evanston IL 60208}

\begin{abstract}
Highly anisotropic interfaces play an important role in the development of material microstructure. Using the diffusive atomistic phase-field crystal (PFC) formalism, we determine the capability of the model to quantitatively describe these interfaces. Specifically, we coarse grain the PFC model to attain both its complex amplitude formulation and its corresponding phase-field limit. Using this latter formulation, in one-dimensional calculations, we determine the surface energy and the properties of the Wulff shape. We find that the model can yield Wulff shapes with missing orientations, the transition to missing orientations, and facet formation. We show that the corresponding phase-field limit of the complex amplitude model yields a self-consistent description of highly anisotropic surface properties that are a function of the surface orientation with respect to the underlying crystal lattice. The phase-field model is also  capable of describing missing orientations on equilibrium shapes of crystals and  naturally includes a regularizing contribution. We demonstrate, in two dimensions, how the resultant model can be used to study growth of crystals with varying degrees of anisotropy in the phase-field limit.
\end{abstract}

\keywords{surface energy; anisotropy; solidification; phase-field crystal; coarse-graining}

\maketitle

\section{Introduction}
Defects define the microstructure of materials.  Surface structures such as interfaces, grain and interphase boundaries often dominate material performance. The result of the various phase transformation processes that occur, e.g. solidification and precipitation reactions, these defect structures determine among other features, the various mechanical, optical and electrical properties of the resultant material. The anisotropic nature of these surfaces, which can lead to so-called faceted structures, has received and continues to receive a great deal of attention.  Of particular interest are microstructures resulting from solidification or vapor deposition of materials with highly anisotropic surface energies that can lead to faceted interfaces. An understanding of the underlying atomistic mechanisms that lead to and determine the characteristics of these surface defects is paramount in our understanding of microstructure and ultimately to the design engineering of materials from the atomistic scale.

Much is known of the underlying mechanisms that lead to highly anisotropic interfaces. Historically, using the example of crystal growth, Gibbs and Wulff established the notion of an equilibrium shape as a consequence of the minimization of the total interfacial solid-liquid free energy of the crystal. The atomistic underpinnings of interface motion and growth can be traced to Burton, Cabrera and Frank (BCF)~\cite{Burton1951}, while the atomistic and thermodynamic treatment of crystal surfaces can be attributed to Herring~\cite{Herring1951} who refined and improved the formalism by Wulff. Turnbull~\cite{Turnbull1950} and Jackson~\cite{Jackson1958}  have both applied a mean field thermodynamic treatment to determine the characteristics of interfaces based on average quantities such as enthalpy and entropy, with the former leading to the Turnbull coefficient which relates the surface energy to the enthalpy production and the latter the so-called $\alpha$-parameter which relates the smooth or rough nature of the interface to entropic effects. In addition, Hoffman and Cahn~\cite{Hoffman1972,Cahn1974} introduced a general thermodynamic theory based on a vector formalism to describe equilibrium crystal shapes and their surfaces. Most recently, this formalism has been extended by Cahn and Carter~\cite{Cahn1996} as a general mathematical foundational basis to describe phase equilibria.

At the microstructural scale, phase-field (PF) methods have provided a unique opportunity to numerically investigate phase transformation kinetics, particularly that of solidification microstructures. The PF modeling formalism is a continuum, mean-field phenomenological theory that can be traced back to the theories of van der Waals, Ginzburg-Landau, Allen-Cahn and Cahn-Hilliard. The method couples a set of uniform order parameters to one or more diffusion fields (i.e., temperature or solute fields), with the kinetics driven by dissipative energy minimization of some postulated free energy functional. Through the gradient terms of the theory, terms proportional to $|\boldsymbol{\nabla}\phi|^2$ (where $\phi$ is the order parameter), the method is amenable to the inclusion of anisotropic functions that describe the characteristics of interfaces and surfaces. Interestingly,  Caginalp~\cite{Caginalp1986,Caginalp1987} and Caginalp and Fife~\cite{Caginalp1986a} were the first to consider inclusion of anisotropic features in Ginzburg-Landau Hamiltonians of systems describing spins. Here, instead of inclusion directly to a gradient term, they considered an underlying lattice, described by reciprocal lattice vectors and a non-local interaction contribution. This became the template for subsequent introduction of anisotropy into PF-type equations. Later, Kobayashi~\cite{Kobayashi1993} included surface tension anisotropy by allowing the gradient energy coefficient to depend on the orientation of the interface into a PF framework to model dendritic growth. This allowed large scale simulation of dendrites, which matched qualitatively to those from experiments. Wheeler et al.~\cite{Wheeler1993} taking the model of Kobayashi~\cite{Kobayashi1993}, performed rigorous numerical calculations, while McFadden et al.~\cite{McFadden1993} extended the approach by including anisotropic mobility and an asymptotic analysis, both allowing the methodology to become a viable quantitative technique for studying dendritic growth conditions. Most PF models to date that include anisotropic surface energy follow the forms introduced in these works. In two-dimensions (2D), it usually takes the form, $1+\delta\cos{4\theta}$, for a crystal with four-fold cubic symmetry, where $\theta$ is the orientation of the interface normal and $\delta$ is the anisotropy strength. 

Faceting can be driven by anisotropic surface energy, anisotropic kinetic processes, or both, occurring at the solid-liquid interface. Sekerka~\cite{Sekerka2005} gives a detailed overview of both theory and approaches to examining anisotropic crystal growth. From the perspective of surface energy, the determination of anisotropy is directly illustrated by the Frank construction. This is an analytical scheme to determine the missing orientations, i.e., those orientations where the surface energy is too large to be represented on the equilibrium crystal shape (ECS), which manifests itself as ``ears'' on the Wulff shape. Given $1/\gamma(\theta)$ (where $\gamma(\theta)$ is the surface energy), the non-convexity of the resultant plot translates to a change in sign from positive to negative of the surface stiffness, $\gamma(\theta) + \partial^2\gamma/\partial\theta^2$, and leads to orientations missing from the ECS.  If $\gamma(\theta) + \partial^2\gamma/\partial\theta^2>0$, and a surface with this orientation is not on the Wulff shape, then it is metastable whereas if $\gamma(\theta) + \partial^2\gamma/\partial\theta^2<0$ the surface is unstable. To describe highly anisotropic surfaces in the PF framework, complex forms, some non-analytical, of the gradient energy description are often necessary. However, field theories such as the PF methodology, can become ill-posed when the surface stiffness becomes negative, and so-called convexification/regularization methods are employed in order to render the formulation well posed. Steps in addressing this shortcoming of PF theories have been proposed. An important contribution was the connection of the Cahn-Hoffman $\xi$-vector, by Wheeler and McFadden~\cite{Wheeler1996,Wheeler1997} and Wheeler~\cite{Wheeler1999},  to PF models which provided a robust mechanism for describing anisotropic interfaces. Numerical implementations to realize faceted surfaces have also been attempted. For example, Eggleston et al.~\cite{Eggleston2001},  Uehara and Sekerka~\cite{Uehara2003}  and Debierre et al.~\cite{Debierre2003}, have developed PF models specifically designed for highly anisotropic crystal growth. The former and latter modeled it through highly anisotropic interfacial energy, while the work of Uehara and Sekerka modeled it through the kinetic coefficient. Eggleston et al.~\cite{Eggleston2001} used a convexified inverse gradient energy coefficient designed to avoid unstable interfacial orientations and thus yielding well-posed evolution equations for the order parameter. The focus in these and other phase-field models used to model missing orientations, leading to edges and facets, have mostly been directed towards determining a feasible and numerically tractable means of regularizing, for convenience, the corresponding anisotropic contributions. 
%
%

The use of regularization methods, however, alters the nucleation physics of the solidifying interface by preventing the formation of new facets~\cite{Dicarlo1992}. A novel idea to circumvent this physical issue is the work by Wise et al.~\cite{Wise2005,Wise2007} and Wheeler~\cite{Wheeler2006} who used a regularizing method based on the addition of the square of the Laplacian of $\phi$, i.e., $(\nabla^2\phi)^2$, effectively the square of the interfacial curvature, to the free-energy functional. The use of regularization methods of this type, can be traced back to the work of Stewart and Goldenfeld~\cite{Stewart1992}, when they considered spinodal decomposition on surfaces, and developed further by Gurtin and Jabbour~\cite{Gurtin2002} all in the context of the thermodynamics of surfaces and sharp interfaces. Recently, Torabi et al.~\cite{Torabi2009} have also presented a new PF model for strongly anisotropic systems that accounts for the Willmore energy by allowing the interface thickness to be independent of the interface orientation.

Growth of faceted crystals have been investigated through various means. From an atomistic point of view, we have gained an understanding of the growth processes involving such features as kinks, steps and dislocations. Models such as those based on the work of BCF, molecular dynamics, Monte Carlo and other such methods have been instrumental in this regard. These, however, have lacked the size and temporal scale to compare and accurately represent those of mesoscale microstructures like those found in experiments. The PF method is capable of capturing the diffusive scale needed for experimental microstructures, and as described in the previous paragraph is also capable of describing certain aspects of anisotropic  surfaces. Despite the utility of the PF method, it inherently lacks a self-consistent atomistic description of the underlying physical mechanisms. 

An atomistic-scale diffusive modeling formalism, the phase-field crystal (PFC) methodology~\cite{Elder2002,Elder2004,Elder2007}, has emerged as an alternative to the numerical simulations of the kinetics of phase transformations. The methodology, unlike the PF method, is a formalism where the free energy functional is minimized by periodic fields. This allows the method to operate on atomistic length scales yet access diffusive time scales. The periodic nature of these fields also allows the self-consistent description of elasto-plastic effects, multiple crystal orientations, grain boundaries and dislocations, all evolving over mesoscopic time scales. It has be shown that the method is derivable from the fundamental classical density functional theory (CDFT)~\cite{Elder2007,Jin2006}, with certain approximations based on the Ramakrishnan and Yussouff~\cite{Ramakrishnan1979} free energy formalism. Since its inception, the method has evolved to model structural phase transformations~\cite{Wu2010,Greenwood2010,Mkhonta2013,Seymour2016}, alloy systems~\cite{Elder2007,Greenwood2011,Ofori-Opoku2013}, multiferroic composite materials~\cite{Seymour2015}, order-disorder systems~\cite{Alster2016}, nucleation and polymorphism~\cite{Tegze2009,Toth2010}, amorphous or glass transitions\cite{Berry2008,Archer2012}, quasicrystals~\cite{Archer2015} and crystal plasticity~\cite{Berry2015}. Through coarse graining procedures, the methodology may be used to generate models that operate on mesoscopic length scales. These, so-called complex amplitude models bear a striking resemblance to PF models, however they retain many of the rich atomistic level phenomena of their more microscopic PFC counterparts. The complex amplitude formalisms have been used to describe crystal-melt interfaces~\cite{Wu2007,Majaniemi2009,Provatas2010,Toth2014}, grain boundary pre-melting~\cite{Mellenthin2008,Spatschek2013}, shear coupled and defect motion in alloys~\cite{Spatschek2010}, and lattice pinning effects on interfaces~\cite{Huang2013,Huang2016}.

In this study, we will use the PFC formalism to investigate surface properties during solidification. Specifically, we coarse-grain a variant of the PFC model to generate a corresponding PF-type model and analyze the surface properties as a function of the surface orientation through surface energy calculations in 1D. Unlike traditional PF models, the gradient energy in this derived model is not postulated to be of a particular form, but rather depends directly on the underlying crystal lattice through the set of reciprocal lattice vectors, similar to the approach of Caginalp and Fife~\cite{Caginalp1986,Caginalp1986a,Caginalp1987}. At the core of this gradient description is the capability to self-consistently describe anisotropy directly from underlying atomistic considerations. The results of surface energy calculations are then used to construct the Wulff shapes from which the ECSs may be inferred. Following this, we perform full dynamic simulations of dendrite growth in 2D.

The paper is organized as follows. We begin by briefly describing the PFC model used in this work and its coarse-grained limit in \cref{pfc-func}. A description of surface energy is given in \cref{surf-ene} and the resulting coarse-grained model is then used to calculate surface energy and Wulff shapes in \cref{results}, as well as 2D crystal growth. Finally we summarize in \cref{summ}.

\section{Phase-field crystal modeling}
\label{pfc-func}
In this section the PFC method is introduced and its equilibrium properties are described. The model is then coarse-grained, generating two variant PF-type models. 

\subsection{Free energy functional}
\label{pfc-func-freeene}
We will use the simplest, so-called standard PFC model~\cite{Elder2002} first proposed as an extension of the Swift-Hohenberg (SH) model~\cite{Swift1977}, which is used to study Rayleigh-B{\'e}nard convection. To put the model on more fundamental footing, it has been derived from CDFT in Ref.\cite{Elder2007}. The dimensionless free energy functional reads,
\beq
\freem =   \int d\mbfr \, \lt\{ \frac{n}{2}\lt[\epsilon + (1+\nabla^2)^2\rt]n + \frac{n^4}{4}\rt\},
\label{eq:fene}
\eeq
where $\epsilon \le 0$ is an effective temperature parameter related to properties of the two-point direct correlation function resulting from the liquid structure factor and $n$ is a dimensionless number density. This functional is minimized by a stripe phase in 1D, and triangular and bcc lattice in 2D and 3D respectively. The equation of motion generally follows from variational principles of conserved systems,
\begin{align}
\pxpy{n}{t} &= -\bm{\nabla}\cdot \bm{J} \nline
& = \bm{\nabla}\cdot \lt (M \bm{\nabla} \mu \rt) \nline
& = \bm{\nabla}\cdot \lt (M \bm{\nabla} \fxfy{\freem}{n} \rt)
\label{eq:eom}
\end{align}
where $\bm{J}$ is the mass flux and $\mu$ is the chemical potential defined by the functional derivative.

\subsection{Equilibrium}
\label{pfc-func-equil}
The equilibrium phase diagram can be determined from standard thermodynamic minimization schemes. These methods culminate in what is known as the common tangent or Maxwell equal area construction. These constructions are the geometrical representation of the thermodynamic statements of equal chemical potential and equal pressure of any two or more co-existing bulk phases as dictated by the Gibbs phase rule. The procedures for calculating phase diagrams for PFC models are well documented  \cite{Elder2002,Elder2007,Greenwood2010,Greenwood2011} and the method used is briefly summarized here.

Considering 2D solid structures, specifically a triangular lattice which varies on the atomic length scale, the density field $n$ is approximated using a single mode, plane wave approximation given by
\beq
n(\mbfr) = \bar{n}(\mbfr) + \sum_{j=1}^{3}A_j(\mbfr)\exp\left({\frac{2\pi}{a} \mathbf{i} \mbfk_{j} \cdot \mbfr}\right) + \text{c.c.},
%
%
\label{eq:dens}
\eeq
where $\bar{n}$ is the average density, a conserved variable, $\{A_j\}$ are the amplitudes of the density oscillations along each reciprocal lattice direction $j$ related to the Fourier components of the structure factor of the solid, $a=4\pi/\sqrt{3}$ is the equilibrium lattice spacing and $\{\mbfk_j\}$ denote the set of reciprocal lattice vectors which describe the triangular structure and $\text{c.c.}$ denotes the complex conjugate. 

We proceed by substituting \cref{eq:dens} into \cref{eq:fene}, where for a bulk crystal we assume that $A_j(\mbfr) \rightarrow \phi, ~\forall\, j$, and integrating over one unit cell, the resultant normalized free energy is calculated for the solid phase as a function of $\bar{n}$ and $\phi$. Taking the values of the amplitudes, which are non-conserved variables, that minimize the free energy  and substituting back into the free energy gives  a free energy landscape $f_{\rm{sol}}(\bar{n})$, where $f_{\rm{sol}}$ represents an amplitude-minimized solid free energy. The liquid phase in this description, denoted by $f_{\rm{liq}}$, is trivially computed by setting $\phi=0$.

\begin{figure}[htb]
\centering
	\resizebox{3.4in}{!}{\includegraphics{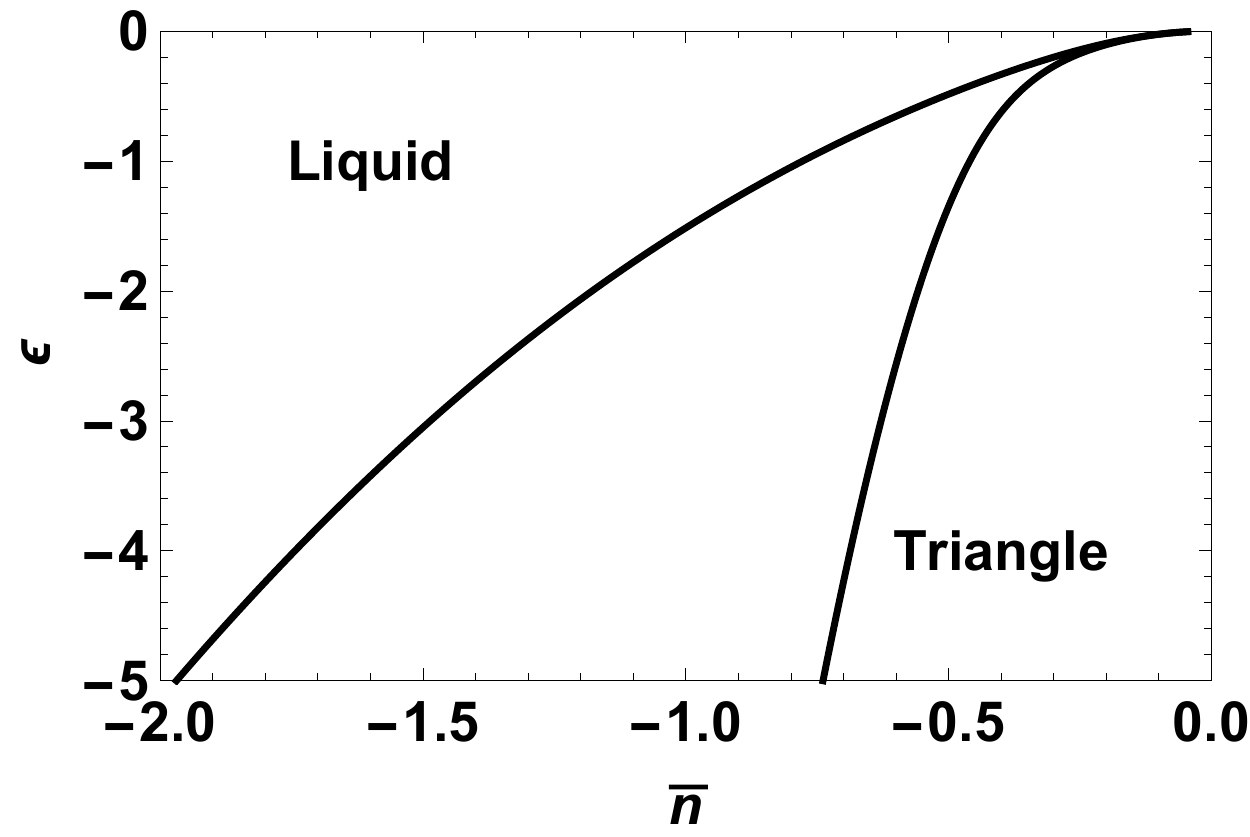}}
    \caption{Phase diagram for the standard PFC model. The dimensionless temperature, $\epsilon$, plotted against the dimensionless density $\bar{n}$. In 2D, the triangular phase is in coexistence with the constant liquid phase.}
\label{fig:phase-diag}
\end{figure}

Having the free energy landscapes of the solid and liquid phases, phase boundaries between the solid and liquid phases for a given temperature, $\epsilon$, are computed by solving the following set of equations,
\begin{align}
&\mu_{\rm{liq}}=\mu_{\rm{sol}}\nline
&\Omega_{\rm{liq}}=\Omega_{\rm{sol}},
\label{eq:coex}
\end{align}
where the last of these, the equality of grand potentials, implies equilibration of pressure, defined explicitly by,
\begin{align}
f_{\rm{liq}}-\mu_{\rm{liq}}\bar{n}_{\rm{liq}} = f_{\rm{sol}}-\mu_{\rm{sol}}\bar{n}_{\rm{sol}},
\end{align}
where $\bar{n}_{\rm{sol}}$ and $\bar{n}_{\rm{liq}}$ represent the equilibrium density values for the solid and liquid respectively. The calculation of \cref{eq:coex} for various temperature values $\epsilon$, leads to \cref{fig:phase-diag}, which shows the equilibrium phase diagram  between the solid triangular phase and the liquid phase. We emphasize that this phase diagram has been extended to effective temperatures far below those of previous studies using this model. Through this extension, we increase the allowable phase space in which we can explore the surface properties of the crystal. It is worth noting that the single mode approximation is strictly valid around the reference point of the PFC expansion, i.e., $\bar{n}=0$, close to $\epsilon=0$. With decreasing $\epsilon$, more modes are necessarily required in order to capture the finer peak structure of the solid that emerges. Admittedly, an extension of the single mode expansion to these low temperature values ignores the small wavelengths that would otherwise emerge in the bulk solid via the peak-to-peak oscillations, which inevitably contributes to the interfacial description. Though this presents a considerable inaccuracy when comparing equilibrium and other emergent properties directly with the microscopic PFC model far from the expansion, we are still able to capture, in the coarse-grained description discussed below,  the salient features of interfacial phenomena. Specifically, we note that interfacial widths do indeed decrease with decreasing temperature. 
%
%
%
%

\subsection{Deriving effective phase-field models through coarse-graining}
The full atomistic model of \cref{eq:fene} has built-in descriptions of nucleation, surfaces and faceting of interfaces, e.g. Ref.~\cite{Tegze2011}. The ability to exploit these smaller length scale physics at larger scales, through coarse-graining methods to produce effective higher length scale models, is one interesting feature of the PFC and other CDFT-type models. Two such model variants can be derived, based on specific approximations and interests. The first, called the complex-amplitude formulation, which retains some of the salient physics from the atomistic level such as defects and elastic-plastic behavior. The other, called the phase-field limit, here referred to simply as an amplitude model, retains less of the underlying atomistic level information, e.g., the description of dislocations, but still is useful in describing a host of phase transformation features. 

The details of coarse-graining PFC and CDFT-type equations are well documented in the literature. Generally, we start by making an {\it ansatz}, the plane-wave, mode approximation of \cref{eq:dens} for the density. It is important to note that this approximation is {\it exact} around the reference density from which the model is expanded, or at small values of $\epsilon$ around the melting temperature of the CDFT model the PFC model is derived from. Similar to the phase diagram calculation, this approximation is then inserted into the free energy functional, \cref{eq:fene}, after which the coarse-graining follows. The success of the For the coarse-graining to be successful, it is tacitly assumed that  $\bar{n}(\mbfr)$, which is the dimensionless average density, is a ``slow'' variable, i.e., smoothly varying on the scale of the periodic atomistic oscillations. $\{A_j(\mbfr)\}$ represents the complex amplitudes describing the height of the oscillation of the density field. Like the dimensionless average density, the amplitudes are also ``slow'' variables. This tacit assumption enables the separation of scales required for successful coarse graining.

Once the density approximation has been inserted into the free energy functional, to zeroth order we want to retain only those terms where the oscillating exponential phase factors vanish. Specifically, under coarse-graining, the functional essentially becomes a series of terms with ``slow'' variables multiplying phase factors of the form $e^{i\Delta K_{l}\cdot\mbfr}$, where $\Delta K_{l}$ are sums or differences in the reciprocal lattice vectors.  The zeroth order approximation results in terms where  $\Delta K_{l}\equiv 0$, referred to as the resonance condition. This constitutes the long wavelength limit.

Coarse-graining \cref{eq:fene} leads to the following complex-amplitude free-energy functional
\begin{align}
\mathcal{F}^{cg} &= \int d\mbfr \Bigg\{\lt(\epsilon +1\rt)\frac{\bar{n}^2}{2} + \frac{\bar{n}^4}{4} + \lt(\epsilon+ 3\,\bar{n}^2\rt)\,\sum_{j}^3|A_j|^2 
\nline
&-6 \bar{n}\lt[A_1A_2A_3 + \text{c.c.} \rt] 
+\frac{3}{2} \sum_{j}^3 |A_j|^2 \lt[ |A_j|^2 + 4\sum_{m>j}^3|A_m|^2 \rt] \nline
&+ \sum_{j}^{3}\lt|\lt(\nabla^2 + 2i\mbfk_j\cdot\boldsymbol{\nabla}\rt)A_{j}\rt|^2 \Bigg\},
\label{eq:compAmp}
\end{align}
where to lowest order, gradients in the average density have been neglected. This free energy functional, for 2D triangular structures, is capable of describing phase transitions and through the complex nature of the order parameters, it can describe the physics of defects, elasticity, plasticity and multiple crystal orientations self consistently. For the bcc structure which minimizes the microscopic PFC free energy functional in 3D, the coarse grained free energy will have the appropriate complex amplitudes, and their coupling, commensurate with the Fourier components and symmetry of that crystal structure. We have traded the single order-parameter theory of the PFC model for the 4-order parameter theory of the complex amplitudes, however we have also gained an increase in length scale, and corresponding time scale, allowing for large microstructure simulations to be tractable. \cref{eq:compAmp} resembles a typical Ginzburg-Landau functional of several order parameters, i.e., multi-order parameter PF models. The difference is that here, the set of order parameters, save for the average density $\bar{n}$, are complex fields. Forgoing that however, we have a $\phi^4$-theory in both the average density (van der Waals and Cahn-Hilliard theories) and the complex amplitudes (Ginzburg-Landau theory) and of course coupling between the order parameters. 

The model can be further coarse-grained by taking the phase-field limit of the complex-amplitude model of \cref{eq:compAmp}. This limit amounts to extracting {\it only} the real portions of the complex amplitudes, i.e., $A_j=\phi_j$ and $A^*_j=\phi_j$. This effectively removes most of the atomistic level information that was inherited from the original coarse-graining, giving us more conventional phase-field-like order parameters. Note however, that this is not a single order parameter theory as we have retained the individual amplitudes for each reciprocal lattice direction. The amplitude model, i.e., effective phase-field model,  reads
\begin{align}
\label{eq:phAmp}
\mathcal{F}^{ph} &= \int d\mbfr \Bigg\{\lt(\epsilon + 1\rt)\frac{\bar{n}^2}{2} + \frac{\bar{n}^4}{4} + \lt(\epsilon+ 3\,\bar{n}^2\rt)\,\sum_{j}^3\phi_j^2 \\
&-12\,\bar{n}\phi_1\phi_2\phi_3 +\frac{3}{2}\sum_{j}^3\phi_j^4 + 6\sum_j^3\sum_{m>j}^3\phi_j^2\phi_m^2 \nline
& +\sum_{j}^{3}\lt[ (\nabla^2\phi_{j})^2 + 4(\mbfk_j\cdot\boldsymbol{\nabla}\phi_j)^2 \rt]\Bigg\}.\nonumber
\end{align}
This Ginzburg-Landau free energy is described in polynomials of the order parameters $\phi_1, \phi_2$ and $\phi_3$, similar to the multi-order parameter phase-field models.  Unlike their traditional phase-field counterpart, these order parameters are not fixed between specific bulk values, but instead take a value of zero in the liquid and some finite nonzero value in the solid, where the value changes as a function of temperature, $\epsilon$, and the average density $\bar{n}$. We now  modify this free energy functional by the introduction of an additional parameter. The free energy now reads
\begin{align}
\label{eq:phAmp2}
\freem^{ph} &= \int d\mbfr \Bigg\{\lt(\epsilon + 1\rt)\frac{\bar{n}^2}{2} + \frac{\bar{n}^4}{4} + \lt(\epsilon+ 3\,\bar{n}^2\rt)\,\sum_{j}^3\phi_j^2 \\
&-12\,\bar{n}\phi_1\phi_2\phi_3 +\frac{3}{2}\sum_{j}^3\phi_j^4 + 6\sum_j^3\sum_{m>j}^3\phi_j^2\phi_m^2 \nline
& +\sum_{j}^{3}\lt[ \beta(\nabla^2\phi_{j})^2 + 4(\mbfk_j\cdot\boldsymbol{\nabla}\phi_j)^2 \rt]\Bigg\},\nonumber
\end{align}
where the coefficient $\beta$ has been introduced. We note that $\beta=1$ represents the exact derivation of the model corresponding to \cref{eq:phAmp}. This parameter controls the strength of the Laplacian-squared term, and can be motivated in a variety of ways. Here, it represents the lack of parameter diversity when the model is reduced to a one-parameter model from the CDFT formulation, and our further reduction of the model through coarse graining. The important thing here is the inclusion of the higher order gradient contribution itself, which will be shown to influence anisotropic features of the model.

The use of gradient contributions to this order is not typical of phase-field models, but have been included in a number of works as a means of regularizing the free energy, specifically in cases where highly anisotropic interfaces are of interest~\cite{Wise2005,Wheeler2006}. The Laplacian squared term provides access to smaller length scale contributions that account for atomic interactions on longer ranges. The parameter $\beta$, characterizes the region of these length scale contributions and is related to the corner energy.  We have for the equations of motion,
\begin{align}
\label{eq:dyn-n-ph}
\frac{\partial \bar{n}}{\partial t} &= M_{\bar{n}} \nabla^2 \fxfy{\freem^{ph}}{\bar{n}} \\
&=  M_{\bar{n}}\Bigg\{\lt(\epsilon + 1\rt)\bar{n} + \bar{n}^3 + 6\,\bar{n}\sum_{j}^3\phi_j^2 - 12\phi_1\phi_2\phi_3 \Bigg\},\nonumber
\end{align}
\begin{align}
\label{eq:dyn-phi-ph}
\frac{\partial \phi_{j}}{\partial t} &= -M_{j}\fxfy{\freem^{ph}}{\phi_j}, \qquad\qquad \forall\,\,j=1,2,3\\
&= -M_j\Bigg\{\lt(\epsilon + 3\,\bar{n}^2\rt)2\phi_j - 12\,\bar{n}\prod_{m \ne j}^{3}\phi_m \nline
&+ 6\phi_j\lt[\phi_j^2 + 2\sum_{m\ne j}^{3}\phi_m^2\rt] +2\beta\nabla^4\phi_{j} - 8(\mbfk_j\cdot\boldsymbol{\nabla})^2\phi_j \Bigg\},\nonumber
\end{align}
where $M_{\bar{n}}$ and $M_j$ are the mobility coefficients.

\section{Calculating surface energy and Wulff shape}
\label{surf-ene}
This section describes our methods of calculating the surface properties from our simulations. We start by describing interfacial excess energy calculations of the proposed model. We follow with the fitting scheme employed here and the Wulff construction.
\subsection{Interfacial excess energy}
Interfacial energy is a thermodynamic excess quantity. Quantitative calculation of surface excess quantities is best performed using the grand potential. The grand potential is defined as $\Omega = - pV$, where $p$ is the pressure and $V$ the volume, in the absence of surfaces and other excess artifacts. In the presence of an interface, the potential reads
\begin{align}
\label{eq:grand-pot}
\Omega =  - pV + \gamma A,
\end{align}
where $\gamma$ is the surface energy and $A$ the area of the interface. The additional term accounts for the work done to create the interface structure. The excess of the grand potential is then nothing more than the difference between the total grand potential of a system and that of a bulk phase. Thus, the surface energy is simply defined as
\begin{align}
\label{eq:surf-omega}
\gamma \equiv \frac{\Omega  - \Omega_{\nu}}{A},
\end{align}
where $\nu$ is either one of the bulk liquid or solid phases. Traditionally, the determination of the interfacial energy requires intimate knowledge of the exact position and size of the interfacial area. To accomplish this, the Gibbs criterion, or Gibbs dividing surface, method has been applied extensively. With the introduction of diffuse interface approaches, the fields employed as order parameters in such theories implicitly possess information about the the position and size of the interfacial area. In our current theory, the grand potential function becomes a functional defined by
%
%
%
%
\begin{align}
\Omega^{ph}\lt[\{\phi_j\},\bar{n}\rt] = \freem^{ph}\lt[\{\phi_j\},\bar{n}\rt] - \mu \int d\mbfr \,\bar{n} (\mbfr),
\end{align}
with $\mu$ the equilibrium chemical potential, the natural variable of the grand potential. Using \cref{eq:surf-omega}, in equilibrium the surface energy is defined by
\begin{align}
\label{eq:surf-ene}
\gamma = \frac{1}{A}\int_V d\mbfr \lt( f^{ph} - f^{ph}_{s} \frac{\bar{n}(\mbfr) - \bar{n}_{\ell}}{\bar{n}_{s}-\bar{n}_{\ell}} + f^{ph}_{\ell} \frac{\bar{n}(\mbfr) - \bar{n}_{s}}{\bar{n}_{s}-\bar{n}_{\ell}} \rt) 
\end{align}
where $f^{ph}$ is the  integrand of \cref{eq:phAmp2}. We have used the tangent rule for the chemical potential and $f^{ph}_{\nu}$ and $\bar{n}_{\nu}$ represent the equilibrium values for the free energy and solid and liquid densities respectively. It is worth mentioning that the above equation can also be used as a general construction to describe the excess energy of interfaces when the constraint of equilibrium is relaxed. In \cref{derive-surfene}, we further explore this derivation and another using a first integral methodology along with a description of the corner energy.

For different orientations of the interface, $\Omega^{ph}$ necessarily assumes a different value since the amplitudes, $\{\phi_j\}$, and the average density, $\bar{n}$, evolve to different profiles. This leads to anisotropy of the surface tension. This is also included naturally in our free energy description through the gradient terms. When considering traditional PF models, specific functions for the gradient coefficient need to be added in order to capture surface energy anisotropy. Our effective phase-field model, \cref{eq:phAmp2}, captures anisotropic surface energy directly by inclusion of atomistic level information,   specifically through the reciprocal lattice vectors. Particularly, we note the nature of the square gradient term in \cref{eq:phAmp2}, which can be written as 
\beq
(\mbfk_j\cdot\boldsymbol{\nabla}\phi_j)^2 = (\mbfk_j\cdot \mathbf{n}_j)^2 |\boldsymbol{\nabla}\phi_j|^2,
\eeq
where $\mathbf{n}_j$ is the normal vector to the interface of each of the amplitudes with respect to a reference frame. We can loosely interpret this as collectively contributing some functional form with $(\mbfk_j\cdot \mathbf{n}_j)^2 \approx w(\mathbf{n})^2$. At once we realize that our gradient energy coefficient, which is based on and derived from lattice symmetry, is akin to gradient energy functions often used in phase-field models to describe anisotropy.
%
%
%
%
\subsection{Fitting surface energy}
After computing the surface energy, we fit the data to a form that is more amenable to mathematical manipulation and analysis. Generally, data of this kind is fit to Kubic harmonics (a linear combination of Spherical harmonics), as a function of the interface orientations. In 3D, these are special functions defined as a series of functions on the surface of a sphere, while in 2D this amounts to a Fourier series defined on a circle. As a function of orientation, i.e., the direction normal to the interface, for a triangularly symmetric crystal (6-fold symmetric), the surface energy can be expanded in the following form
\beq
\label{eq:se-expan}
\gamma(\theta) = \gamma_0\lt( 1 + \sum_{u=1}^N \epsilon_{6u}\cos{\lt(6u\,\theta\rt) } \rt),
\eeq
where $\gamma_0$ is the isotropic surface energy for a planar interface, $N$ is the order of the fit and $\epsilon_{6u}$ is the anisotropy parameter for the given harmonic $6u$. In approaches where the amplitude formalism has been used to determine surface energy, particularly where the anisotropic form is sought as input into the phase-field method~\cite{Majaniemi2009,Provatas2010}, only the first order term $N=1$ from \cref{eq:se-expan} has been used. The work of Jugdutt et al.~\cite{Jugdutt2015}, in a study of solute effects on anisotropy, is the only work known to the authors where higher order terms have been considered. In this study, we will consider an order, $N$, necessary to capture the full breadth of the surface energy. Once a form of \cref{eq:se-expan} has been determined, the stiffness immediately follows,
%
%
%
%
%
\beq
\label{eq:stiff-expan}
\Gamma(\theta) = \gamma(\theta) + \gamma^{\prime\prime}(\theta).
\eeq

Using the surface energy and stiffness measurements, we can determine the Wulff shape which minimizes the free energy. The shapes in conjunction with the surface energy and stiffness can be used as a metric to determine if the system undergoes a transition to missing orientations and eventually complete faceting under the right conditions. For example, a condition for missing orientations is given by the development of cusps in the polar plot of the surface energy. The appearance of cusps means that $\gamma^{-1}(\theta)$ is no longer convex and indicates the existence of facets at those orientations. The non-convexity that leads to these cusps and the initiation of facet formation can also be determined through a stiffness criterion, i.e.,  $\Gamma(\theta) < 0$~\cite{Sekerka2005}. Lastly, we can also visually inspect the Wulff shape, as the appearance of ``ears'' in the Wulff shape is an indication of missing orientations.

\subsection{Wulff construction}
The Wulff construction begins by first plotting the surface energy, $\gamma(\theta)$. On every point on the $\gamma$-surface, a plane is drawn through the point and perpendicular to the radius vector. The Wulff construction is then the inner envelope of all such planes, so-called Wulff planes. In 3D, $\xi$-vector of Hoffman and Cahn~\cite{Hoffman1972,Cahn1974} has greatly improved the ease by which the Wulff construction is done~\cite{Sekerka2005,Wheeler2006}. In 2D Cartesian coordinates, which our results will be presented in, the Wulff construction has a simple mathematical form since the surface energy depends only on a single angle value. To derive this form, we begin with the Gibbs-Thomson equation, which reads
\beq
\label{eq:gibbs-thom}
\widetilde{\Delta \mu} = \lt( \gamma(\theta) + \gamma^{\prime\prime}(\theta) \rt) {\cal K},
\eeq
where $\widetilde{\Delta \mu}$ is a dimensionless difference of chemical potentials between the solid and liquid phases, and for a pure material in the limit of small undercooling is $\sim (T_m-T)/T_m$, and ${\cal K}$ is the Gaussian curvature defined as 
\beq
\label{eq:curv}
{\cal K} = -\frac{\partial_{\theta\theta}s}{\lt(1+ \lt( \partial_{\theta}s\rt)^2\rt)^{\frac{3}{2}}},
\eeq
where $s$ is the arc length and maps out the equilibrium crystal shape. This equation can be solved to yield the following parametric solutions for the interface shape~\cite{Burton1951,Voorhees1984,Saito1998}
\begin{align}
\label{eq:wulff-cart}
x(s) & = \gamma(\theta)\cos{\theta} - \gamma^{\prime}(\theta)\sin{\theta}\nline
y(s) & = \gamma(\theta)\sin{\theta} + \gamma^{\prime}(\theta)\cos{\theta},
\end{align}
where $x$ and $y$ are the Cartesian coordinates of the arc length. For a sufficiently low temperature, the Wulff shape assumes a polygonal structure corresponding to the lattice symmetry, while at larger temperatures close to the melting temperature it assumes a more isotropic, circular shape. Results indicating this behavior are discussed below.

\section{Results}
\label{results}
In this section, we apply the amplitude model of \cref{eq:phAmp2,eq:dyn-n-ph,eq:dyn-phi-ph} to study the surface properties of the effective phase-field model derived from PFC. Using 1D calculations, as a function of temperature, the surface energy will be determined using \cref{eq:wulff-cart}, from which the Wulff shapes will be calculated. Combined with the criterion for the convexity of the surface stiffness $\Gamma(\theta)$~\cite{Godreche1991,Sekerka2005}, we shall determine the transition to missing orientations and formation of facets in this system. Following the 1D calculations of surface energy and Wulff shapes, we perform 2D simulations to corroborate the surface energy calculations in a regime where dendritic growth is observed.

\subsection{1D surface energy calculations}
\label{results-1D}
Examining the surface energy, and subsequently the Wulff shapes, will allow us to determine the anisotropic features of this system leading to missing orientations. While the transition is driven by changing temperature, two additional effects of the effective phase-field model are worth exploring. They are the effects of the regularization parameter, $\beta$, and length scale resolution.

The regularization parameter, $\beta$, was added phenomenologically, in order to afford us some control over the corner energy and its possible effects on the transition to the missing-orientation regime. We note that numerically, the stability and equilibrium properties of this coarse-grained model, like other PF-type methodologies, can be considered to be a balancing effect between bulk and surface contributions. Having, in an {\it ad hoc} manner, altered the inherent contribution of one of these terms requires exploration. In this regard, we have considered various corner energy gradient coefficients which are given in \cref{tab:params-beta-dx}.

\begin{table}[ht]
\centering
  \caption{The set of regularization parameters and grid resolutions (with $a=4\pi/\sqrt{3}$) examined in our study.}
  \begin{tabular}{ c @{\hskip 0.25in} c@{\hskip 0.15in} }
  \hline\hline
  Corner energy ($\beta$) & Resolution ($\Delta x$) \\
  \hline
$\beta_1 = 0$ & $\Delta x_1 = a/16$ \\
$\beta_2 = 1 \times 10^{-6}$ & $\Delta x_2 = a/12$\\
$\beta_3 = 1 \times 10^{-4}$ & $\Delta x_3 = a/8$ \\
$\beta_4 = 1 \times 10^{-2}$ & $\Delta x_4 = a/4$ \\
$\beta_5 = 0.1 $ & $\Delta x_5 = a/2$ \\
$\beta_6 = 0.2$ \\
$\beta_7 = 0.5$ \\
$\beta_8 = 1$ \\
    \hline
  \end{tabular}
  \label{tab:params-beta-dx}
\end{table}

The motivation for moving to a coarse-grained model description was the access to larger length scales to explore microstructural features at relevant scales. Given the coarse-grained nature of the model, use of larger numerical grid resolutions than that of the original PFC model for simulations would be natural. However, the physics of interest here is multiscale in nature, ranging from the pseudo-atomistic to the mesoscale. We therefore must investigate what resolution is sufficient to capture the full length scale range involved to best quantitatively capture the associated physics. To do so, we will consider various numerical grid spacings, $\Delta x$, presented in \cref{tab:params-beta-dx}. We note that the aim here is in exploring in what regime we can safely use coarser resolutions and conversely where finer resolutions are necessary in order to capture the essential physical features.

Simulations are performed by initializing profiles of the density, $\bar{n}$, and the amplitudes $\{\phi_j\}$, in a 1D domain at the equilibrium values of the solid and liquid phases as dictated by the  numerically computed phase diagram [see \cref{pfc-func-equil}] for a given temperature $\epsilon$. \Cref{eq:dyn-n-ph,eq:dyn-phi-ph} are then solved to correct for any artifacts of the initial conditions until equilibrium is reached, which is determined by convergence of the grand potential, $\Omega^{ph}\lt[\{\phi_j\},\bar{n}\rt]$, over time. Once equilibrium is attained, the surface energy is calculated per \cref{eq:surf-ene} for each given orientation. In the following subsections we evaluate the consequences of these simulations, considering first the surface energy and Wulff construction and finally the stiffness.

\subsubsection{Surface energy and Wulff shapes}
Once the surface energy is determined from the 1D simulations, we perform a fit of the data to \cref{eq:se-expan}, where the order of the fit is determined on the basis of minimizing the standard error. Once we have the fit, it is straight forward to determine the stiffness viz. \cref{eq:stiff-expan}.
\begin{figure*}[h]
\centering
\begin{tabular}{ccccc}
	\includegraphics[width=30mm]{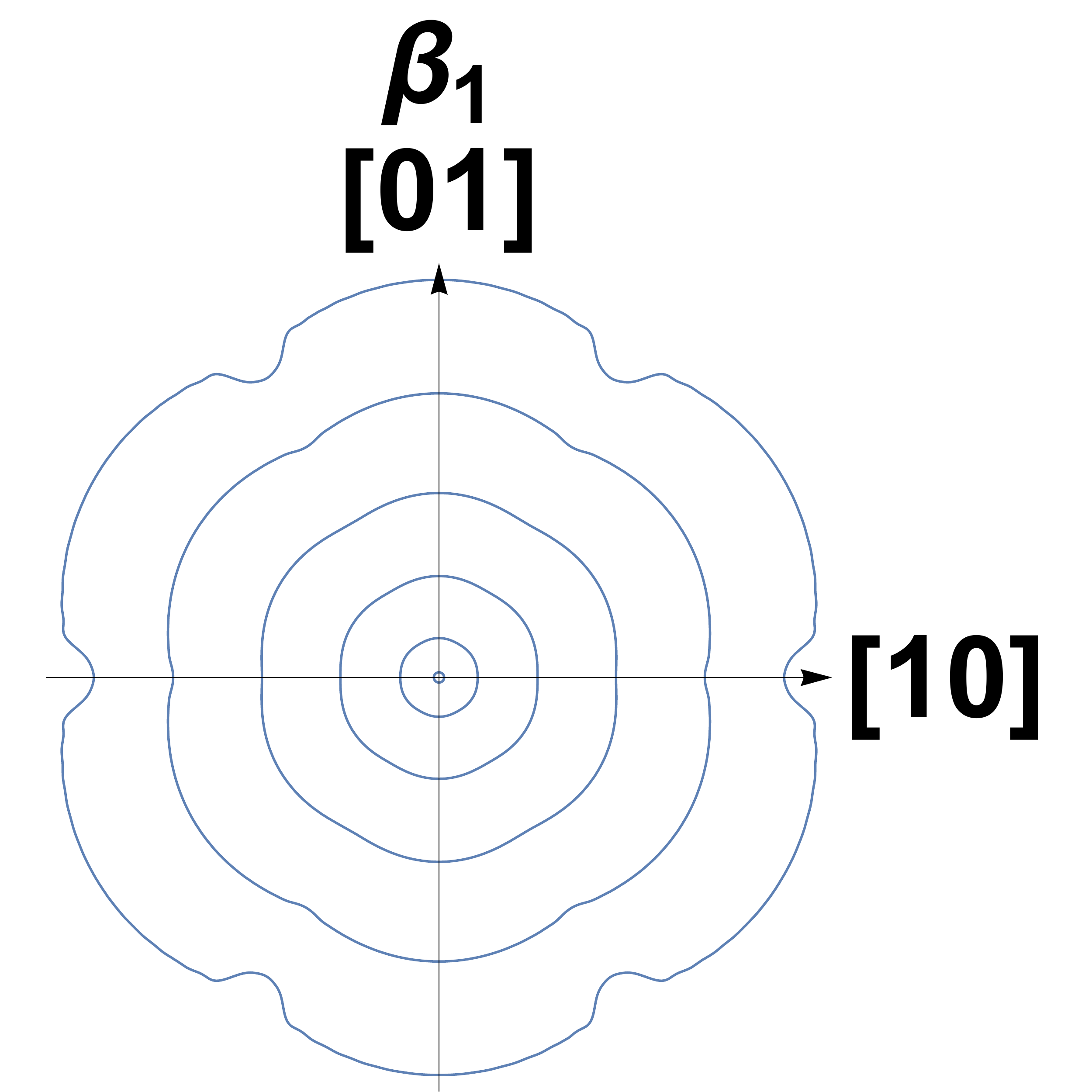} & \includegraphics[width=30mm]{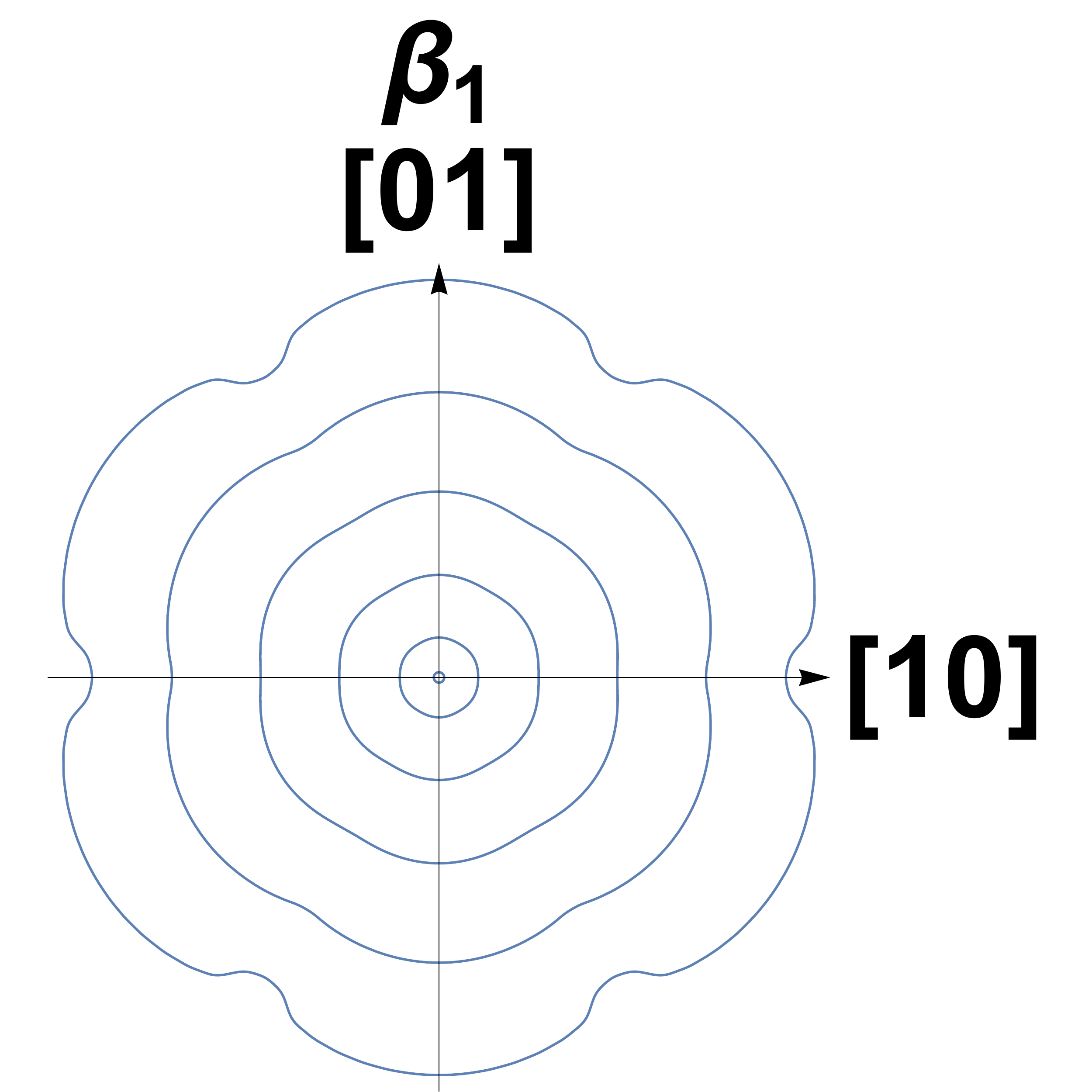} & \includegraphics[width=30mm]{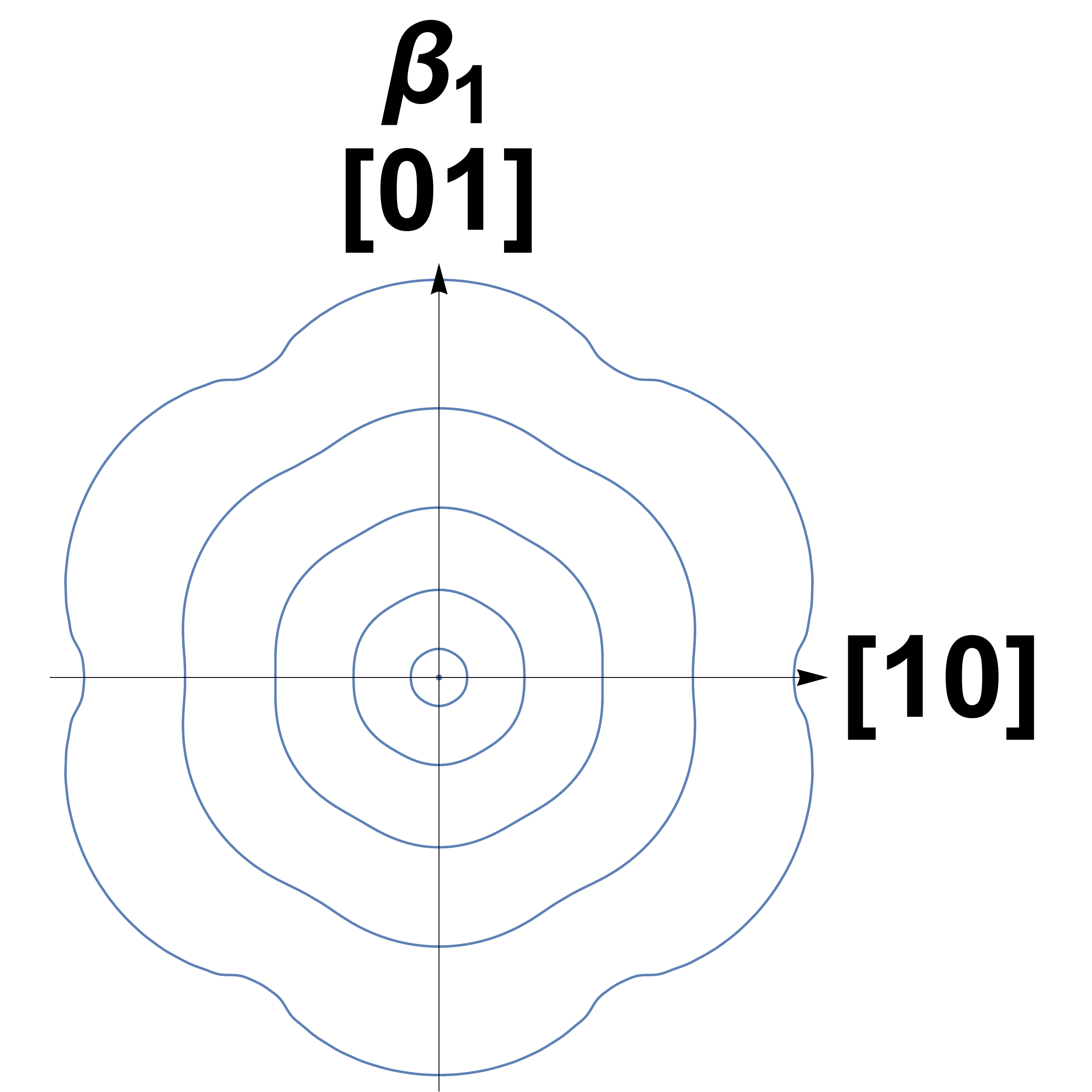} & \includegraphics[width=30mm]{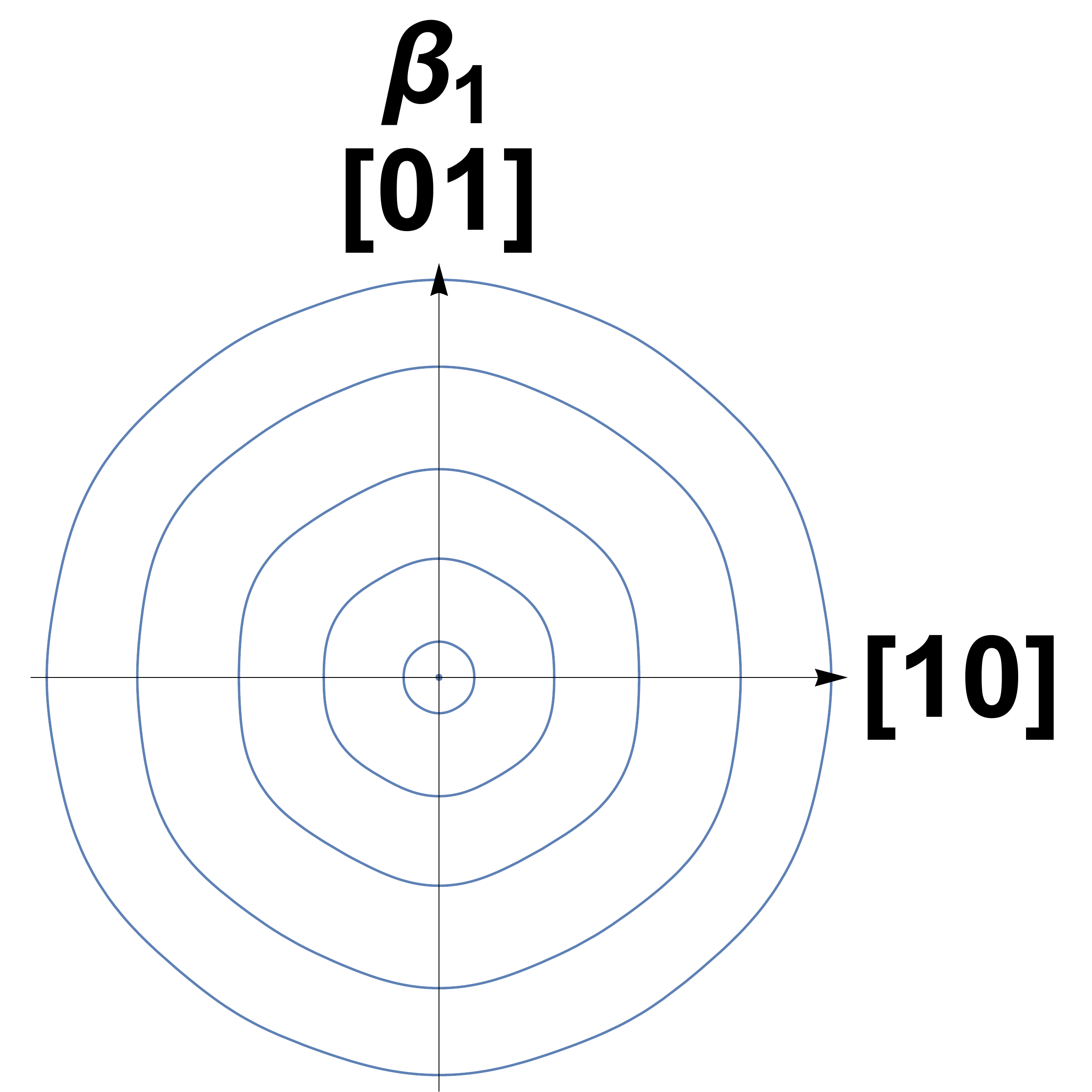} \\
	\includegraphics[width=30mm]{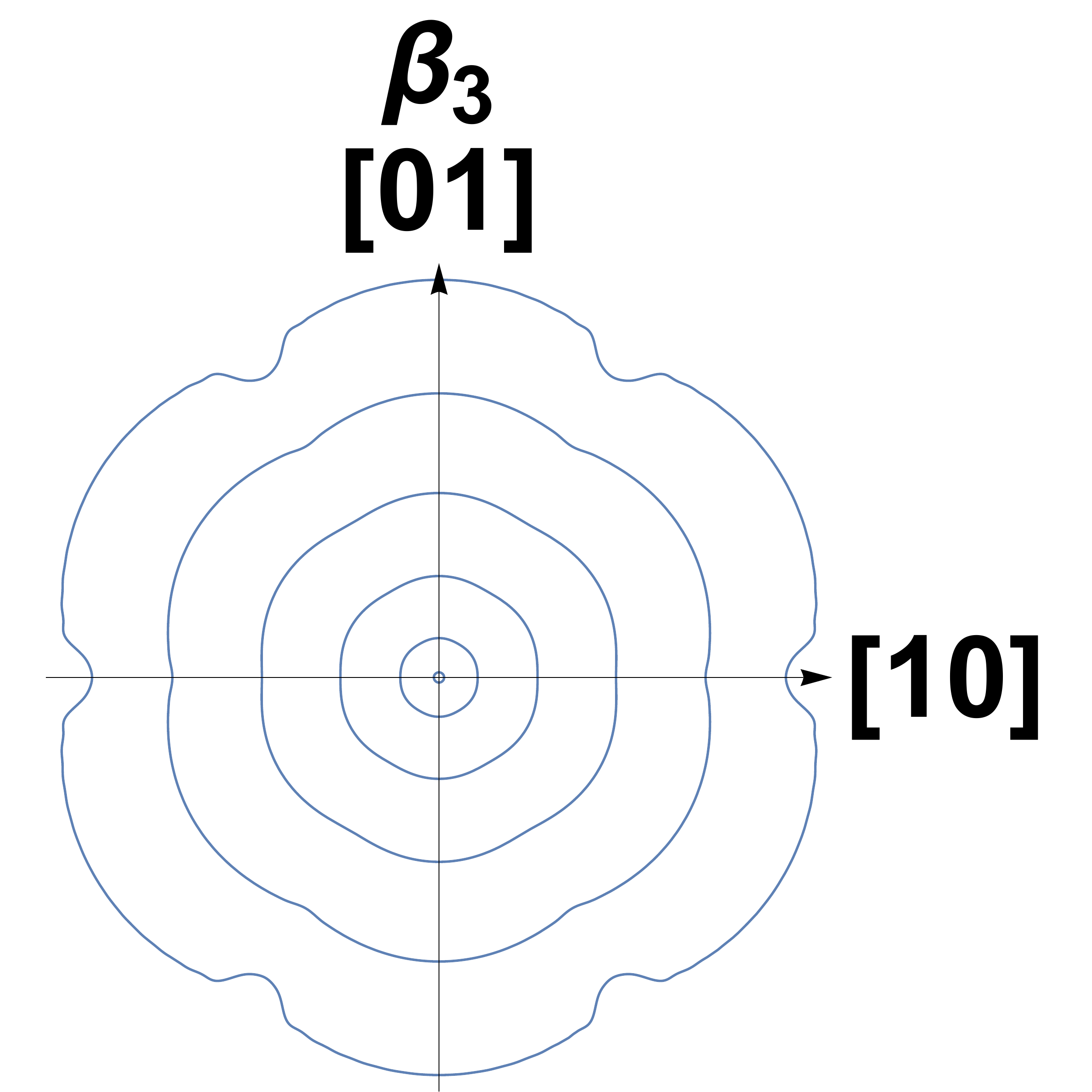} & \includegraphics[width=30mm]{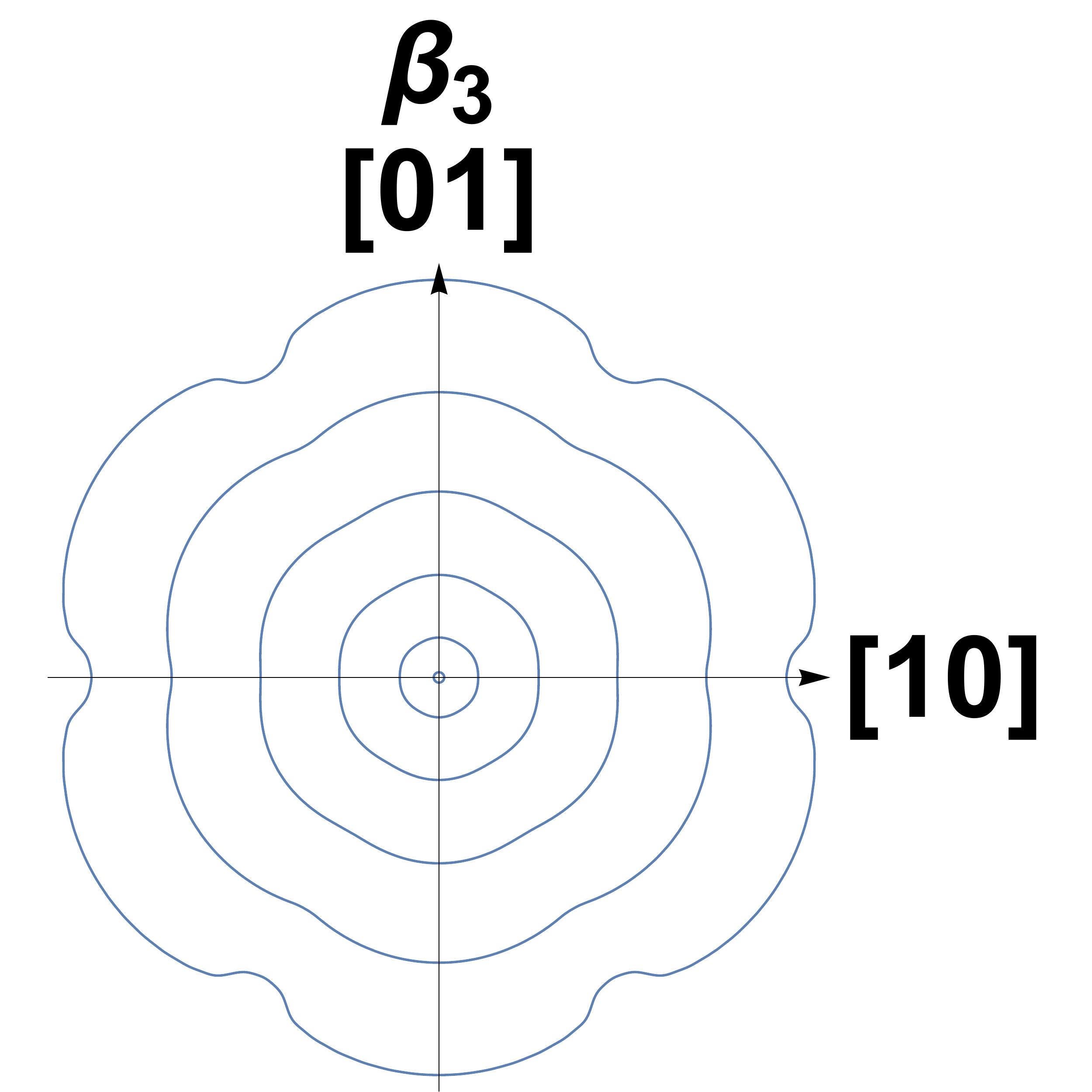} & \includegraphics[width=30mm]{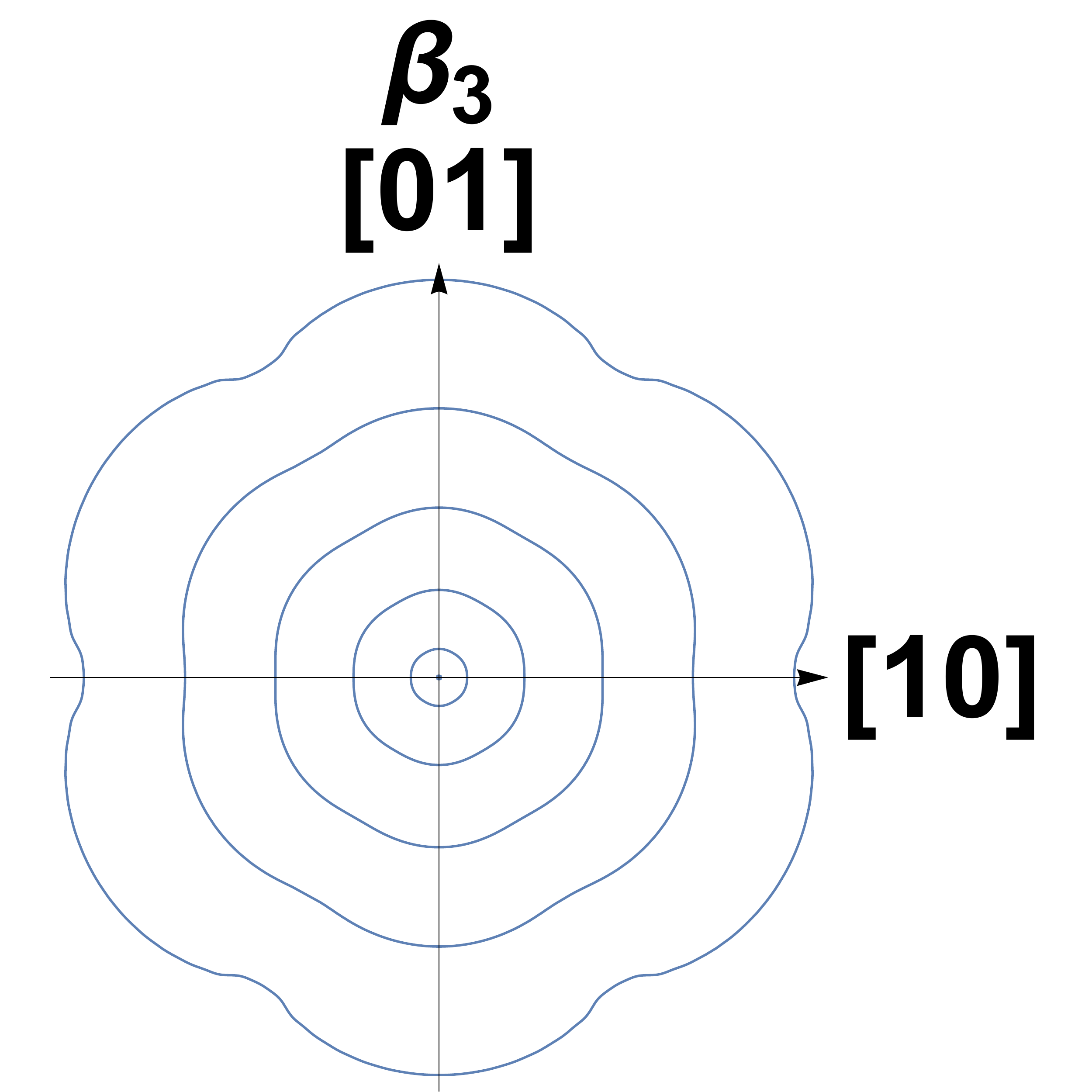} & \includegraphics[width=30mm]{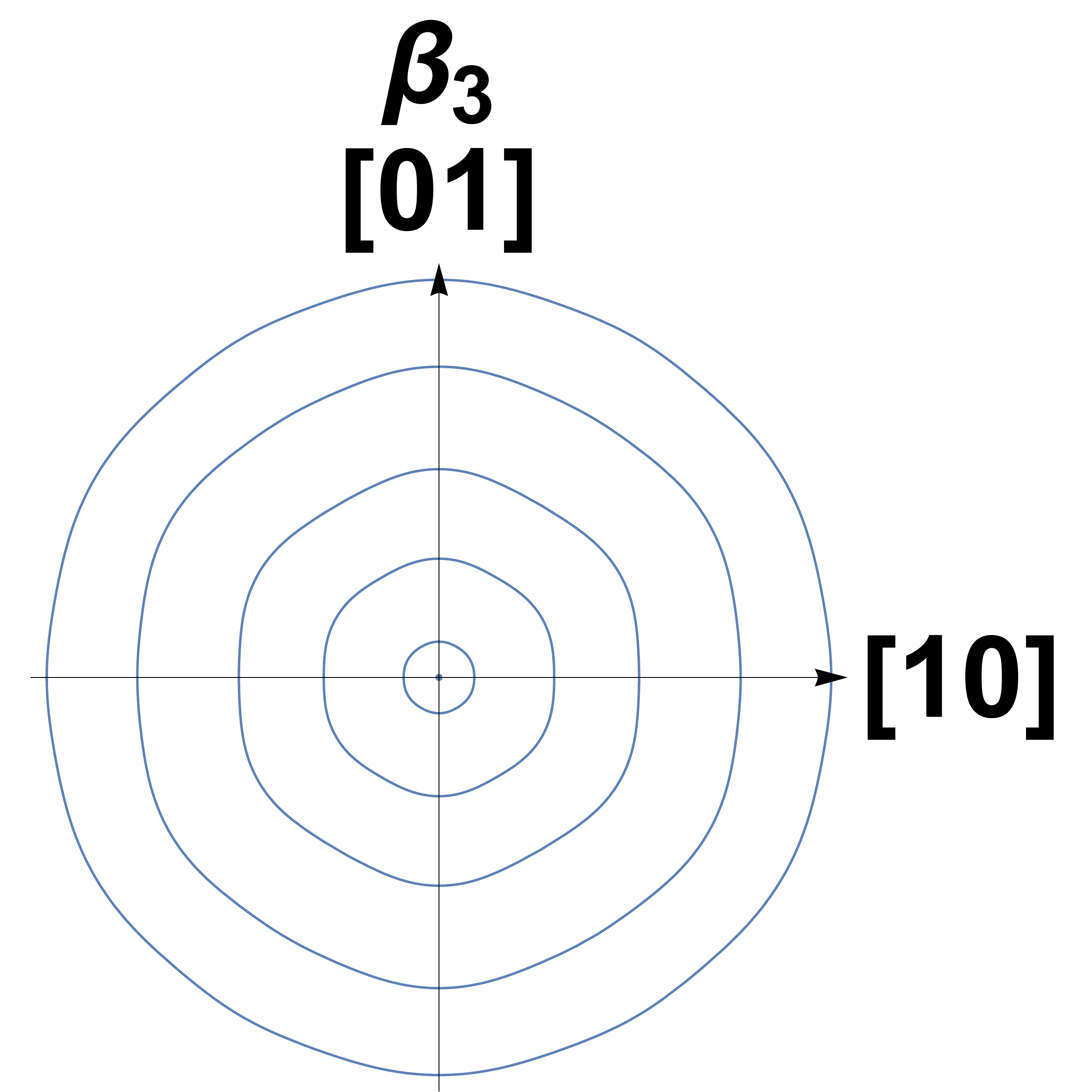} \\
	\includegraphics[width=30mm]{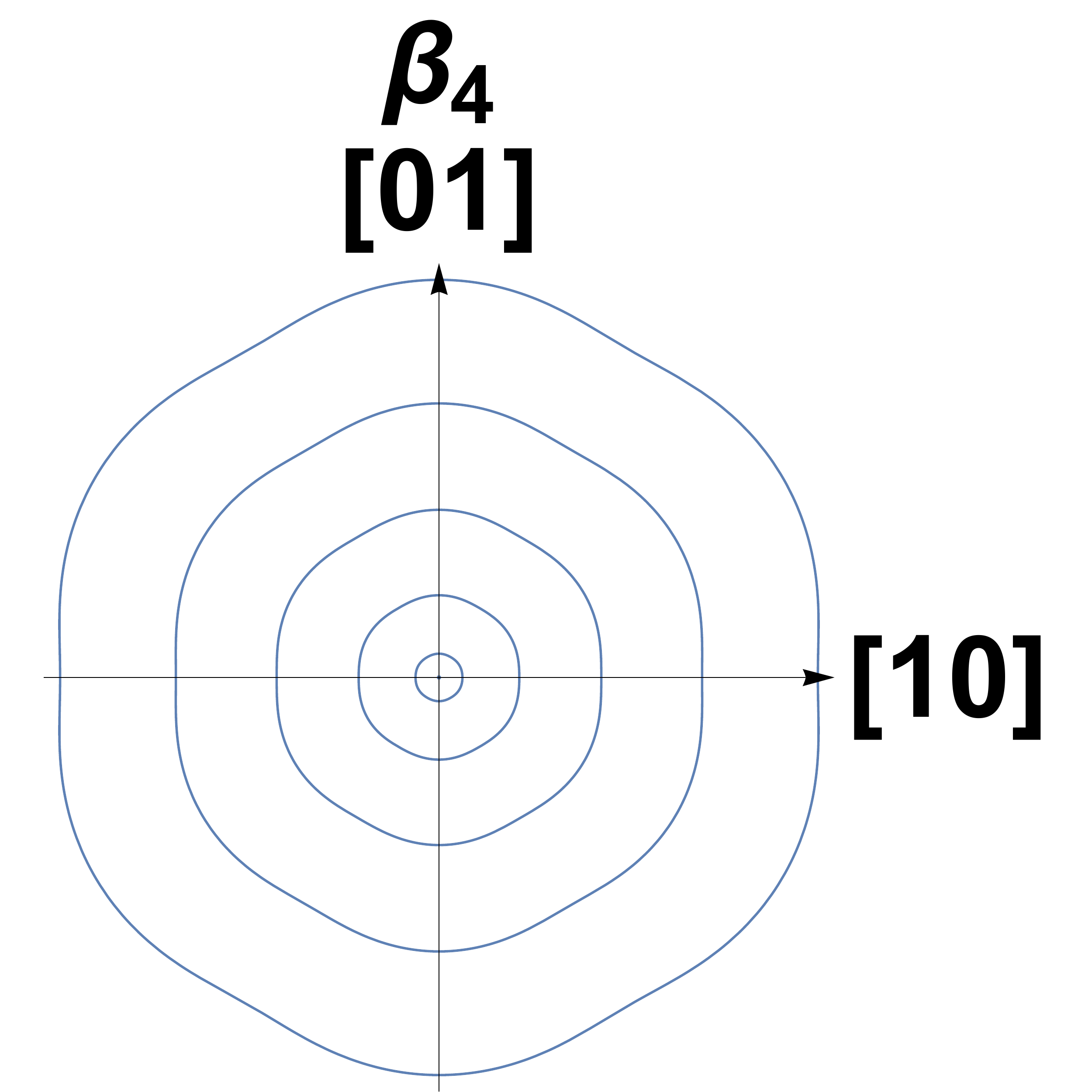} & \includegraphics[width=30mm]{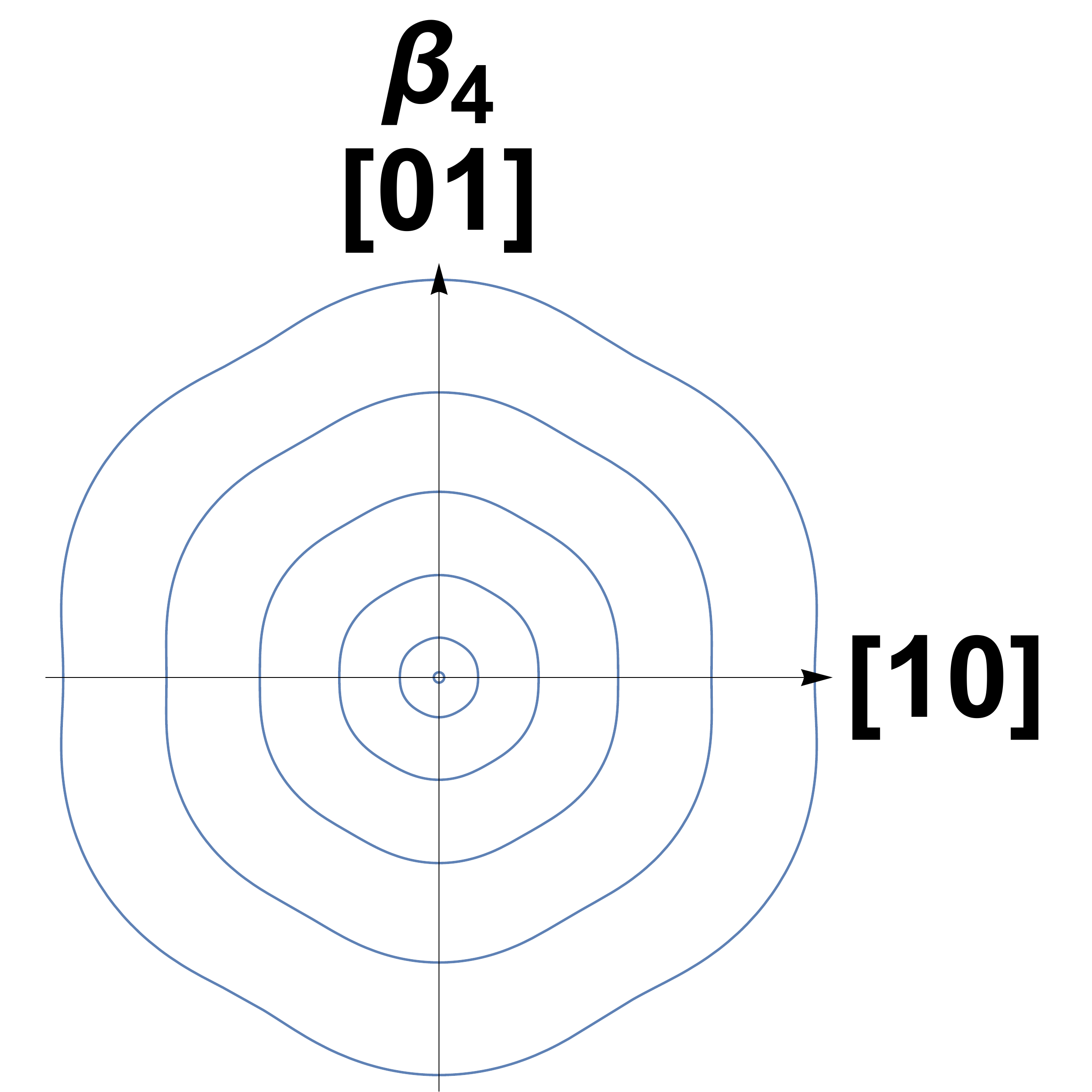} & \includegraphics[width=30mm]{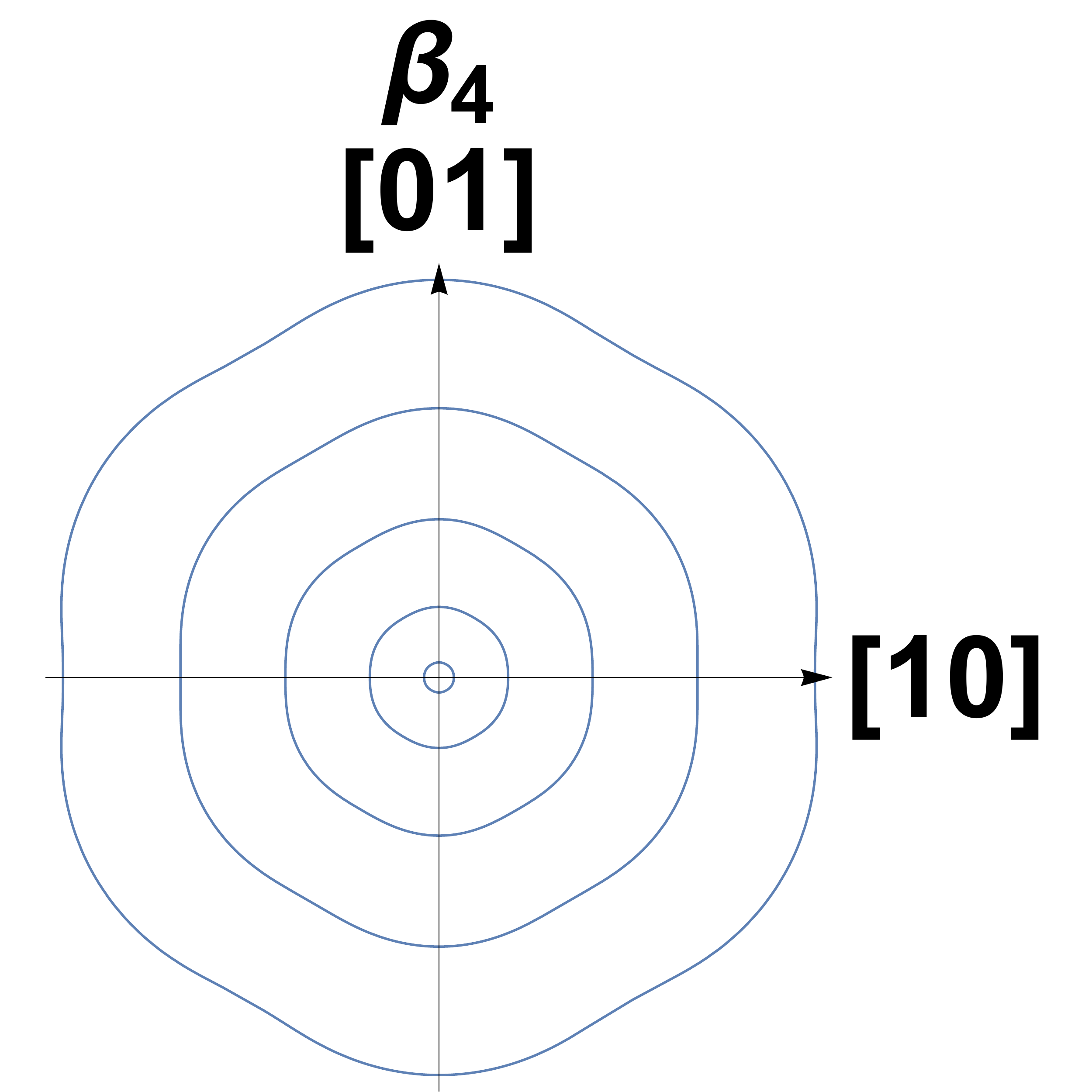} & \includegraphics[width=30mm]{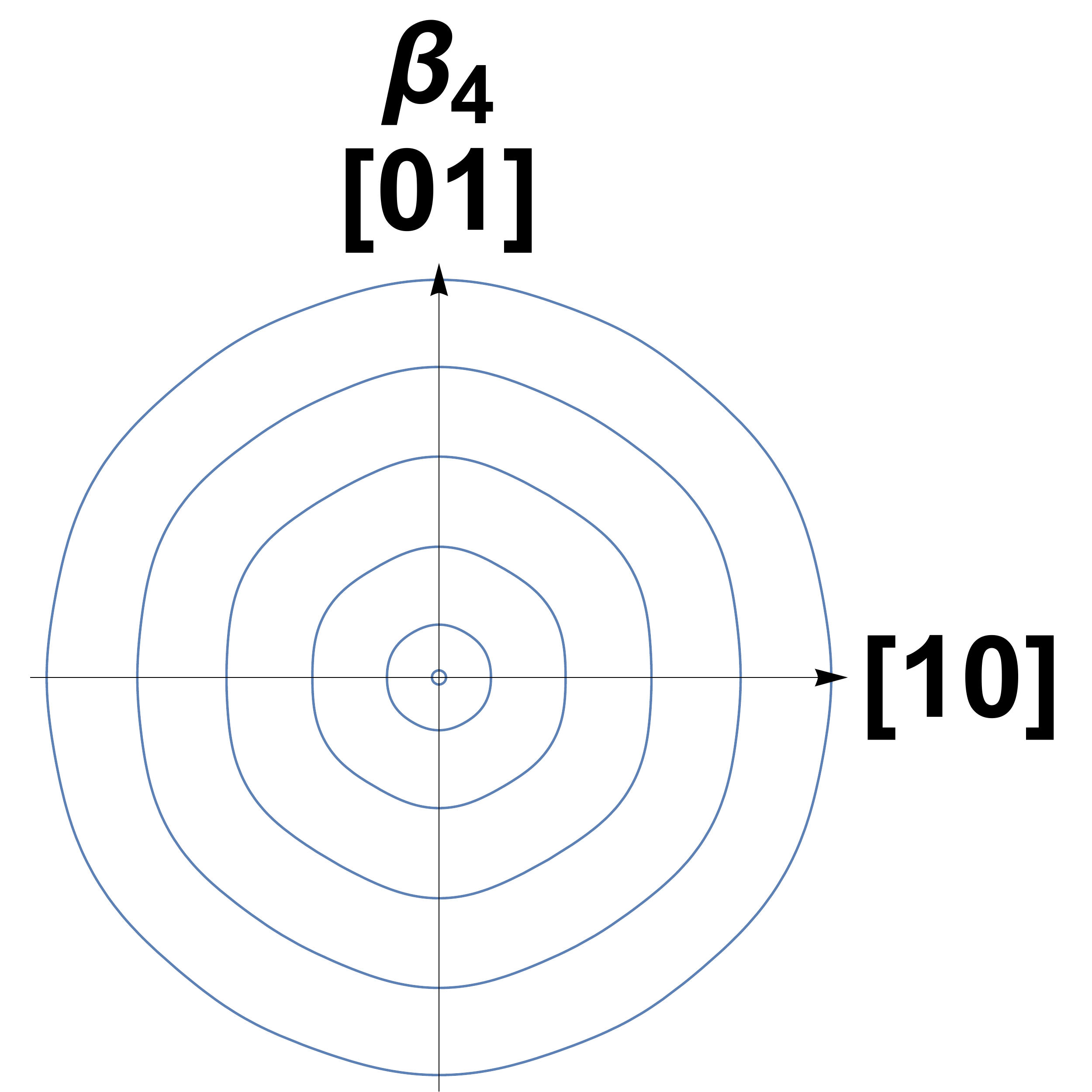}  \\
	\includegraphics[width=30mm]{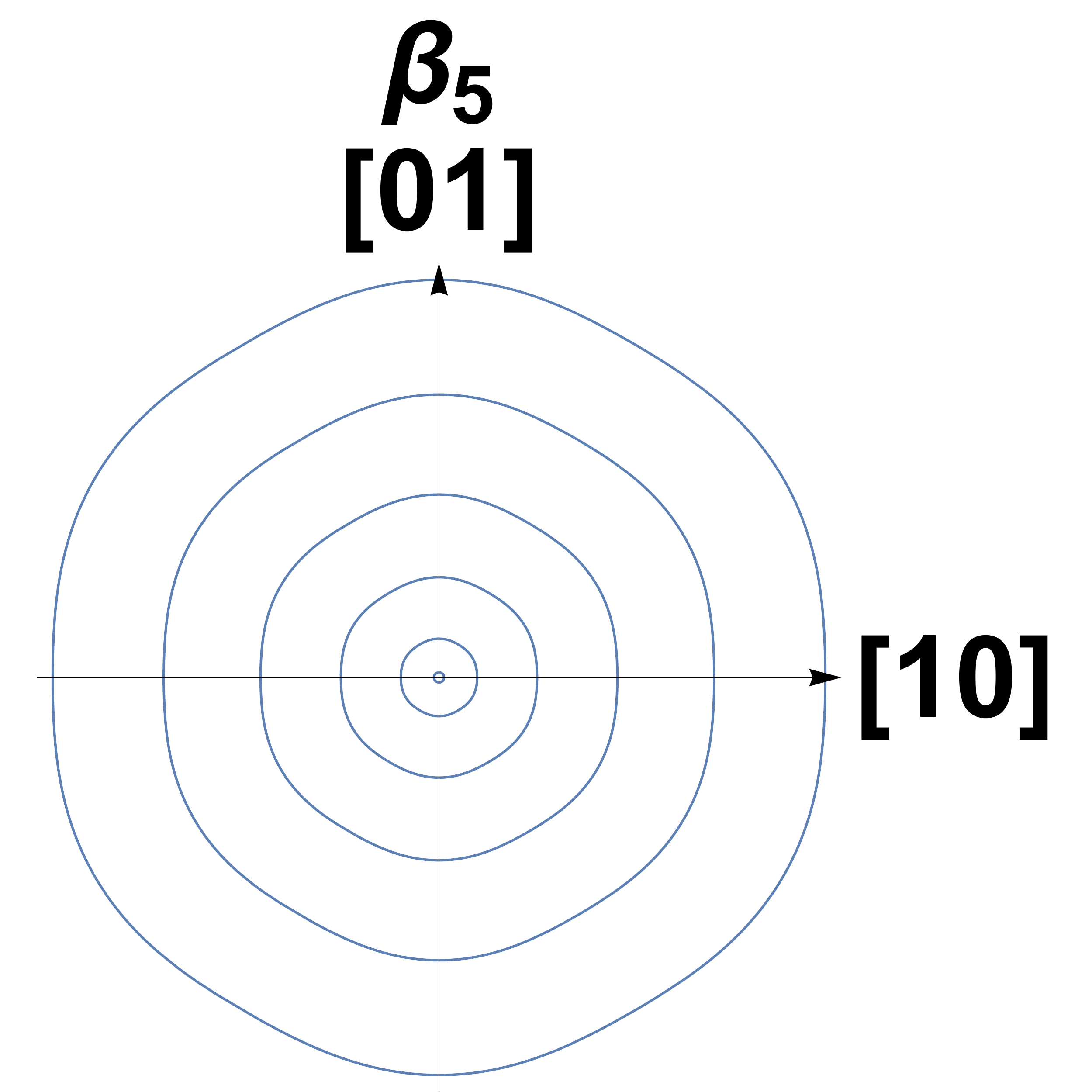} & \includegraphics[width=30mm]{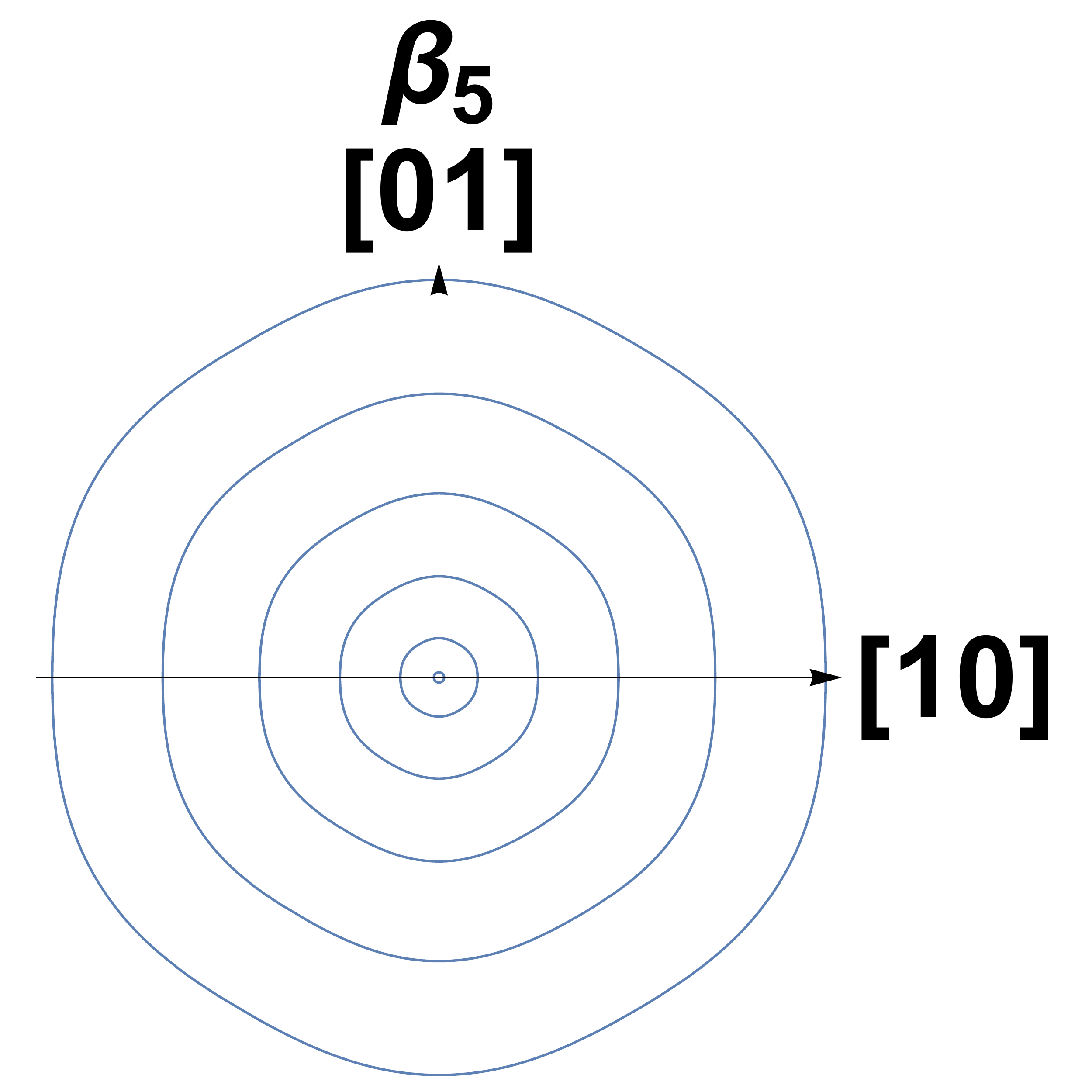} & \includegraphics[width=30mm]{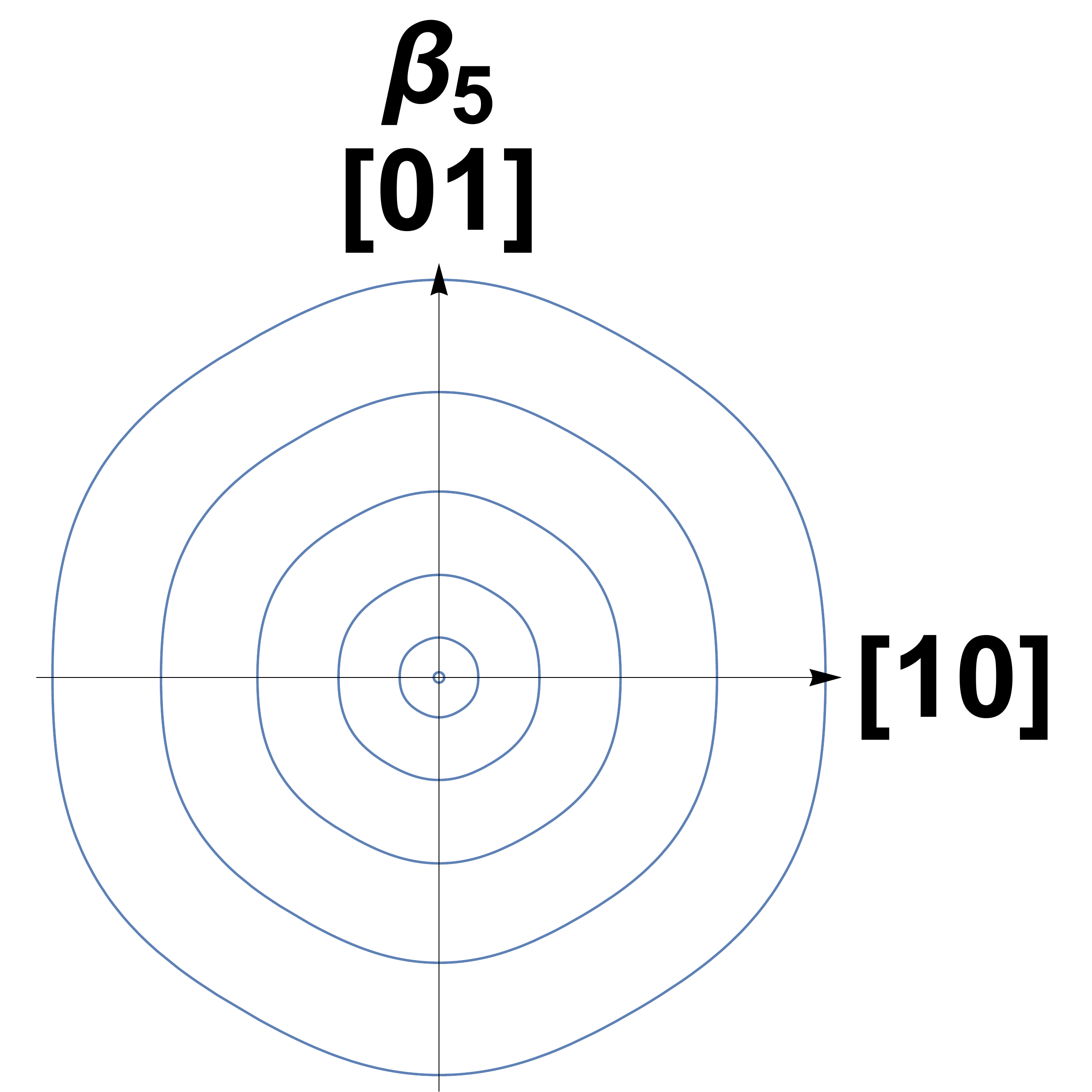} & \includegraphics[width=30mm]{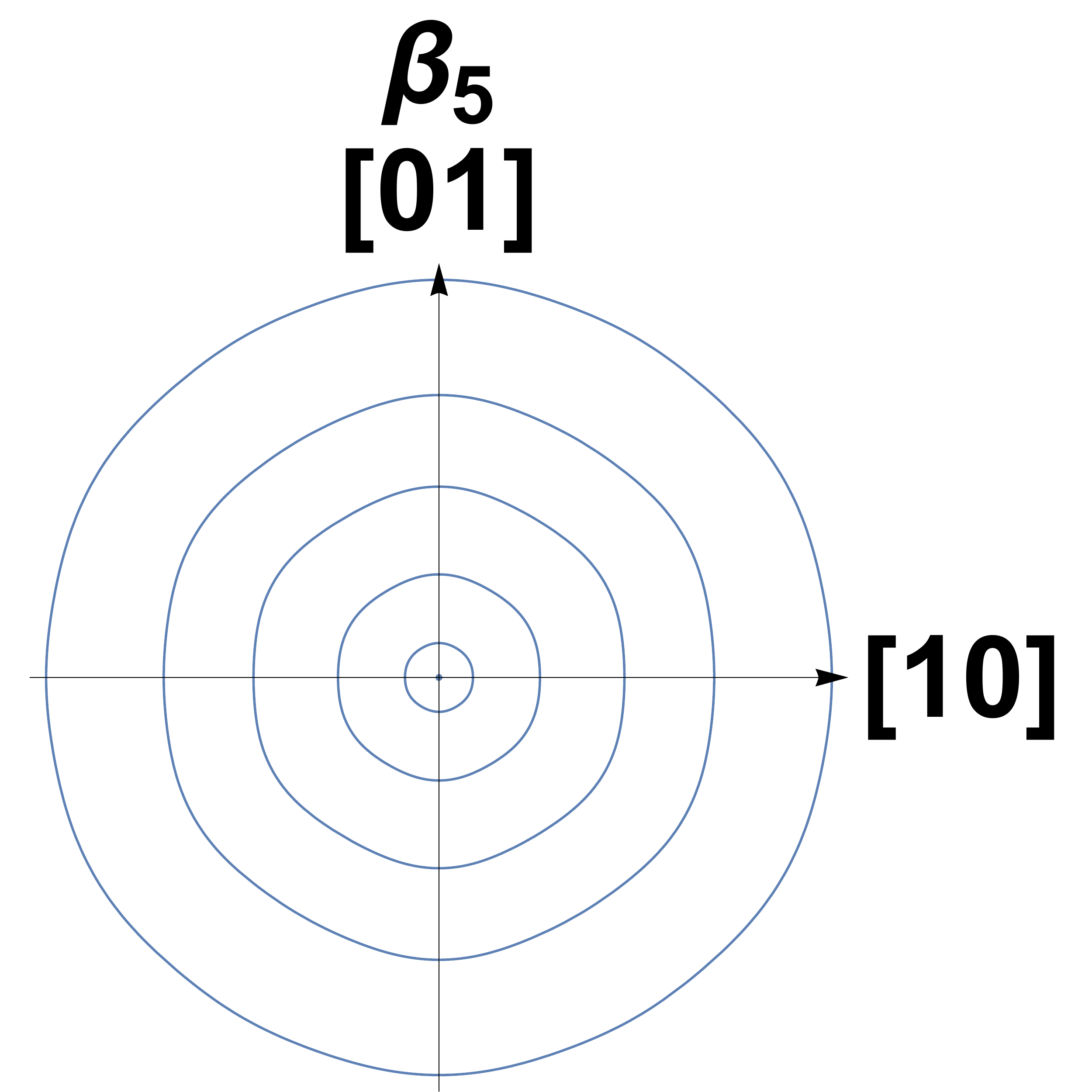} \\
	(a) $\Delta x_1$ & (b) $\Delta x_2$ & (c) $\Delta x_3$ & (d) $\Delta x_4$ \\
\end{tabular}
\caption{Polar plots of the surface energy, as calculated using 1D simulations and fitted using \cref{eq:se-expan}. $\beta$ increases from top to bottom, while $\Delta x$ increases from left to right. In the panels, temperature decreases from the center going outward, with the outer plots representing successively lower temperatures.}
\label{fig:surf-polar}
\end{figure*}

\begin{figure*}[ht]
\centering
\begin{tabular}{ccccc}
	\includegraphics[width=30mm]{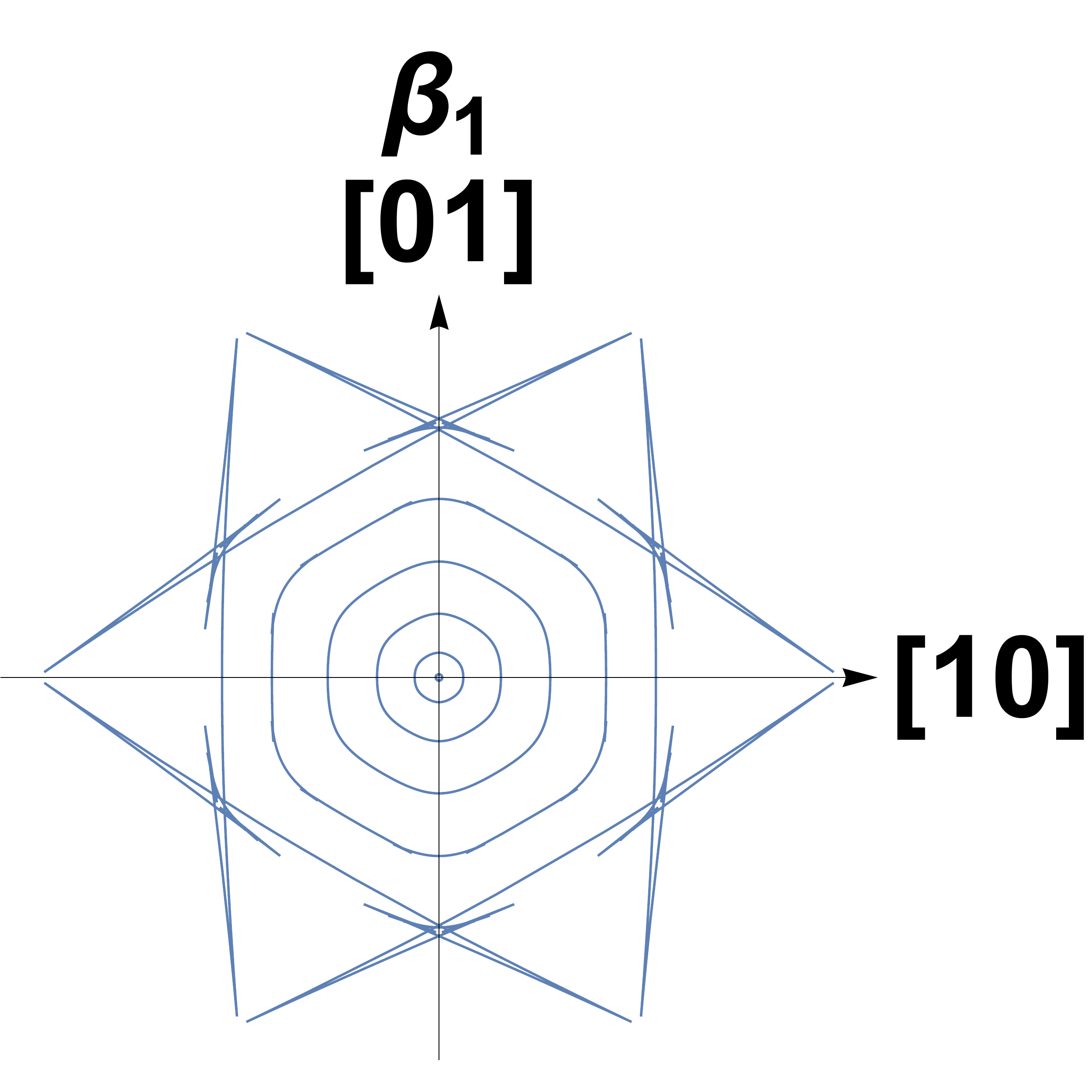} & \includegraphics[width=30mm]{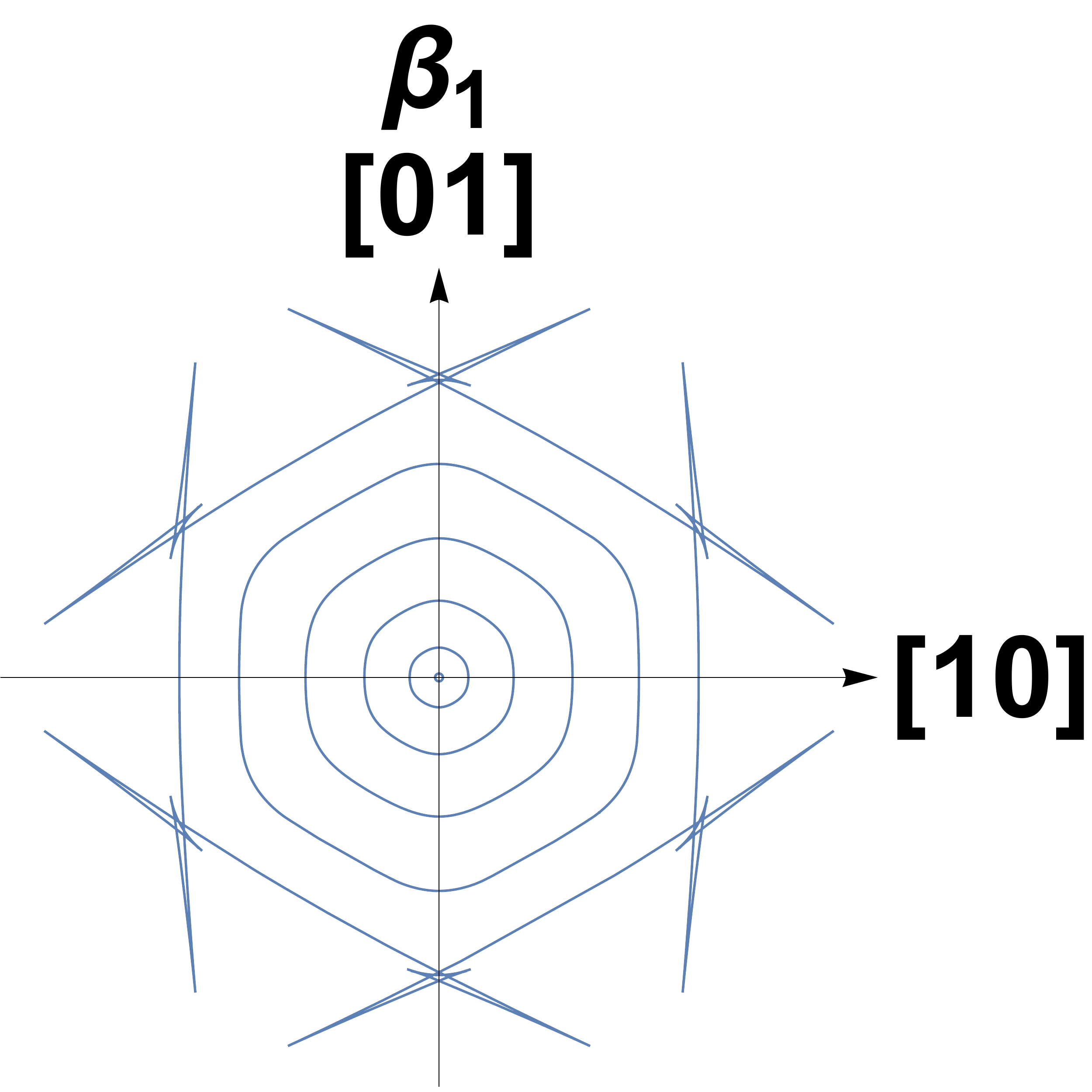} & \includegraphics[width=30mm]{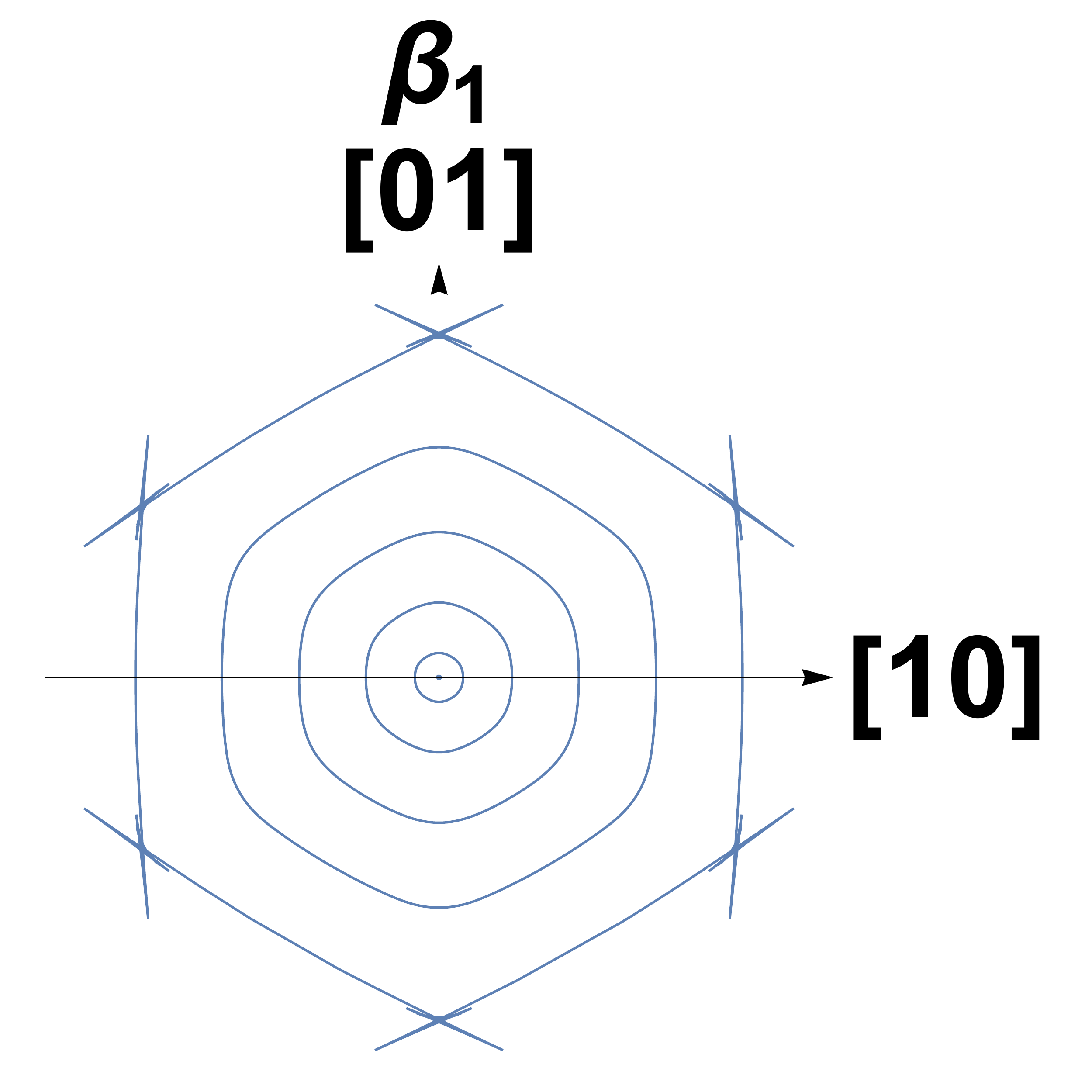} & \includegraphics[width=30mm]{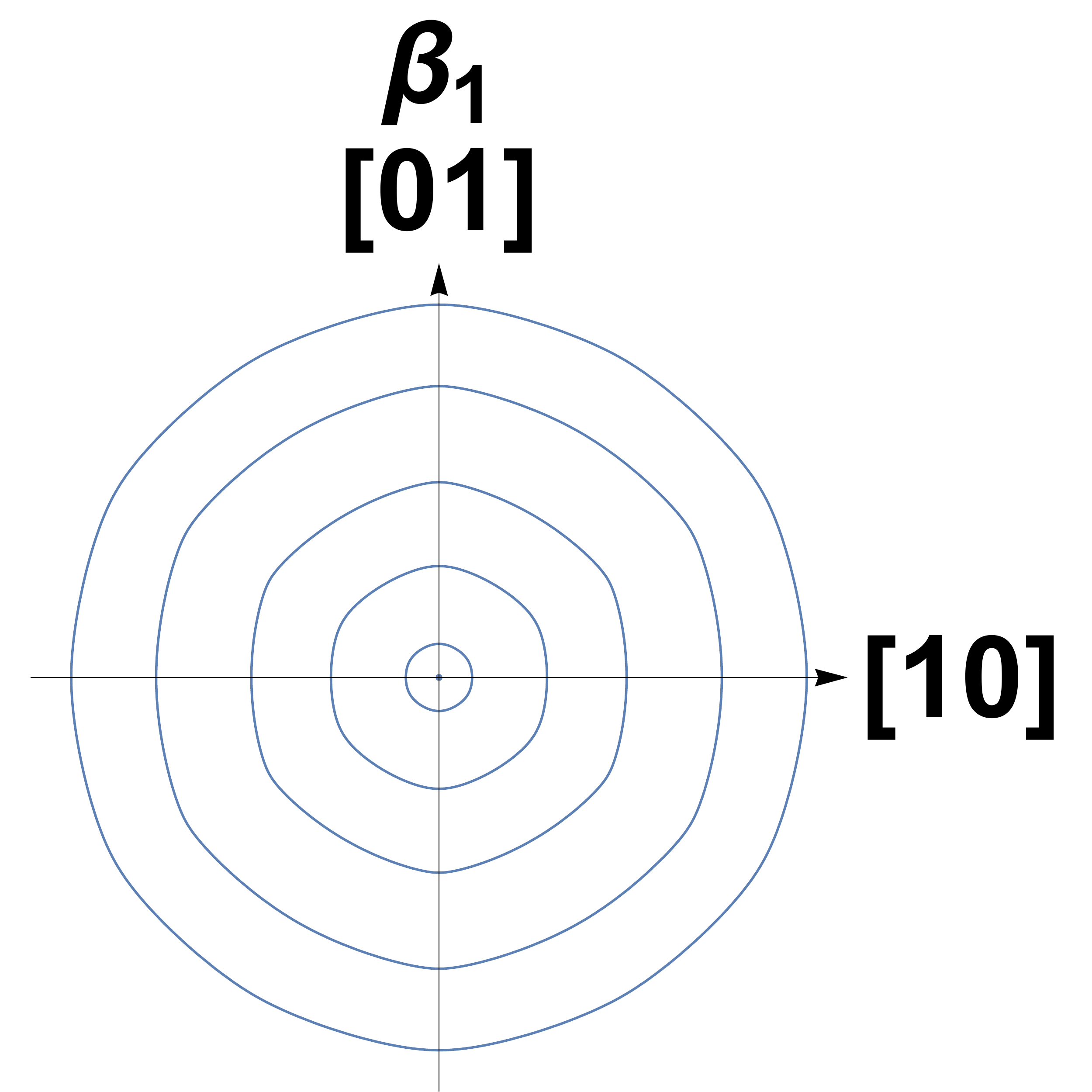} \\
	\includegraphics[width=30mm]{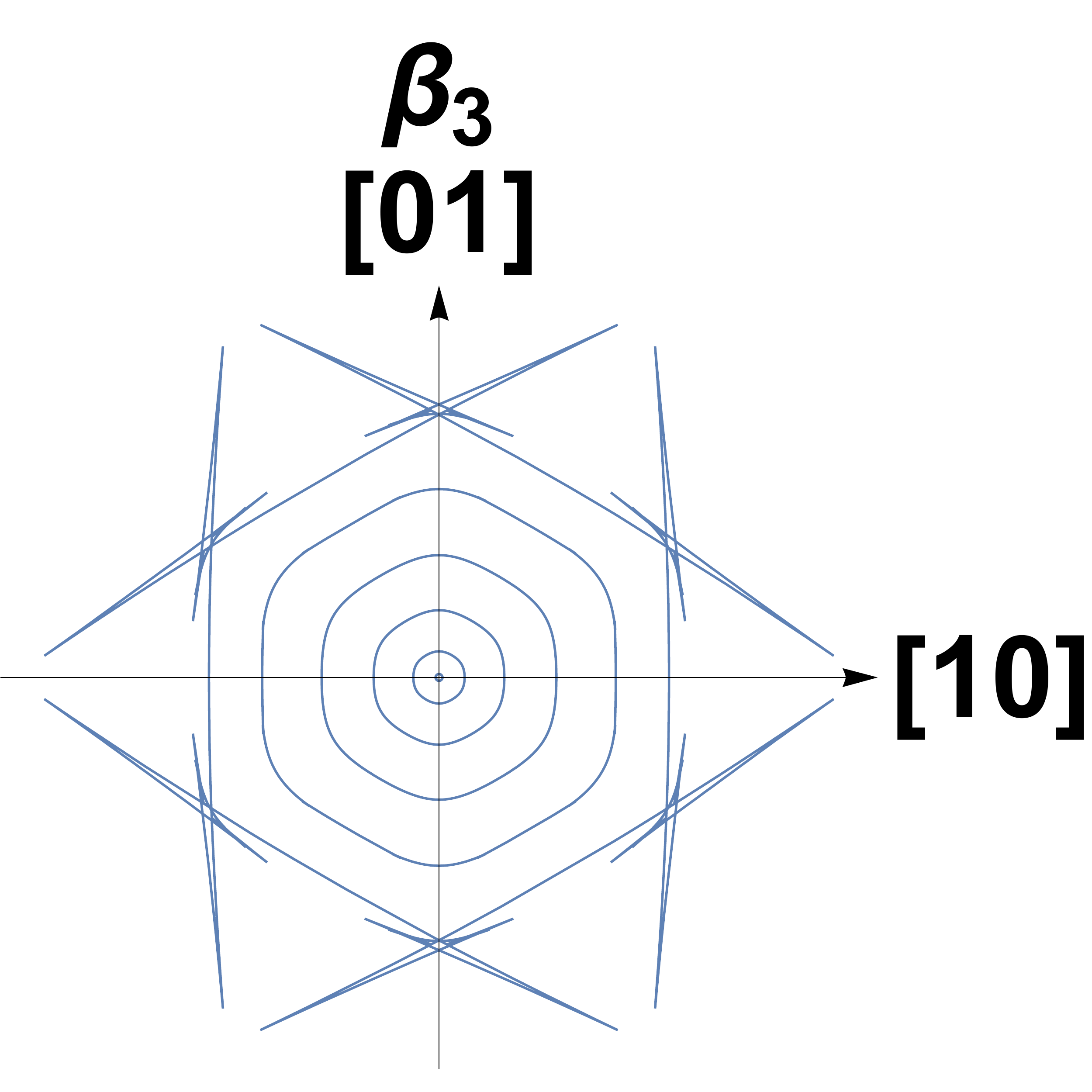} & \includegraphics[width=30mm]{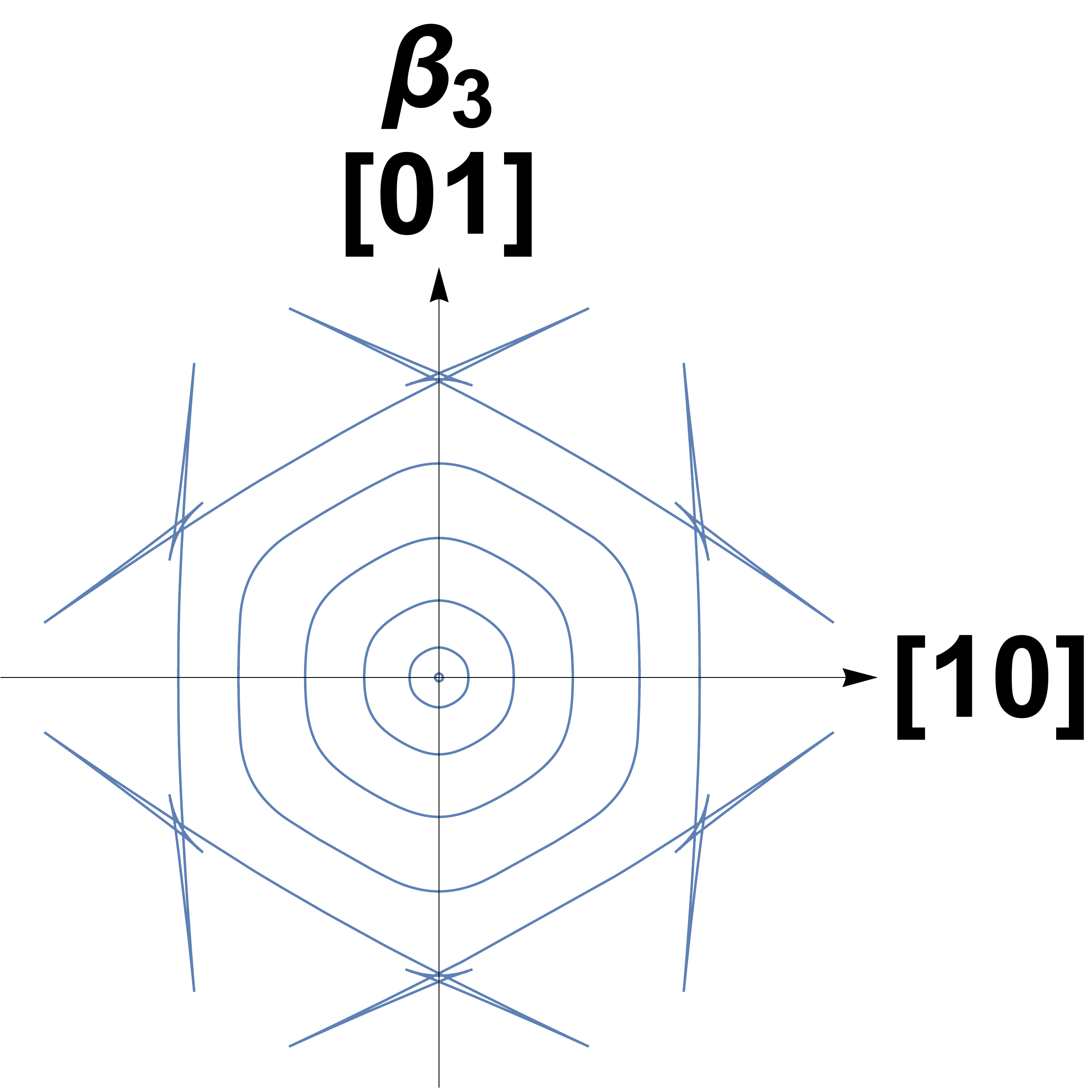} & \includegraphics[width=30mm]{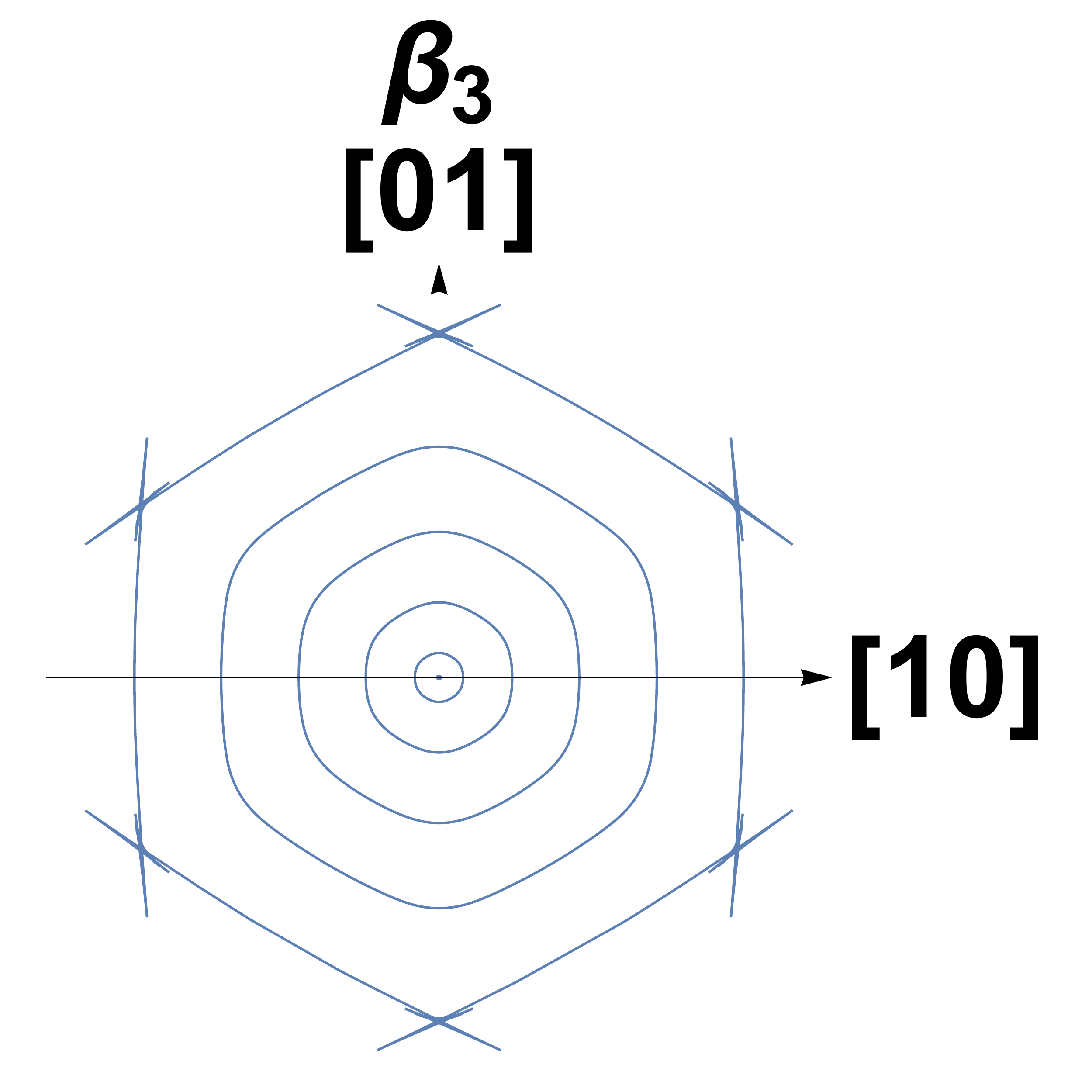} & \includegraphics[width=30mm]{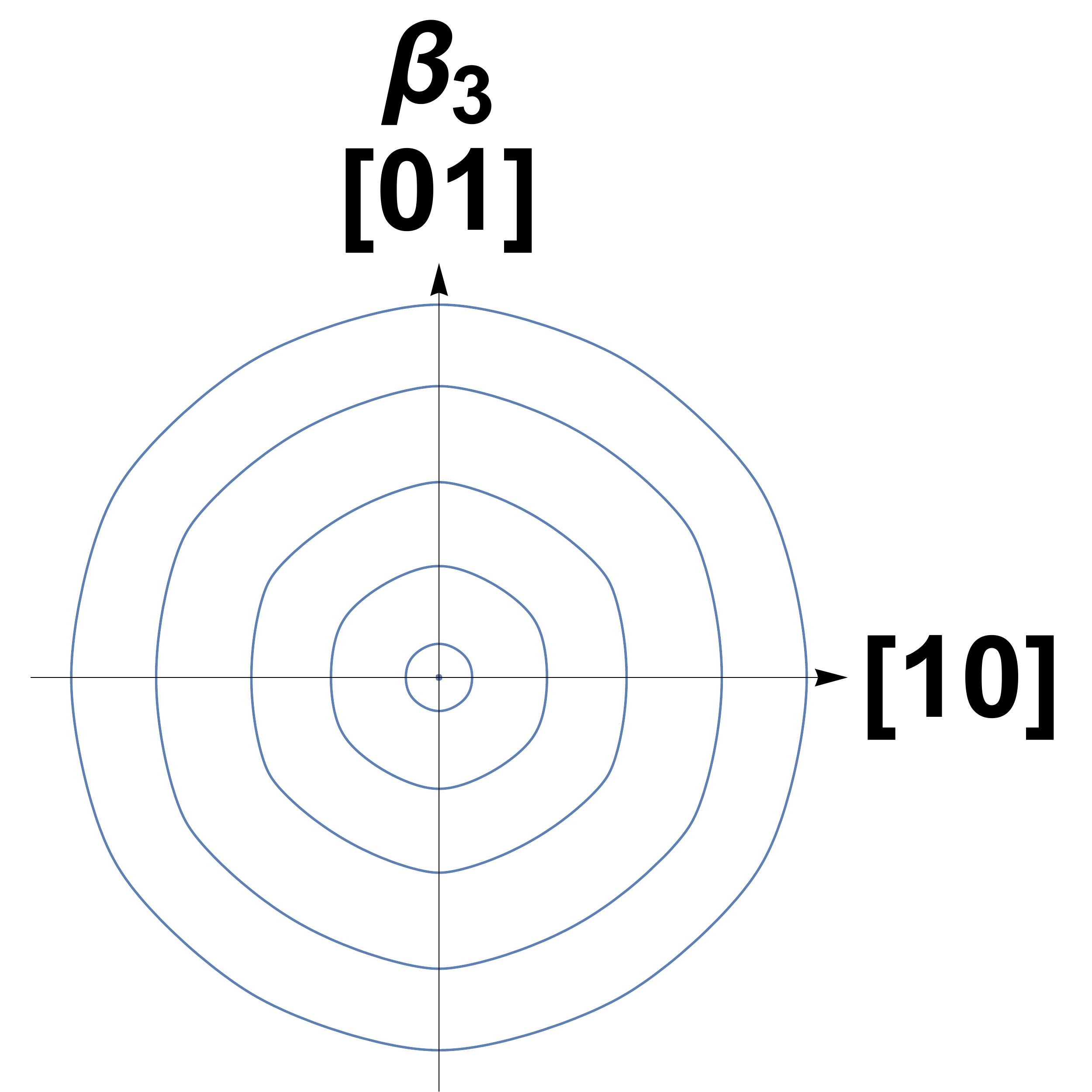} \\
	\includegraphics[width=30mm]{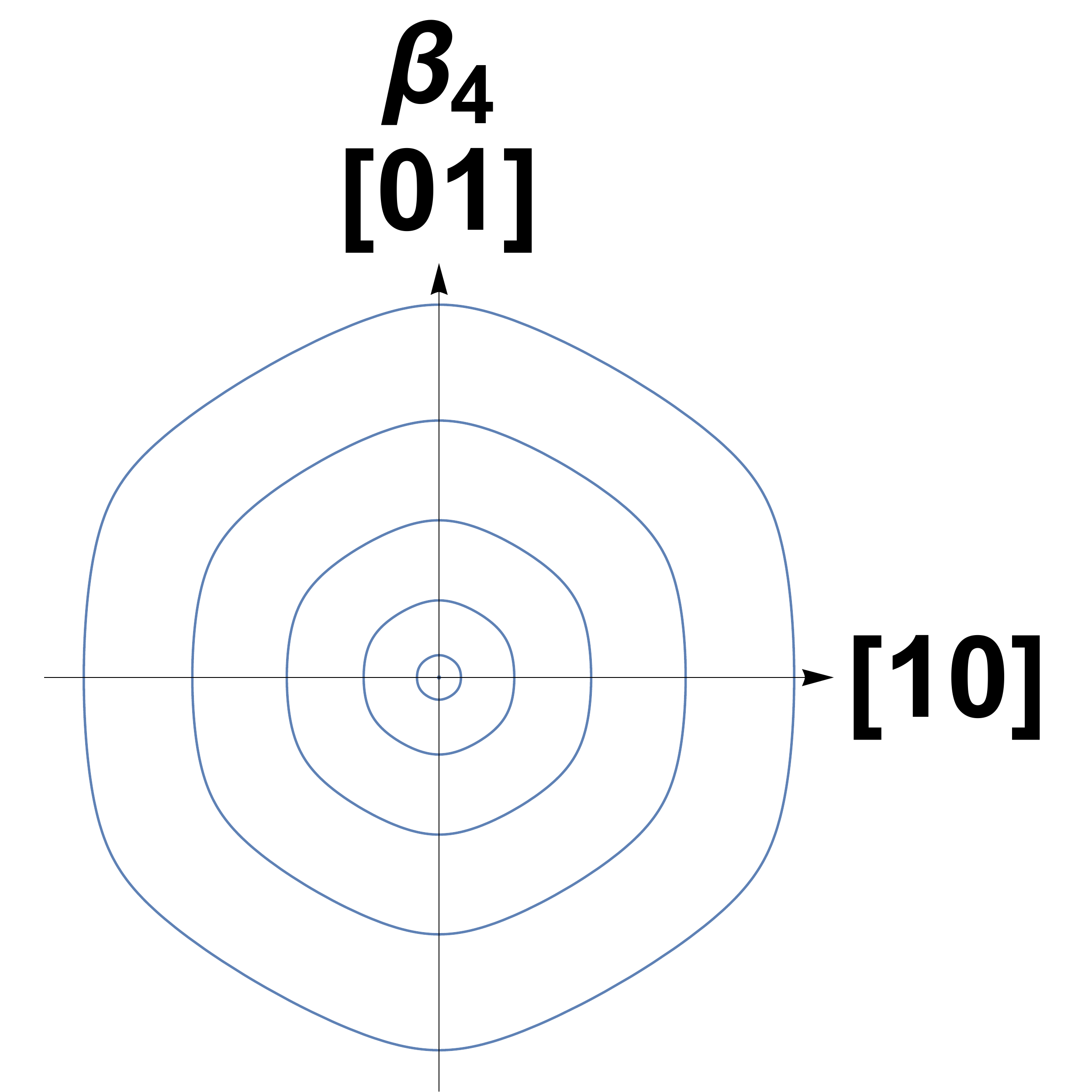} & \includegraphics[width=30mm]{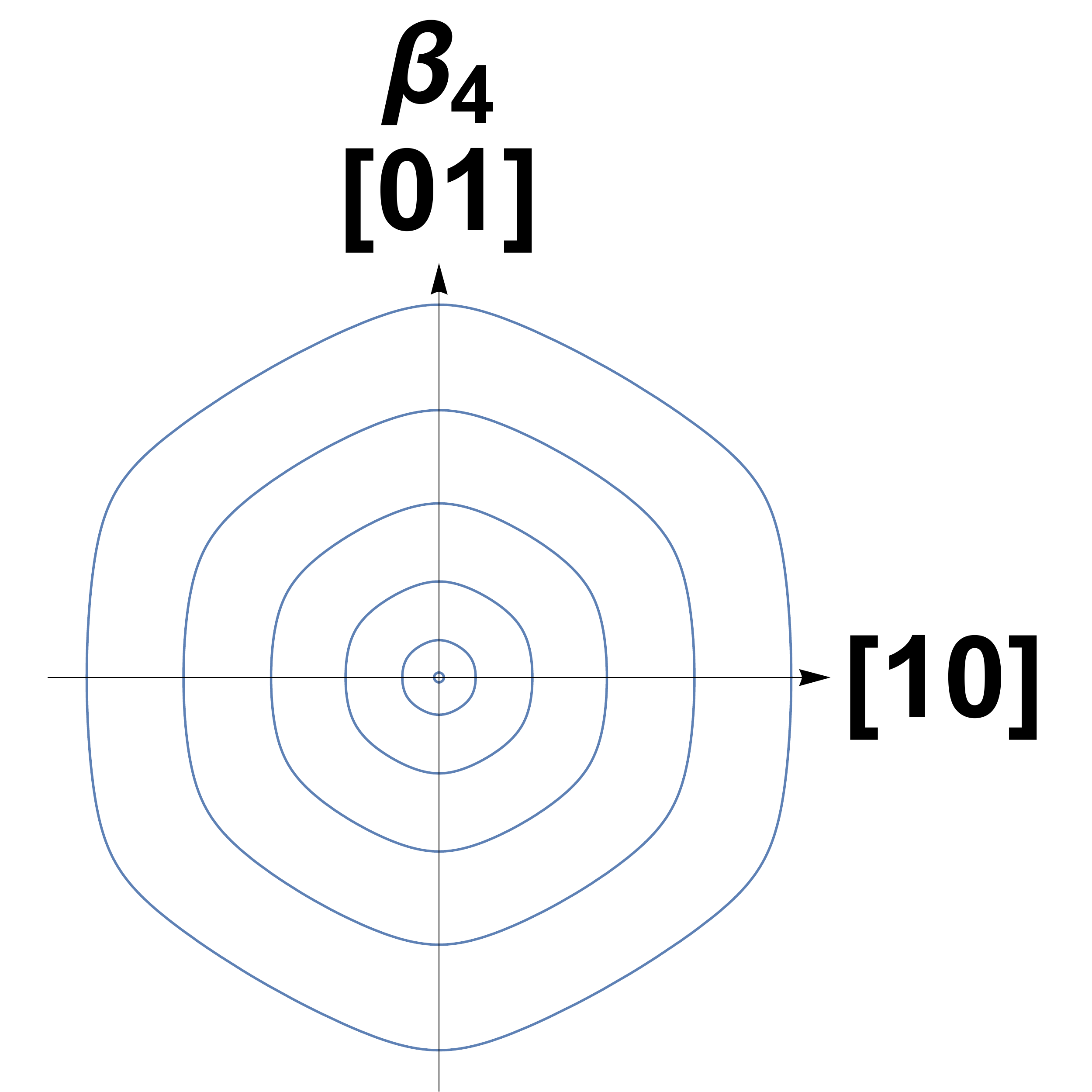} & \includegraphics[width=30mm]{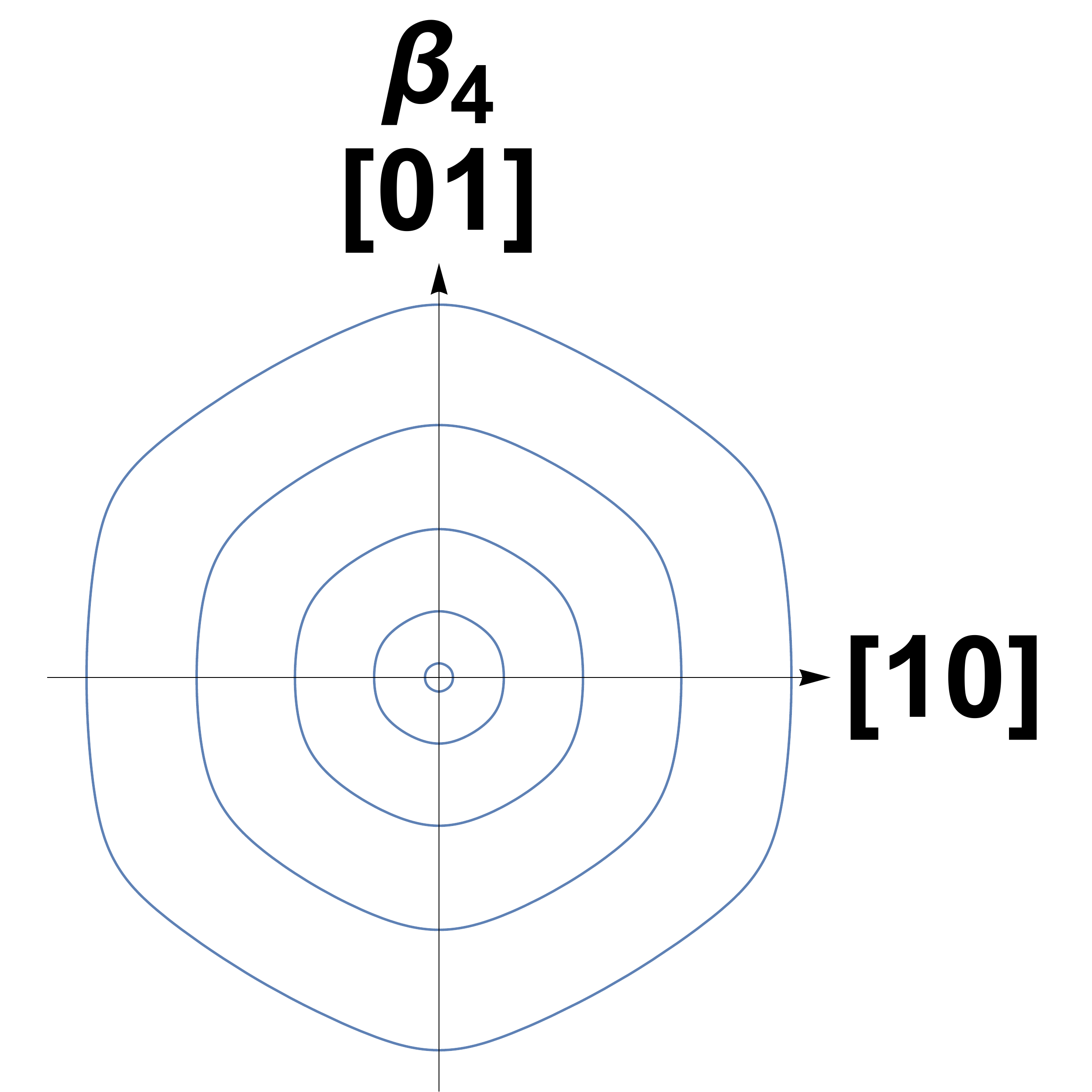} & \includegraphics[width=30mm]{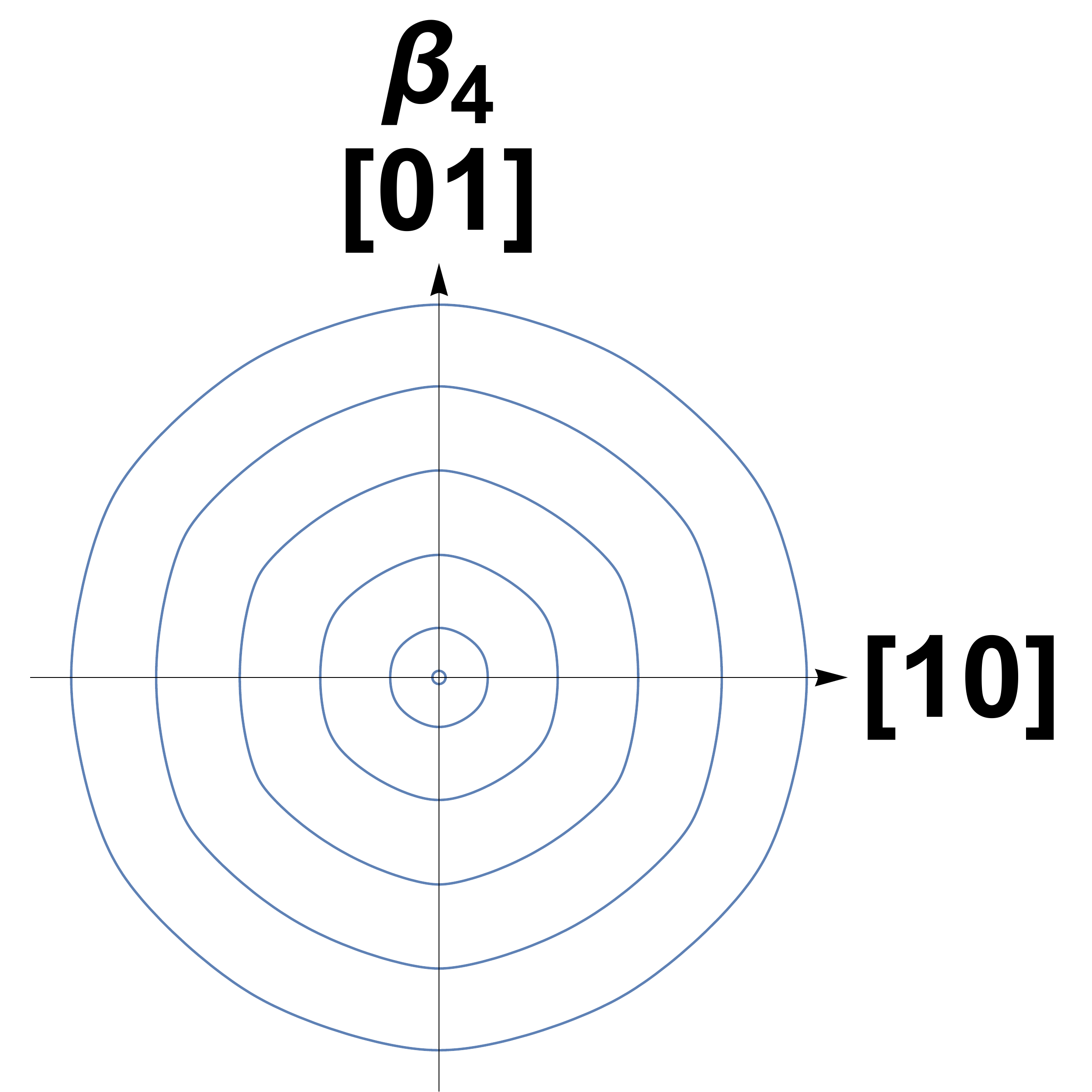} \\
	\includegraphics[width=30mm]{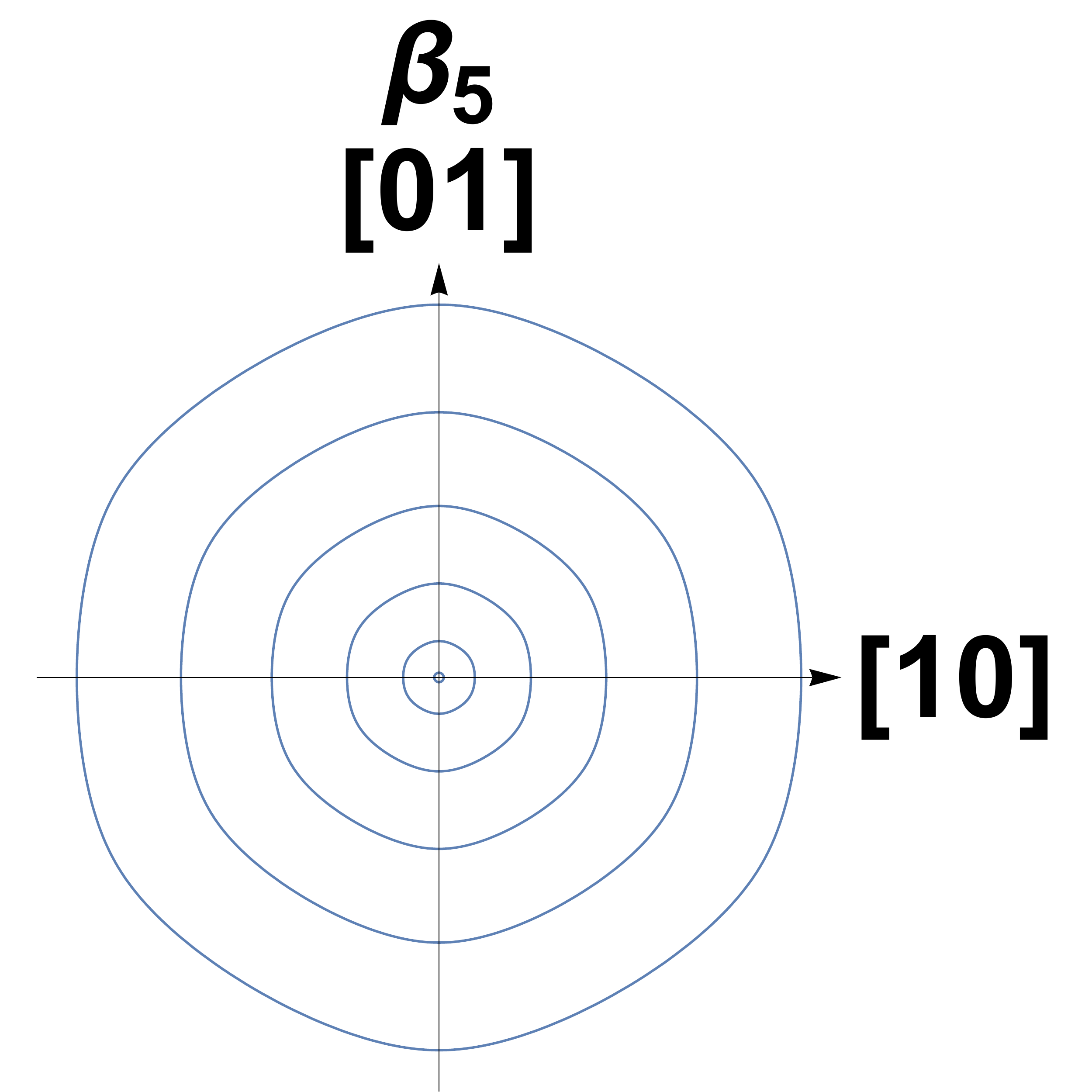} & \includegraphics[width=30mm]{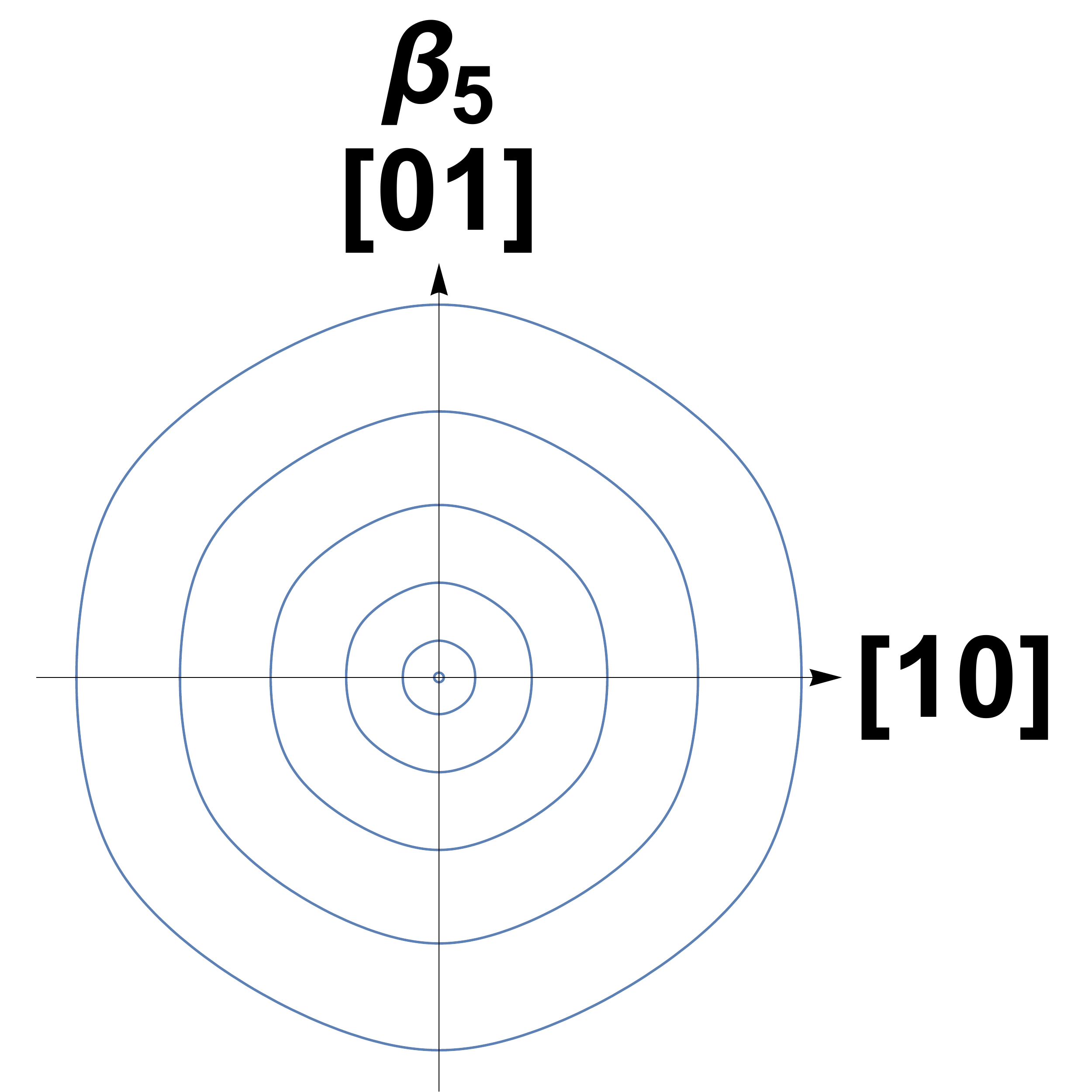} & \includegraphics[width=30mm]{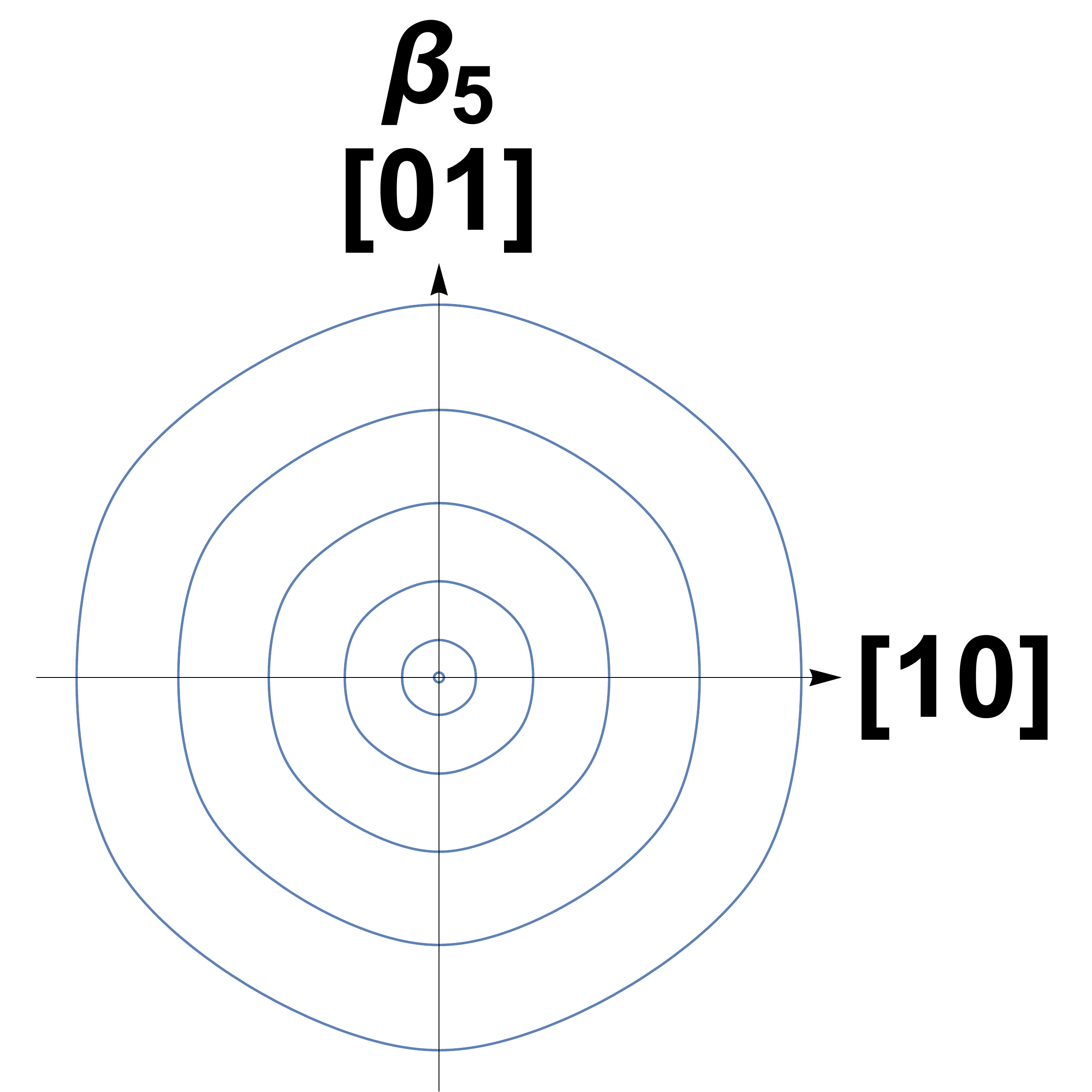} & \includegraphics[width=30mm]{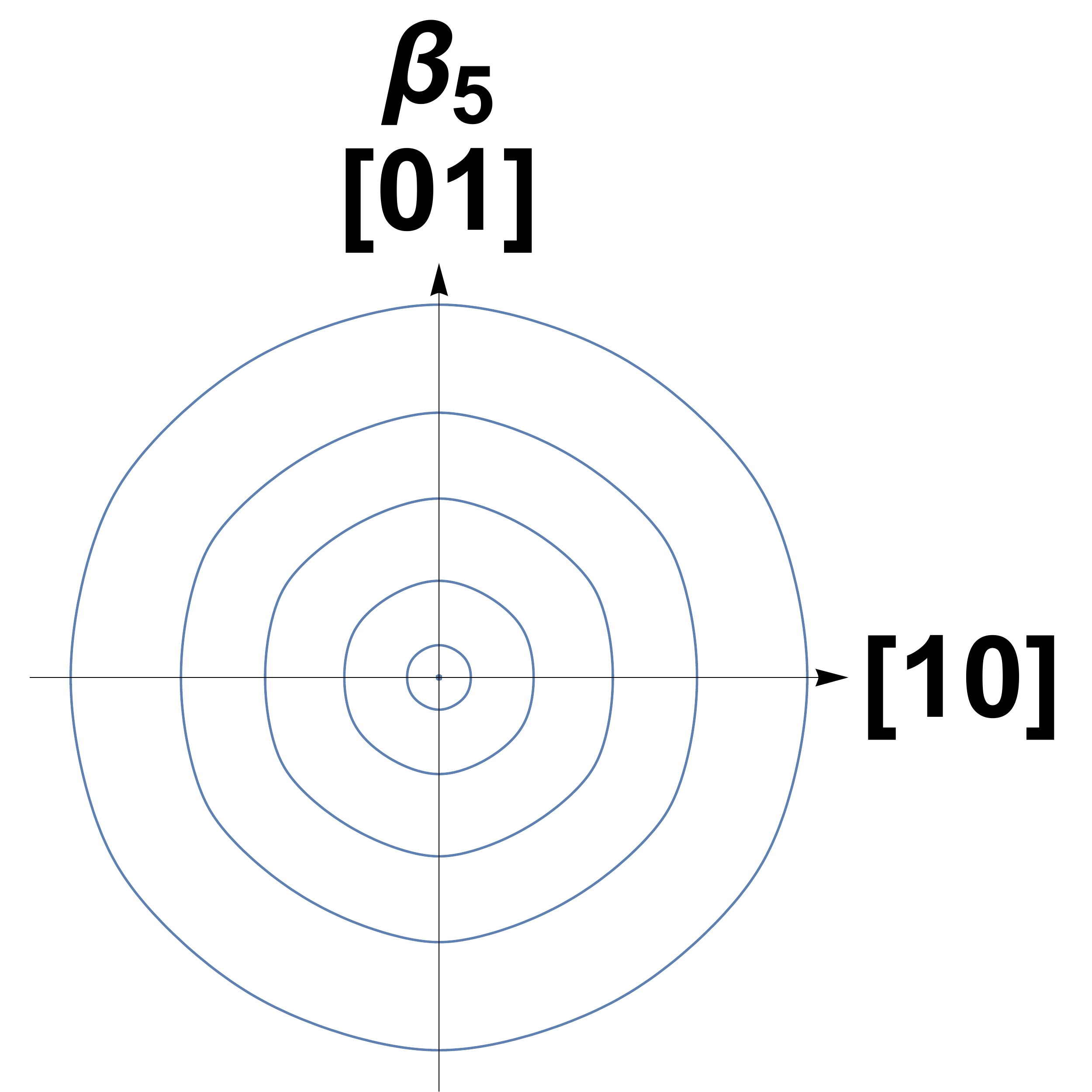} \\
	(a) $\Delta x_1$ & (b) $\Delta x_2$ & (c) $\Delta x_3$ & (d) $\Delta x_4$ \\
\end{tabular}
 \caption{Wulff shapes as a function of temperature. Note the interior, convex region of the Wulff shape gives the equilibrium crystal shape. $\beta$ increases from top to bottom, while $\Delta x$ increases from left to right. Temperature decreases from the center of each panel going outward, with the outer curves representing successively lower temperatures in the figure panels.}
\label{fig:shapes-polar}
\end{figure*}
In \cref{fig:surf-polar}, for various different values of $\beta$ and $\Delta x$, we plot the surface energy in polar form as fit by \cref{eq:se-expan}. In the figure, we have $\beta$ increasing from top to bottom, i.e., increasing corner energy, and $\Delta x$ increasing from left to right, i.e., decreasing resolution. Within each panel in the figure, temperature decreases outward from the center. That is, the surface energy plots further from the center correspond to successively lower temperatures, $\epsilon$. Following the same scheme, \cref{fig:shapes-polar} displays the Wulff shapes for the same parameter values. We note that we have not shown plots for $\beta_2$ and $\beta_6$-$\beta_8$ as the results for those values do not substantially contribute to the discussion to follow. 

There are several interesting features worth noting in the figures. Firstly, we note that, generally, both the surface energy and Wulff shapes become isotropic for increasing $\beta$, i.e., increasing corner energy. There are some slight variations, which are only perceptible when the stiffness is plotted (not shown).  We also see that for the coarsest resolution plotted, $\Delta x_4$ in \cref{fig:shapes-polar}(d), the Wulff shapes are seemingly isotropic for all temperatures and $\beta$, also evident in \cref{fig:surf-polar}(d) for the surface energy. As the resolution becomes finer, decreasing $\Delta x$, and for each $\beta$ as $\epsilon$ decreases, we see the development of high curvature regions in the surface energy plots of \cref{fig:surf-polar}, which translates to Wulff shapes assuming more hexagonal like shapes in \cref{fig:shapes-polar}. Eventually we see the development of ``ears'' indicating we have crossed the threshold to the regime of missing orientations and a transition to highly anisotropic shapes. This is supported in the surface energy plots of \cref{fig:surf-polar}, where depressions and rounded cusp regions appear, coinciding with the emergence of ``ears''. It is important to note here that for the plots in \cref{fig:surf-polar,fig:shapes-polar}, especially those parameters where pronounced ``ears'' are evident, we have plotted the data up to the transition temperature (as determined in the following section) for clarity.
\begin{figure}[!htb]
\centering
\begin{subfigure}[b]{.5\textwidth}
 \includegraphics[width=1.25in]{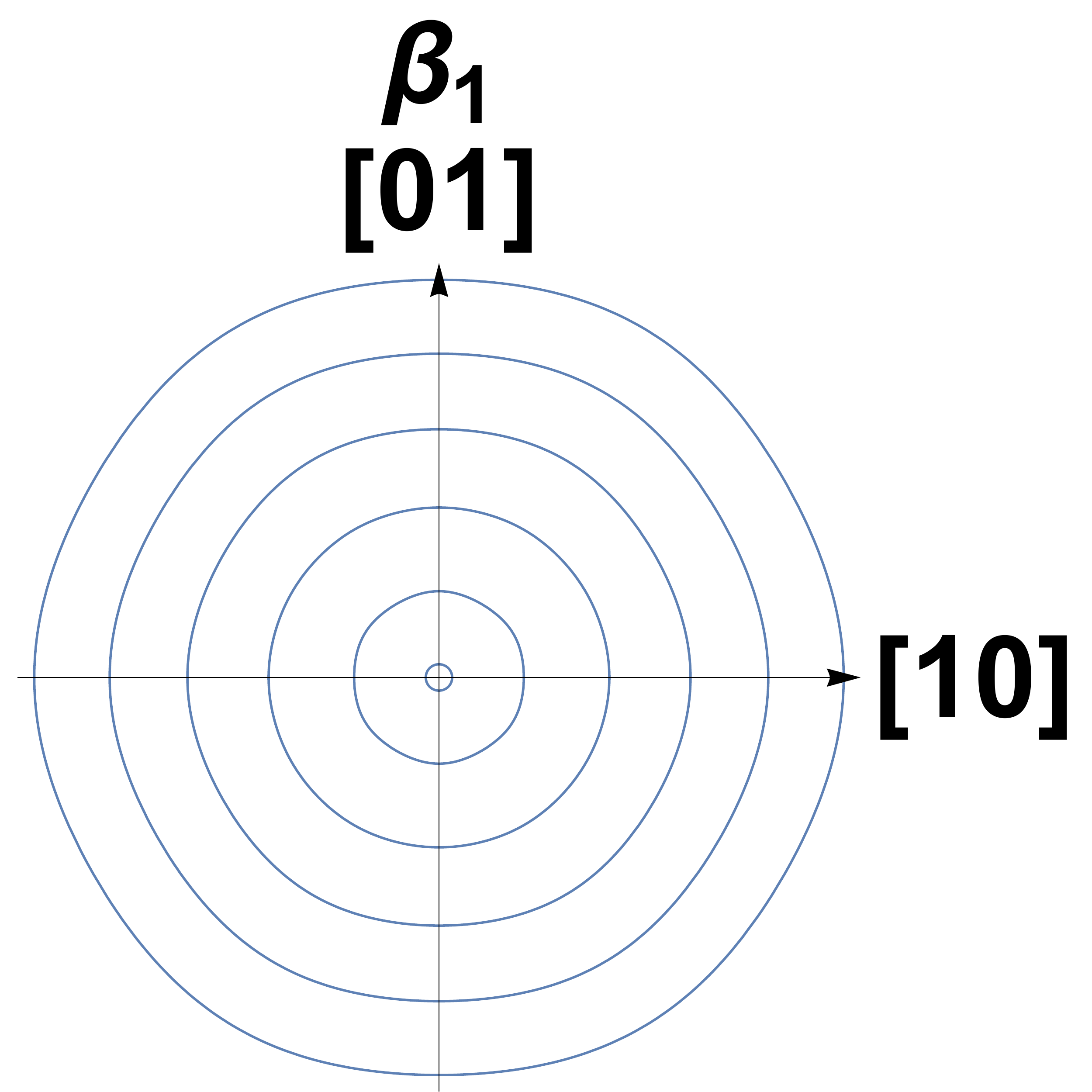} \hspace{1ex}  \includegraphics[width=1.25in]{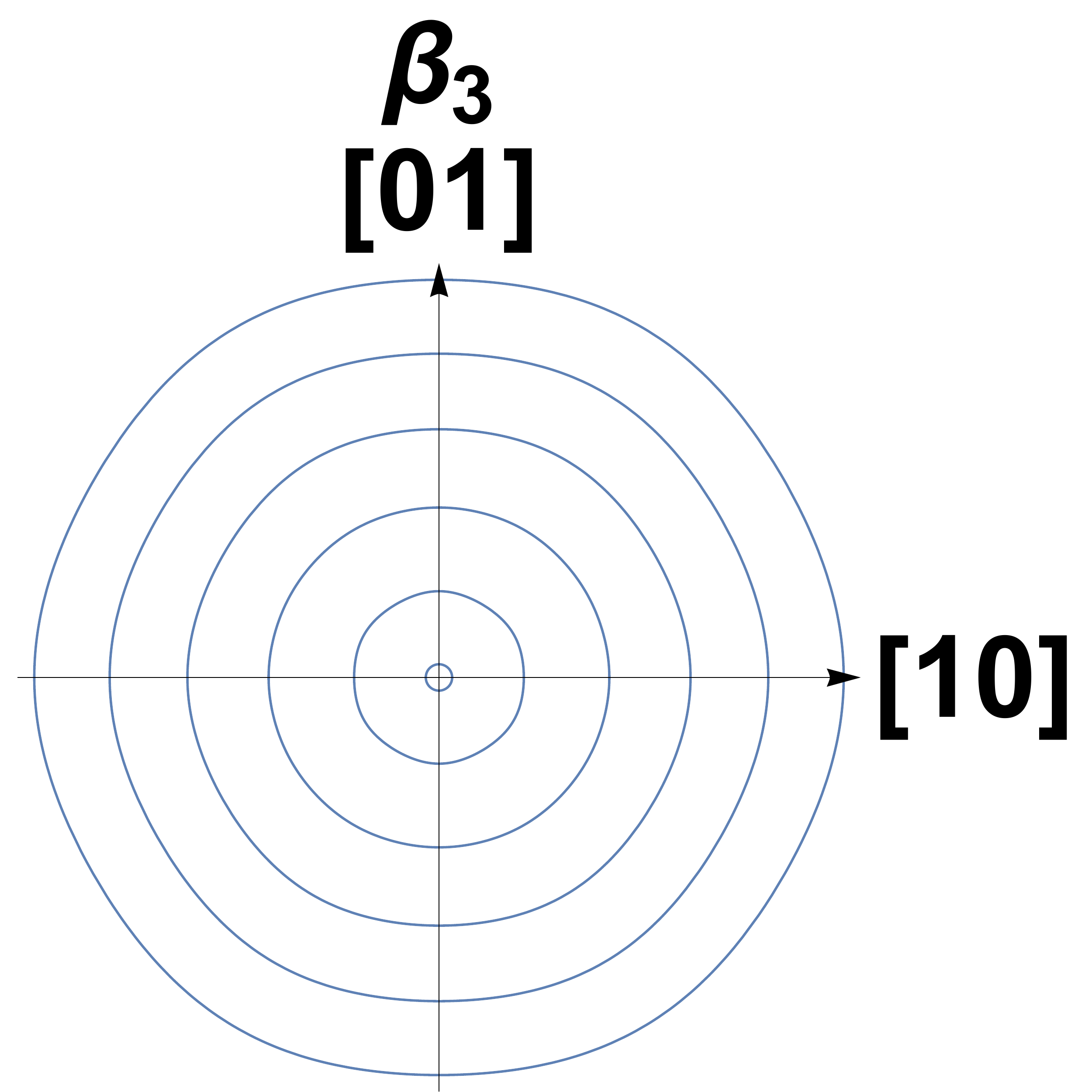}

\vspace{2ex}

 \includegraphics[width=1.25in]{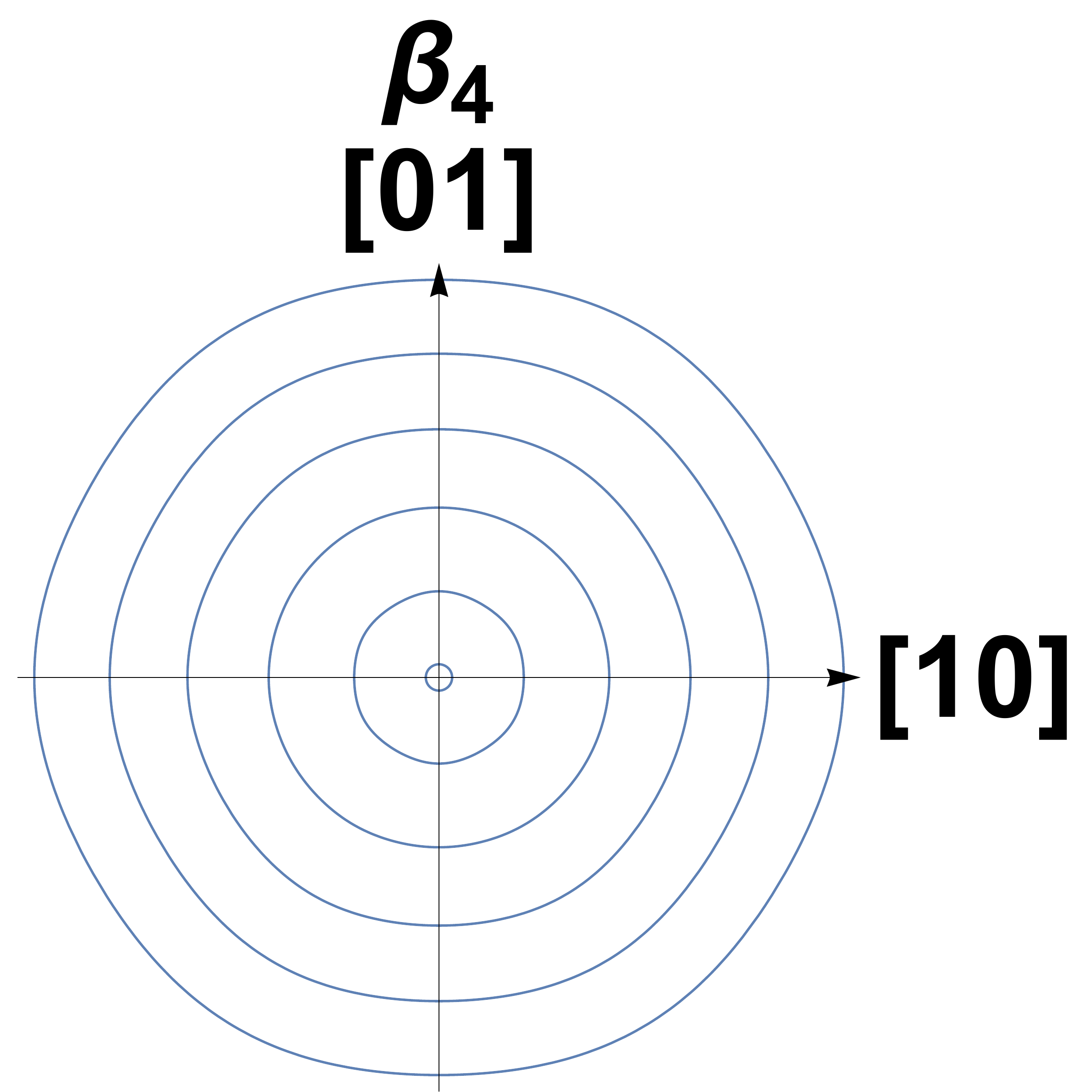} \hspace{1ex}  \includegraphics[width=1.25in]{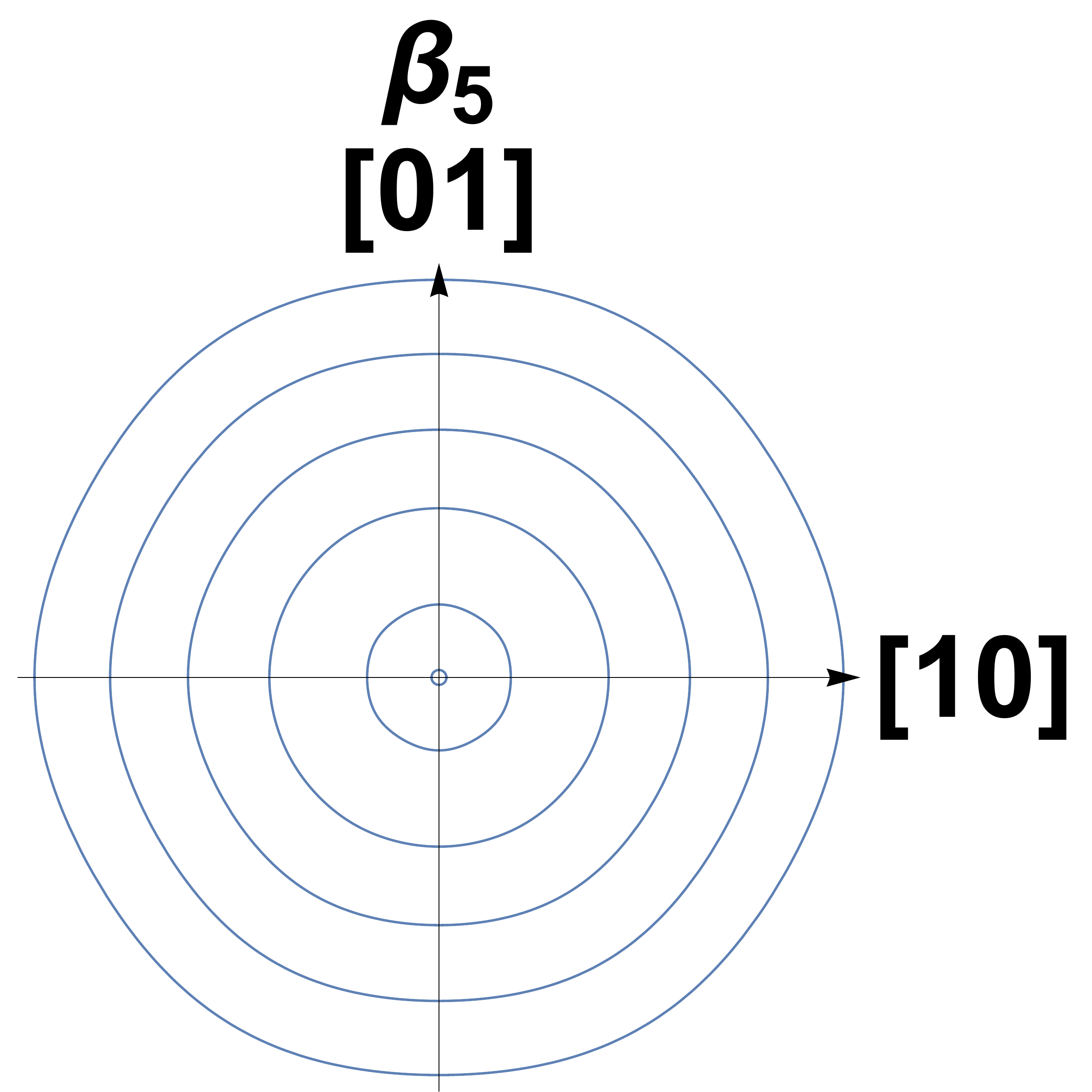}
\caption{Surface energy}
\end{subfigure} 
\begin{subfigure}[b]{.4\textwidth}
\resizebox{2.5in}{!}{\includegraphics{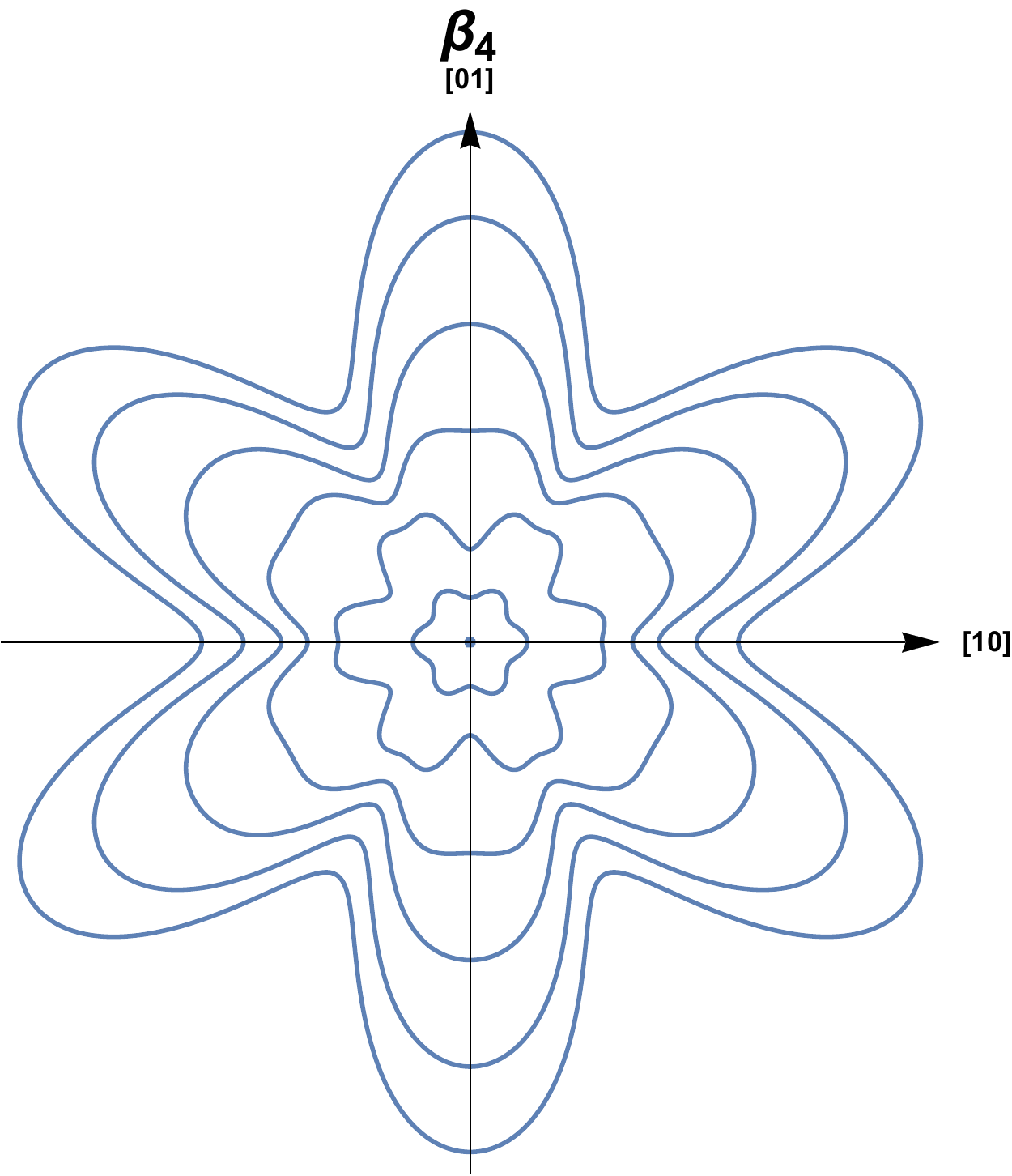}}
\caption{Stiffness}
\end{subfigure}
\caption{Polar plots of surface energy and stiffness. (a) surface energy for $\Delta x_5$ and various values of $\beta$. (b) stiffness for $\Delta x_5$ and $\beta_4$ clearly displaying a change in anisotropy as a function of decreasing temperature. Temperature decreases moving from the center outward.}
\label{fig:stiff-anis}
\end{figure}


Now, we wish to comment on another trend exhibited in the Wulff shapes and surface energy plots, specifically for $\Delta x_4$ and $\Delta x_5$. For these resolution parameters, we have noticed that both the surface energy, and consequently the Wulff shapes undergo a change in the anisotropic direction as a function of temperature. This is more evident in the case of $\Delta x_5$, \cref{fig:stiff-anis}(a). As a function of temperature we see the easy axis direction, i.e., maximum surface energy direction, change from the  $[01]$ direction. In the surface energy and Wulff plots, this phenomena is almost imperceptible. However, since the stiffness is more anisotropic (by at least an order of magnitude),  in \cref{fig:stiff-anis}(b) we show a plot of the stiffness for a representative combination of the grid resolution and corner energy parameters. In the figure, as a function of decreasing temperature, i.e., going from the center outward in the panel, we clearly observe the minimum stiffness direction changing from the [01] direction. This change in anisotropy  can be physical and has been reported in pure materials and alloys both experimentally and numerically~\cite{Warren2006,Haxhimali2006,Friedli2013,Dantzig2013}, although this is the first it has been reported for 2D systems to the authors knowledge. In the case presented here however,  it is an artifact of the system being under-resolved, as evident by the lack of such a transition in the cases where finer resolution parameters are utilized, i.e., $\Delta x_1$, $\Delta x_2$ and $\Delta x_3$ and thus {\it caution should be exercised} when conducting simulations. It is also worth noting that the value of $\Delta x$, i.e., $\Delta x_3$, where this artificial transition arrests is comparable to the minimum resolution often used in the simulation of regular atomistic PFC simulations. An indication that perhaps the sufficient resolution required to capture the salient physical features of highly anisotropic surfaces, in the limit of missing orientations and facet formation, is similar to that required for full PFC model.
\begin{figure*}[h]
\centering
\begin{tabular}{cc}
    \resizebox{3.1in}{!}{\includegraphics{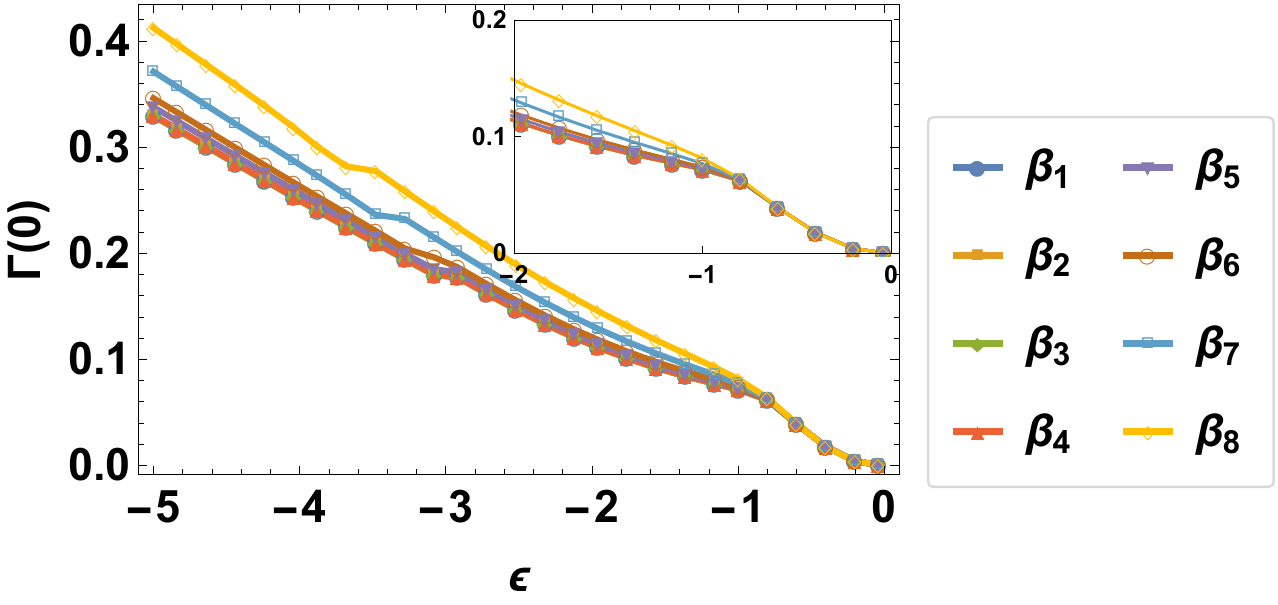}} & \resizebox{3.1in}{!}{\includegraphics{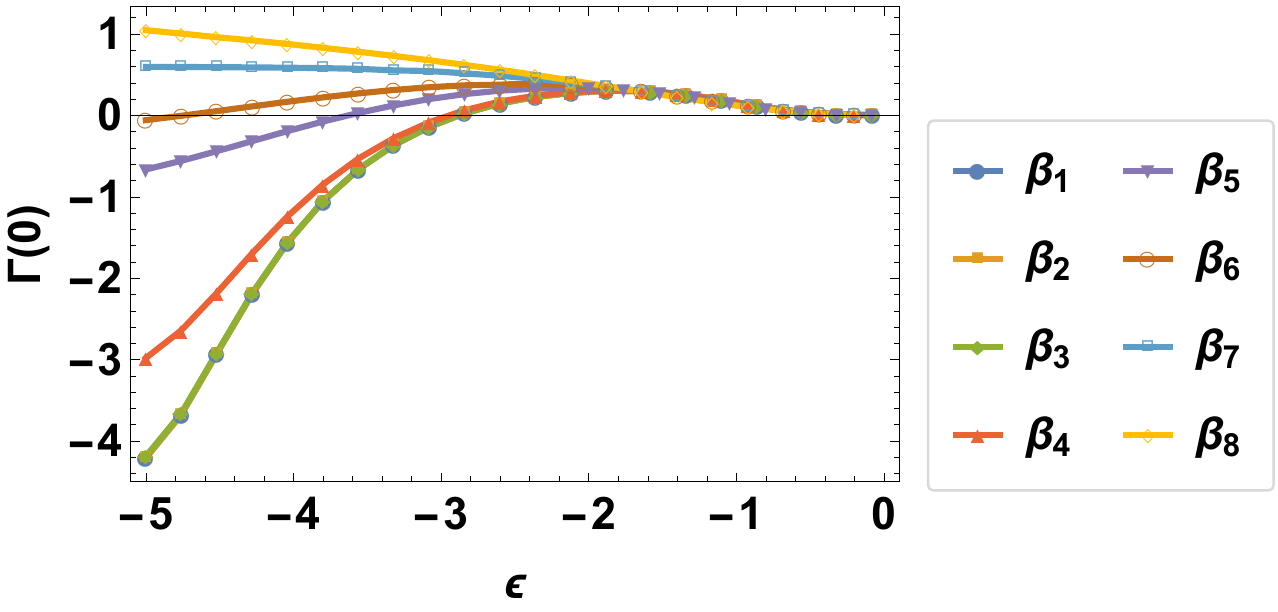}} \\
	(a) $\Delta x_5$ & (b) $\Delta x_4$  \\ \\
	\resizebox{3.1in}{!}{\includegraphics{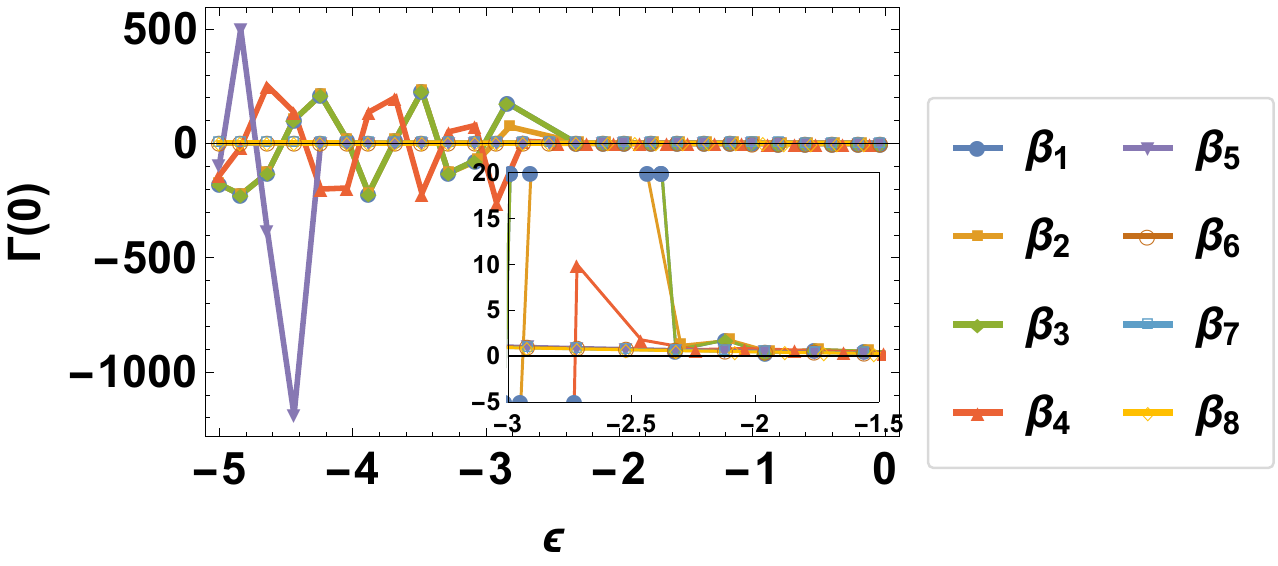}}  & \resizebox{3.1in}{!}{\includegraphics{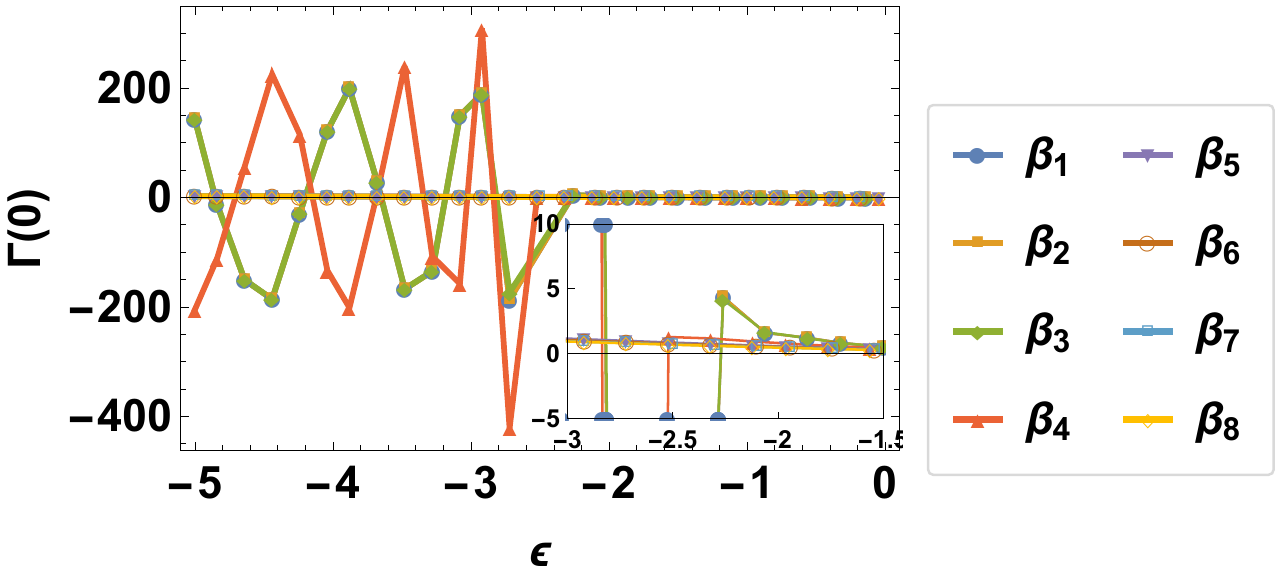}} \\
	(c) $\Delta x_3$ & (d) $\Delta x_2$  \\ \\
	\resizebox{3.1in}{!}{\includegraphics{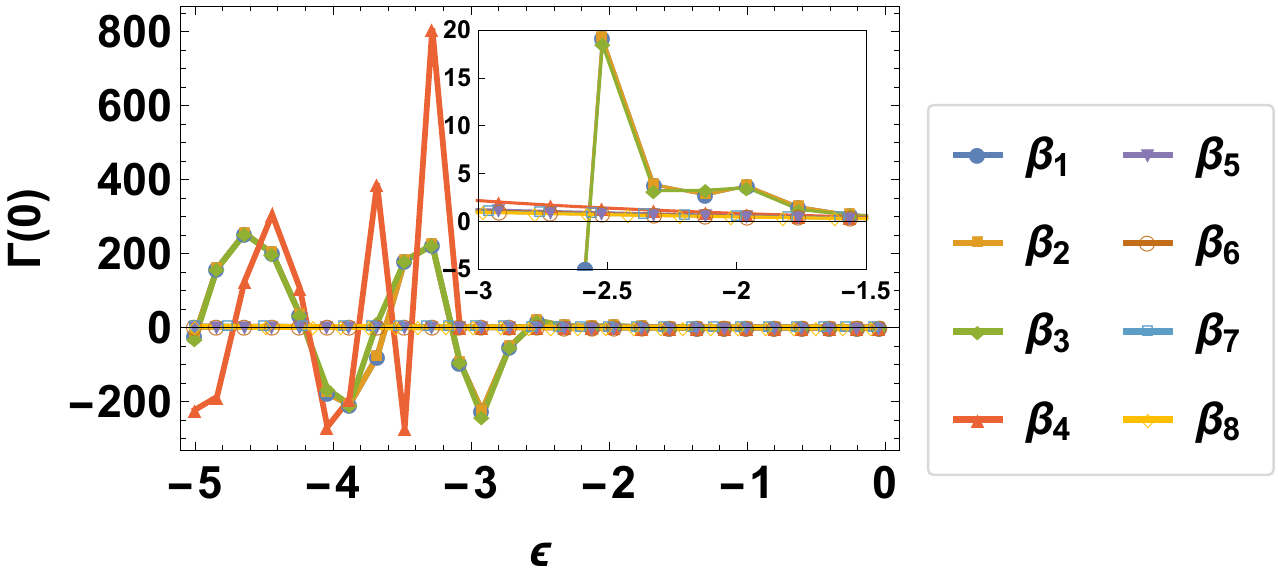}} \\
	(e) $\Delta x_1$  \\
\end{tabular}     
\caption{Plots of the stiffness ($\Gamma$), evaluated at $\theta=0$, versus temperature, $\epsilon$, for the resolution parameters considered (a)-(e). Note that temperature decreases from right to left. The stiffness criterion states that orientations where the stiffness is negative corresponds to those missing from the ECS. In the Wulff construction (\cref{fig:shapes-polar}) this is indicated by the emergence of ``ears''.}
\label{fig:sec-derv}
\end{figure*}

\subsubsection{Convexity of Stiffness}
The plots presented above of surface energy and the Wulff constructions were illuminating. However a more quantitative measure is required in determining the transition to missing orientations. To this end, we also consider the stiffness criterion to determine at which temperature missing orientations, i.e., ``ears'', start appearing in the Wulff shapes and therefore the possibility of a faceting transition. We evaluate the stiffness at the normal direction that corresponds to the closed-packed $[01]$ direction which below the transition temperature is expected to be the initial missing  orientation. We have plotted this value for the stiffness for the various parameters of $\beta$ and $\Delta x$ as a function of the temperature $\epsilon$. The results are shown in \cref{fig:sec-derv}. For $\Delta x_5$, \cref{fig:sec-derv}$(a)$ clearly indicates that the stiffness criterion is not met for any value of $\beta$, as $\Gamma(0)$ is a monotonic function of $\epsilon$. This is expected given behavior exhibited by the surface energy and Wulff plots. The inset shows a magnification of the data, where an inflection is visible. This inflection point indicates the temperature at which the change in anisotropy occurs, as discussed in the previous section and further discussed in \cref{anis-transition}. 

At the next finer resolution, $\Delta x_4$,  \cref{fig:sec-derv}(b), there are three features worth noting. The first being that several values of $\beta$ exhibit negative stiffness values below select temperatures, indicating the transition to missing orientations. Secondly, we observe that this transition does not occur for the same temperature for all $\beta$ values, although $\beta_1$-$\beta_4$ seem to cluster around a similar transition temperature. And thirdly, there is evidence in this plot as well of the anisotropy transition as indicated by the change in curvature of the plot for decreasing values of $\beta$. 

For the remainder of the plots, \cref{fig:sec-derv}(c)-(e), $\Gamma(0)$ for $\beta_5$-$\beta_8$ all show monotonic behavior as a function of temperature. This indicates that for the temperature range considered, there is no transition to missing orientations for any of the corner energy coefficients and notably no anisotropy transition. This is supported by their surface energy and Wulff plots from \cref{fig:surf-polar,fig:shapes-polar}. The remainder of the corner energy parameters do however show an abrupt change at $\epsilon \sim 2.6$, the instance where the $\Gamma(0)$ initially becomes negative, beyond which it fluctuates. The insets in the figure show enlarged regions around this temperature. We interpret this as the transition temperature to missing orientations and possibly the beginning of a faceting transition. 

\begin{table}[ht]
\centering
  \caption{Select temperature values and the corresponding equilibrium values for the solid phase.}
  \begin{tabular}{ c@{\hskip 0.15in} c@{\hskip 0.15in} c@{\hskip 0.15in} }
  \hline\hline
    $\epsilon$ & $\bar{n}_s$ & $\phi_{eq}$ \\
    \hline
    -3.08 & -0.61819 & -0.48167\\
    -5 & -0.73887 & -0.64374\\
    \hline
  \end{tabular}
  \label{tab:eqlm-prop}
\end{table}

\begin{figure}[!htb]
\centering
\begin{subfigure}[b]{.4\textwidth}
\resizebox{2.75in}{!}{\includegraphics{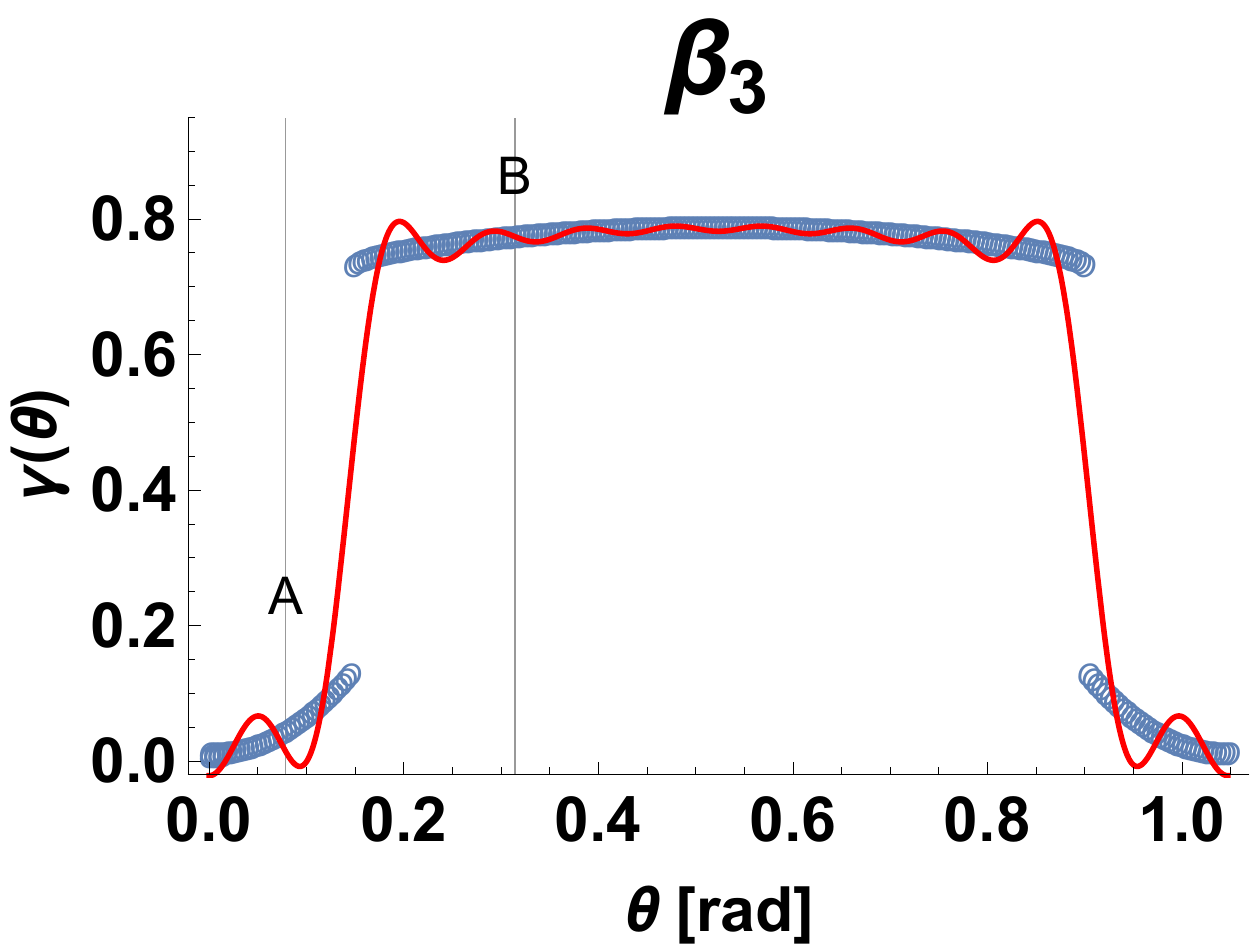}}
\caption{Surface energy $\beta_3$}
\end{subfigure}
\begin{subfigure}[b]{.5\textwidth}
 \includegraphics[width=1.5in]{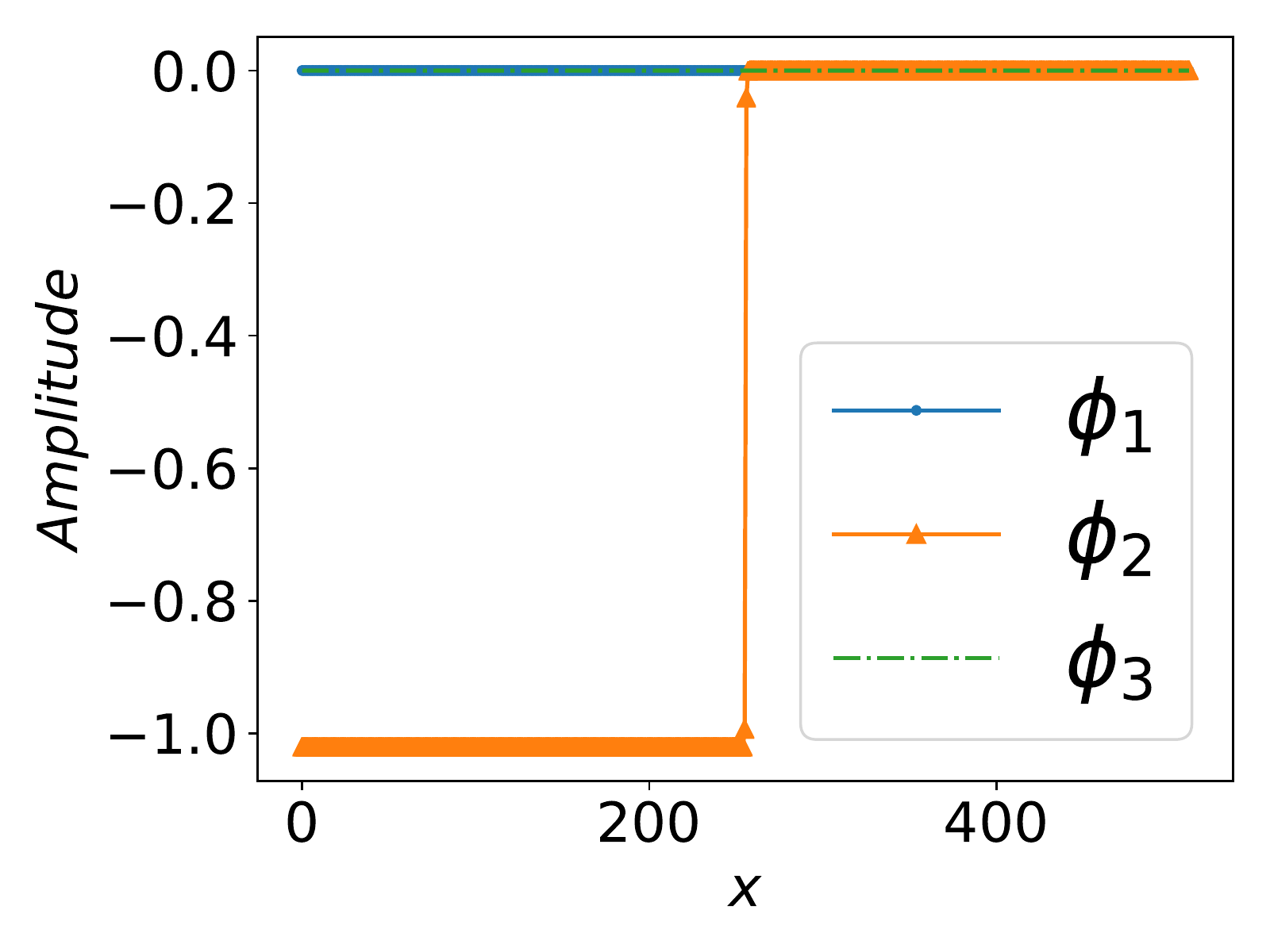} \hspace{0ex}  \includegraphics[width=1.5in]{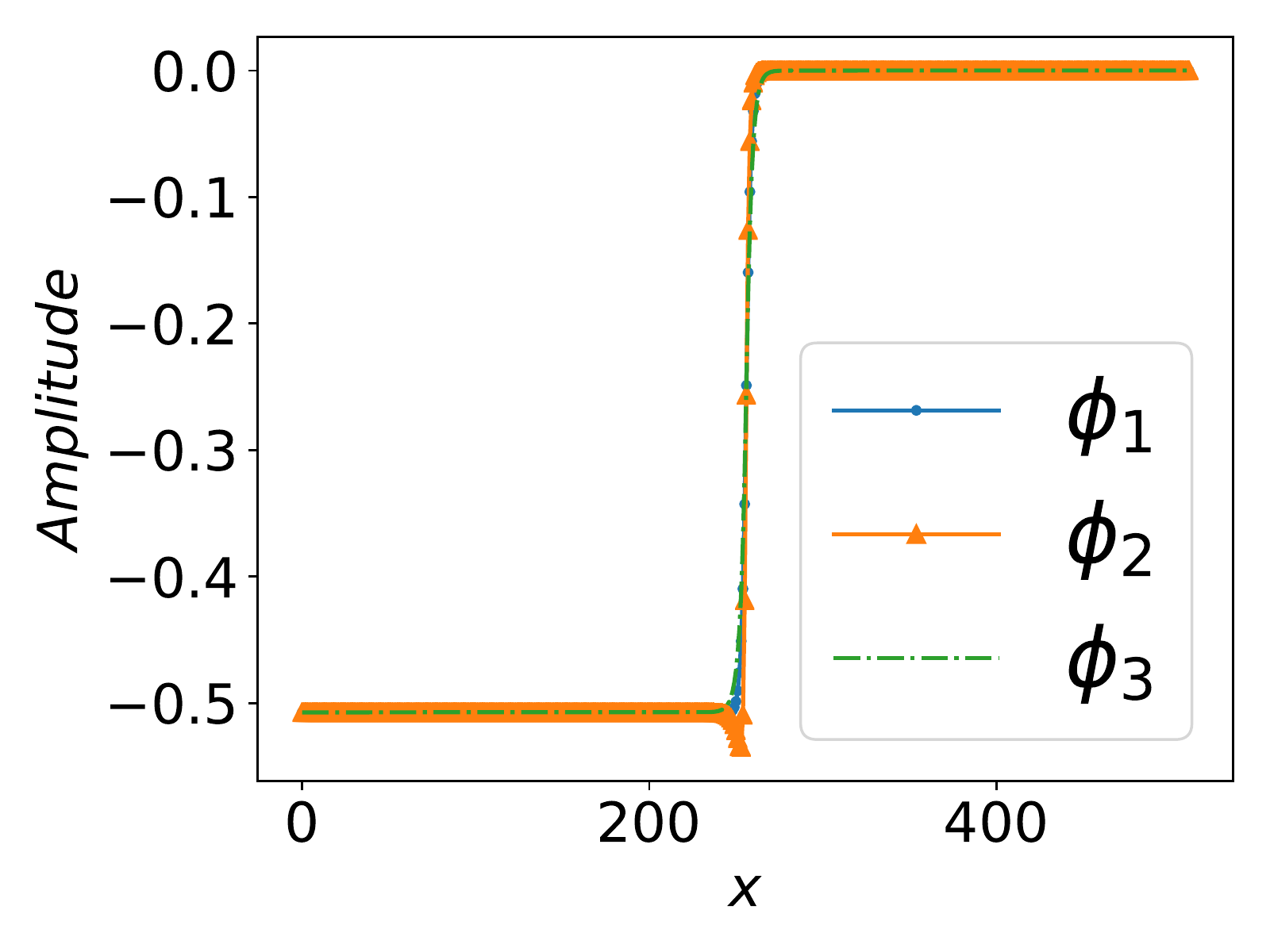}
\vspace{0ex}
\includegraphics[width=1.5in]{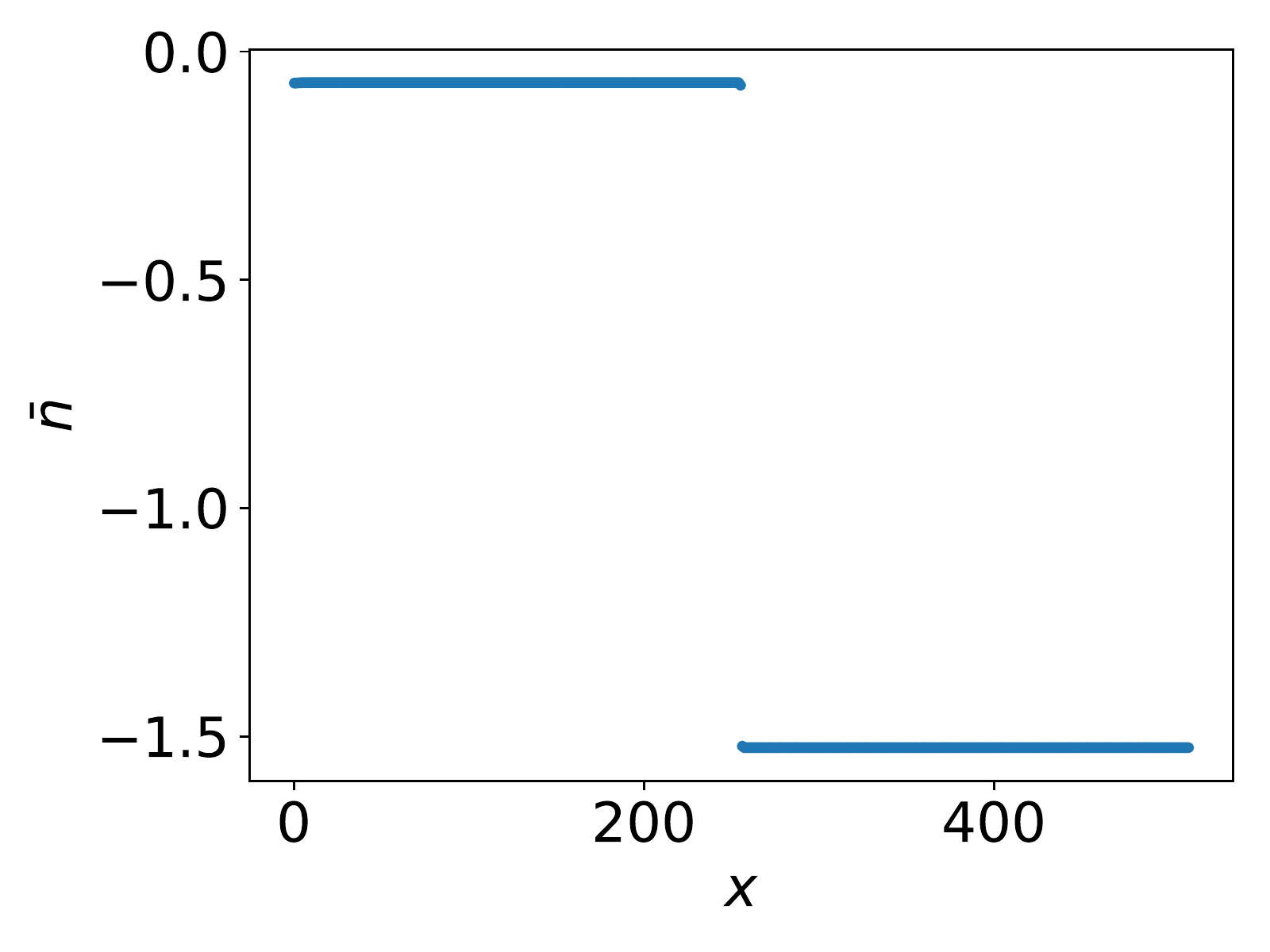} \hspace{0ex}  \includegraphics[width=1.5in]{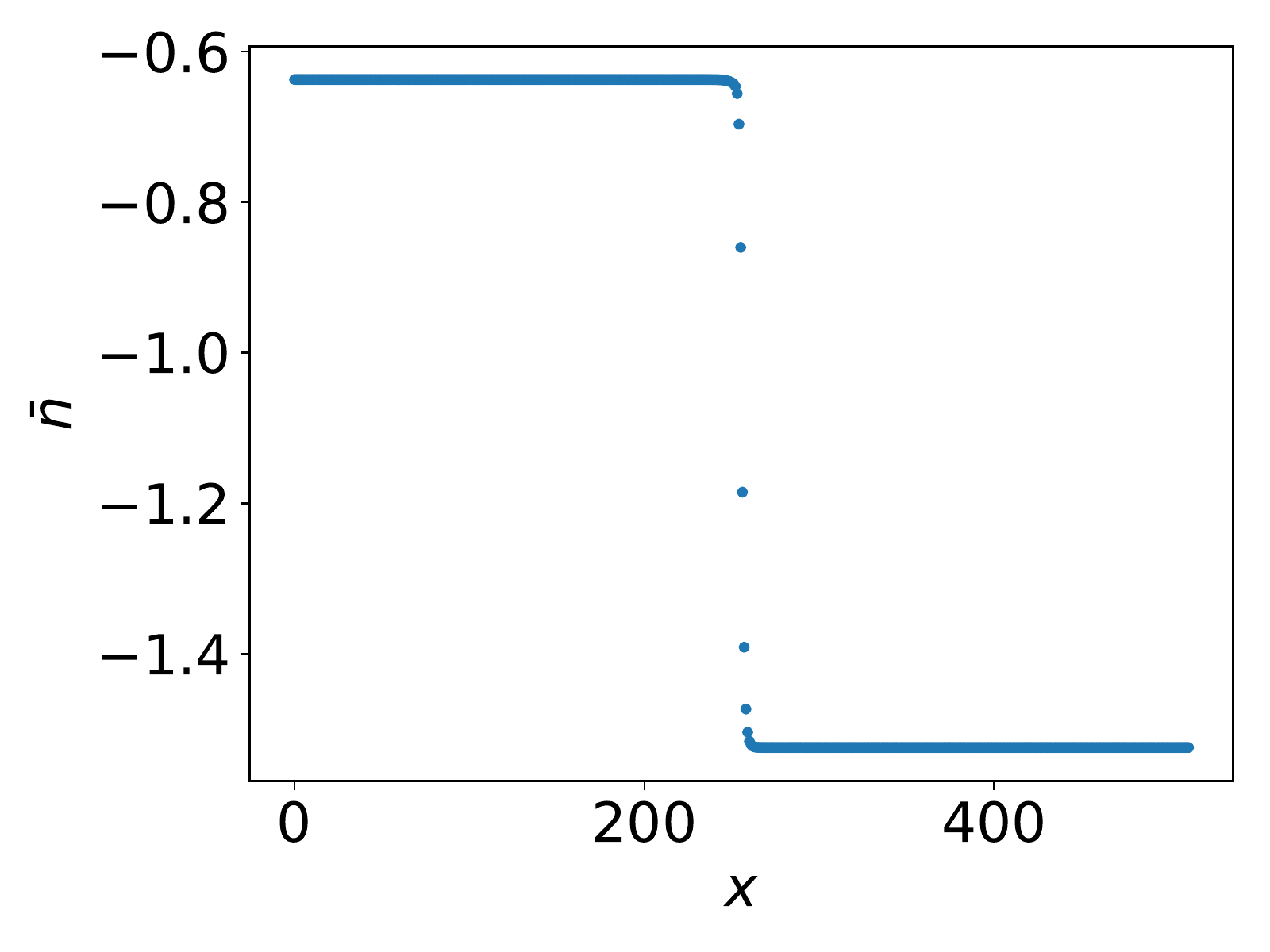}
\caption{Order parameter profiles $\beta_3$}
\end{subfigure} 
\begin{subfigure}[b]{.4\textwidth}
\resizebox{2.75in}{!}{\includegraphics{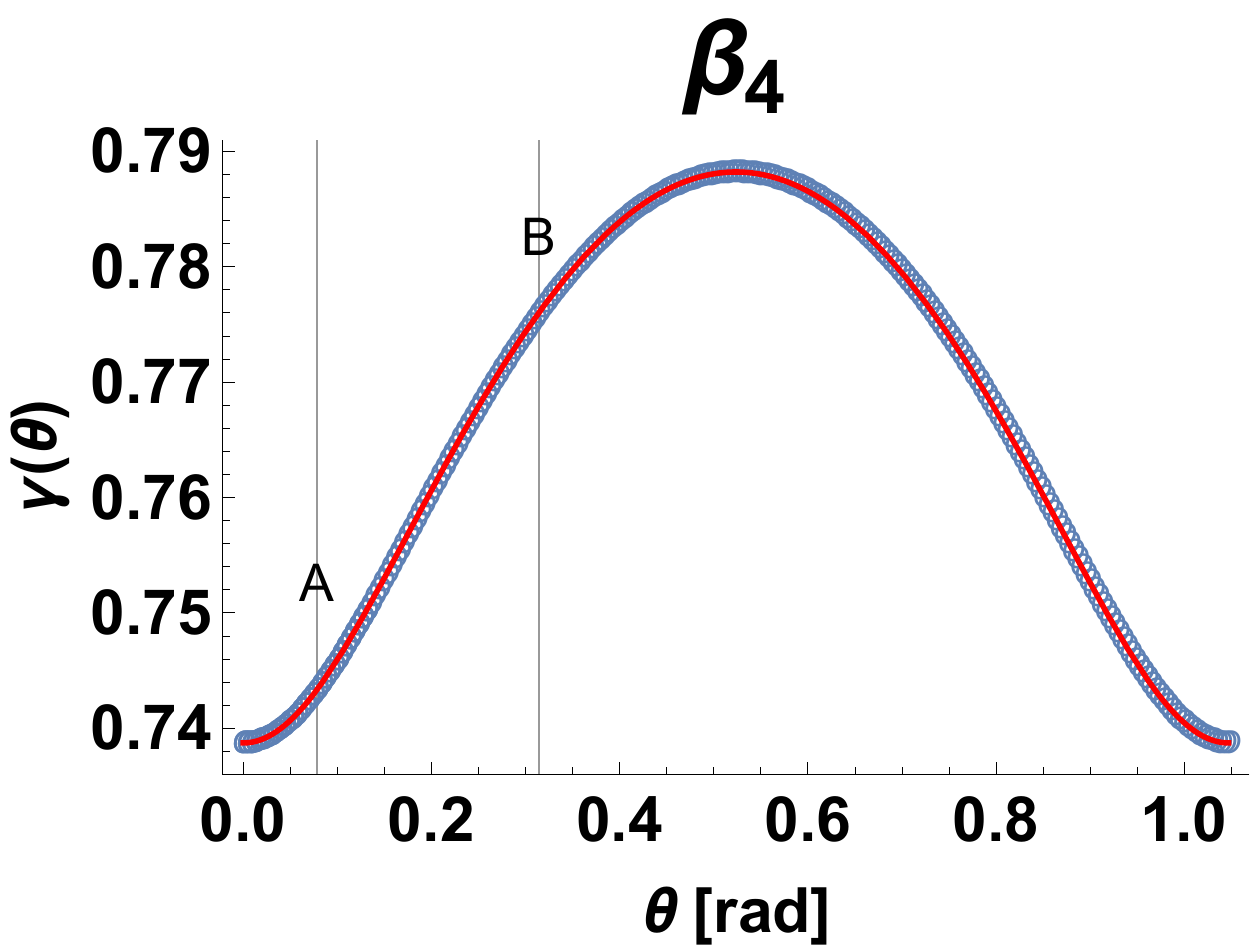}}
\caption{Surface energy $\beta_4$}
\end{subfigure}
\begin{subfigure}[b]{.5\textwidth}
 \includegraphics[width=1.5in]{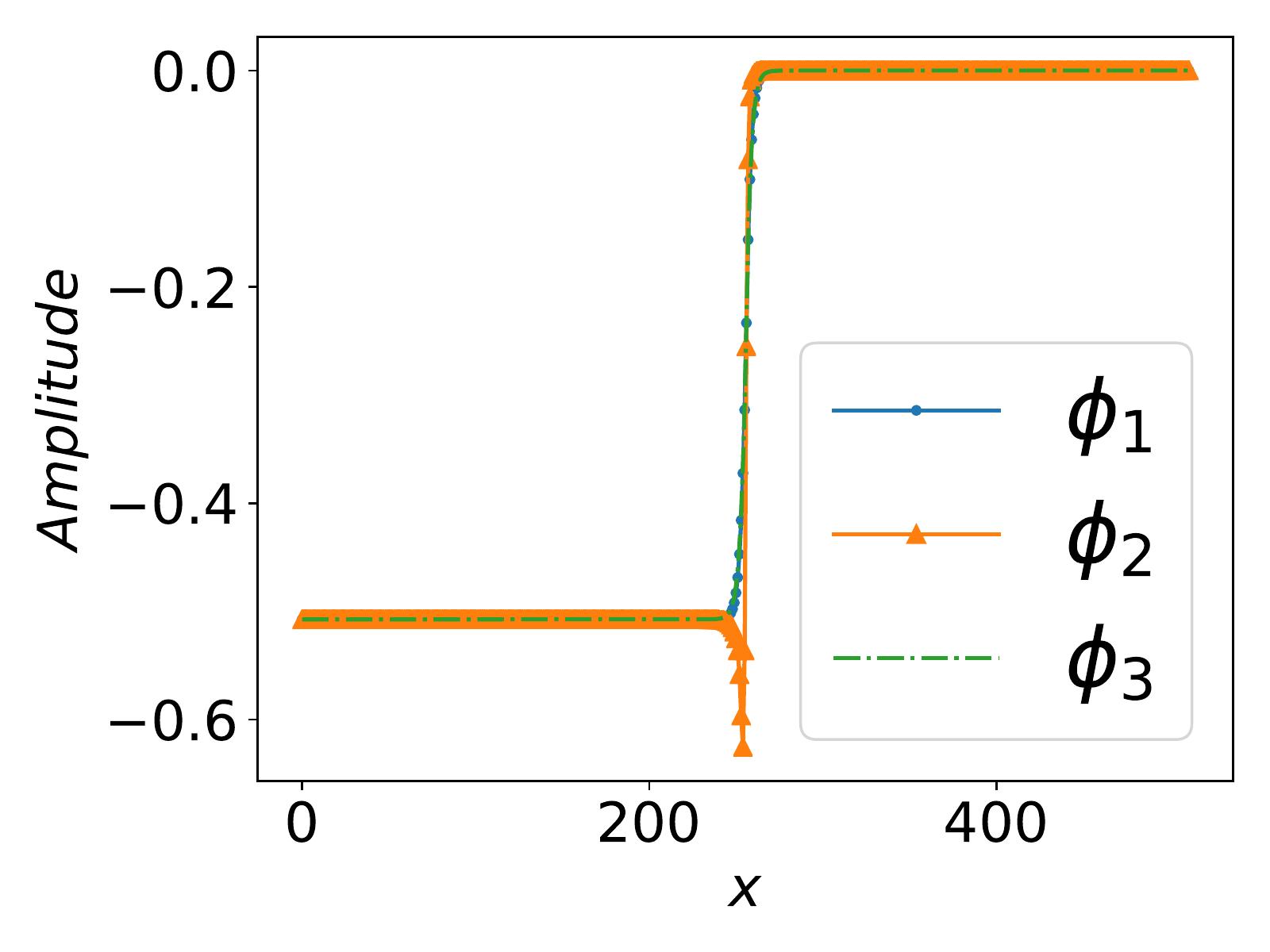} \hspace{0ex}  \includegraphics[width=1.5in]{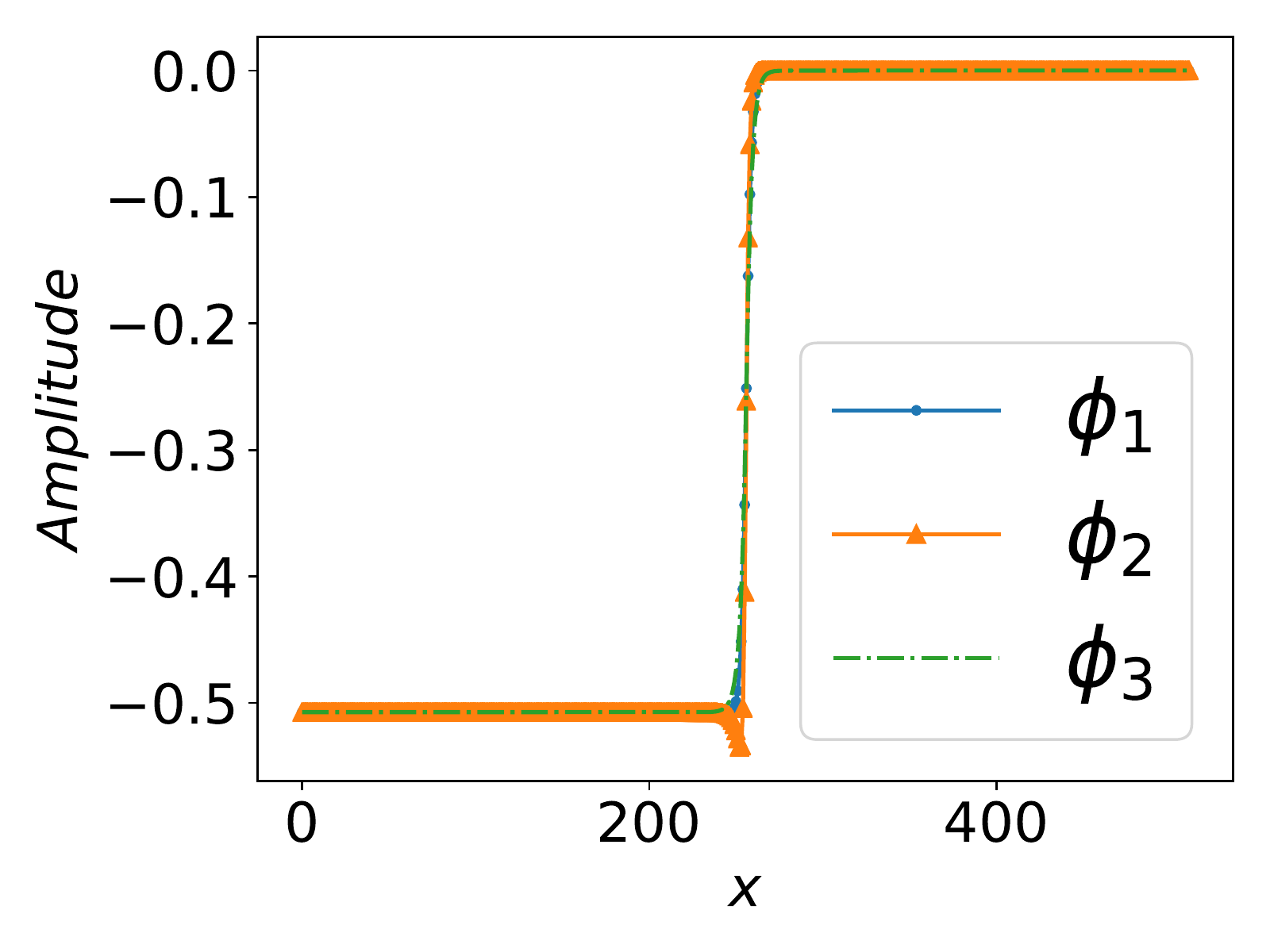}
\vspace{0ex}
\includegraphics[width=1.5in]{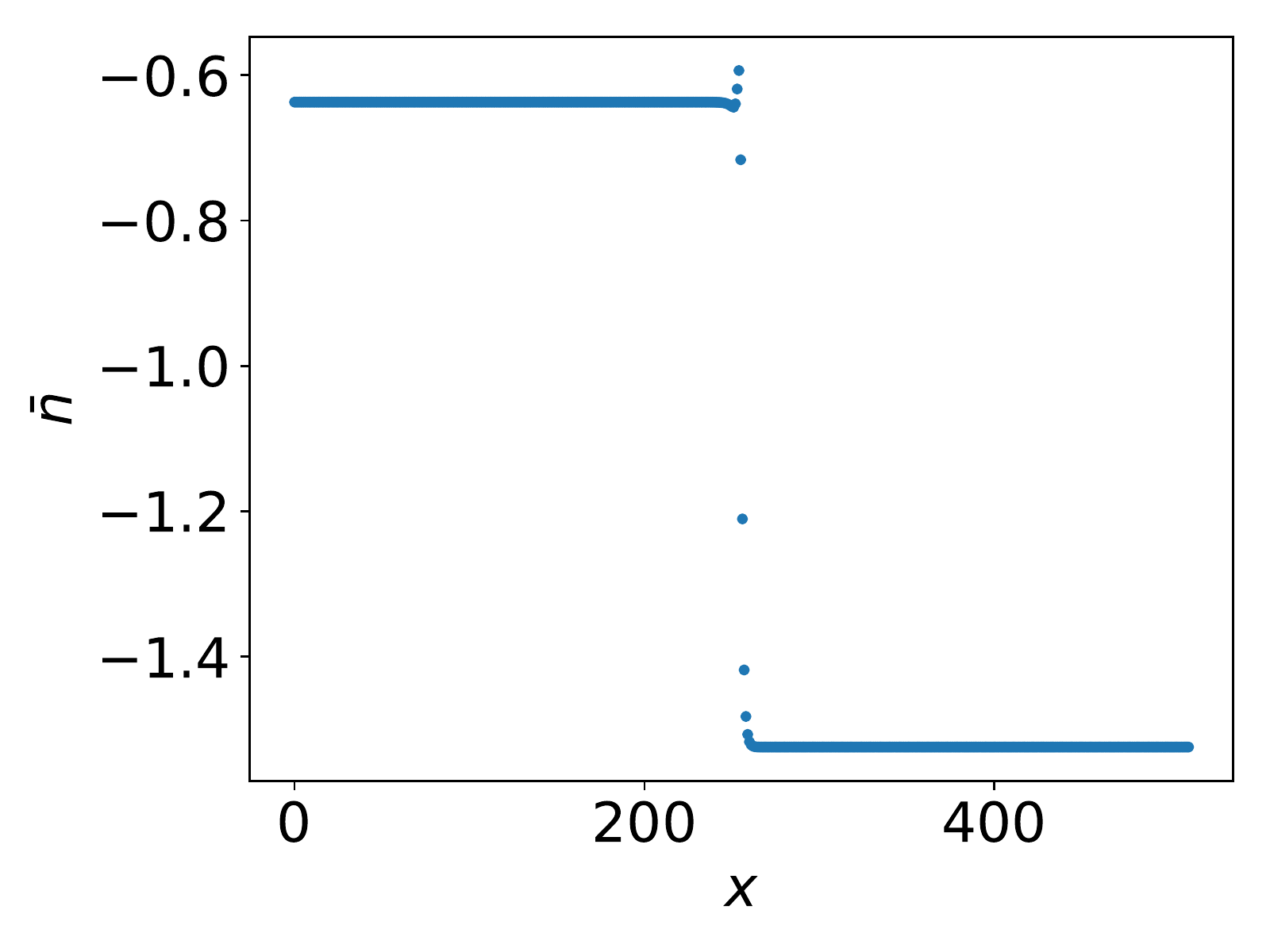} \hspace{0ex}  \includegraphics[width=1.5in]{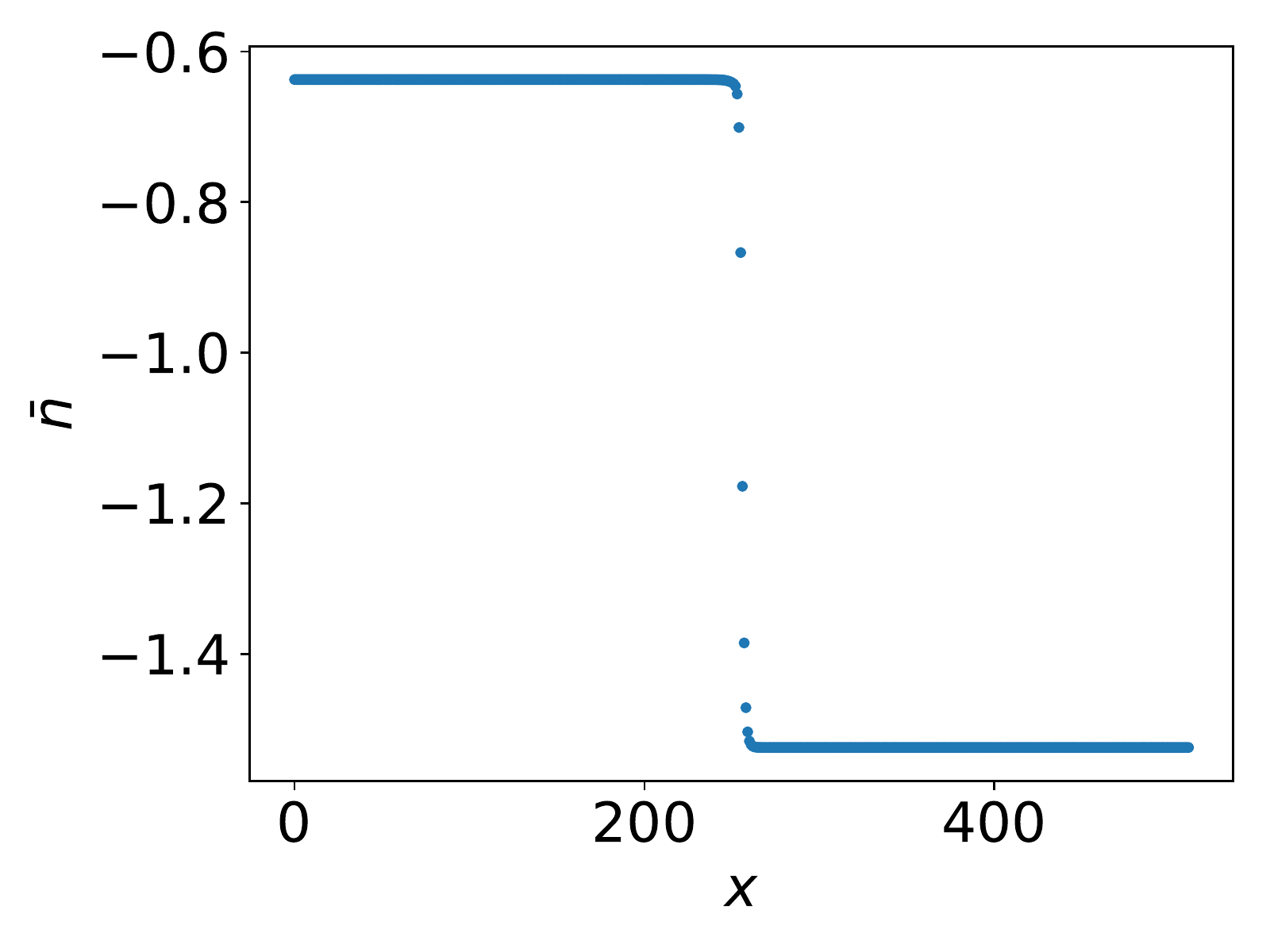}
\caption{Order parameter profiles $\beta_4$}
\end{subfigure} 
\caption{(Color online) Order parameter profiles and surface energy for $\Delta x_1$ at $\epsilon = -3.08$. (a) and (b)  $\beta = 1\times 10^{-4}$ while for (c) and (d) $\beta = 1\times 10^{-2}$.  Surface energy plots, (a) and (c) have raw simulation data in blue, while the red curve is a fit to \cref{eq:se-expan}. For the order parameter profiles, (c) and (d), amplitudes are shown (top row) and the density (bottom row) at the respective orientations $\theta = 0.025\pi$ and $0.1\pi$ marked by the lines A (left column) and B (right column) respectively.}
\label{fig:se-dx1-T2}
\end{figure}

To understand why beyond this transition temperature we see oscillations of the stiffness, we interrogate the surface energy and order parameter profiles for some of these parameters around this transition. We examine the profiles at the finest resolution, $\Delta x_1$, for corner energy coefficients, $\beta_3$ and $\beta_4$, and for temperatures $\epsilon = -3.08$ and $\epsilon = -5$ below the transition. As reference for the discussion, we have listed in \cref{tab:eqlm-prop} the equilibrium properties for the solid phase for the aforementioned temperature values.

\begin{figure}[!htb]
\centering

\begin{subfigure}[b]{.4\textwidth}
\resizebox{2.75in}{!}{\includegraphics{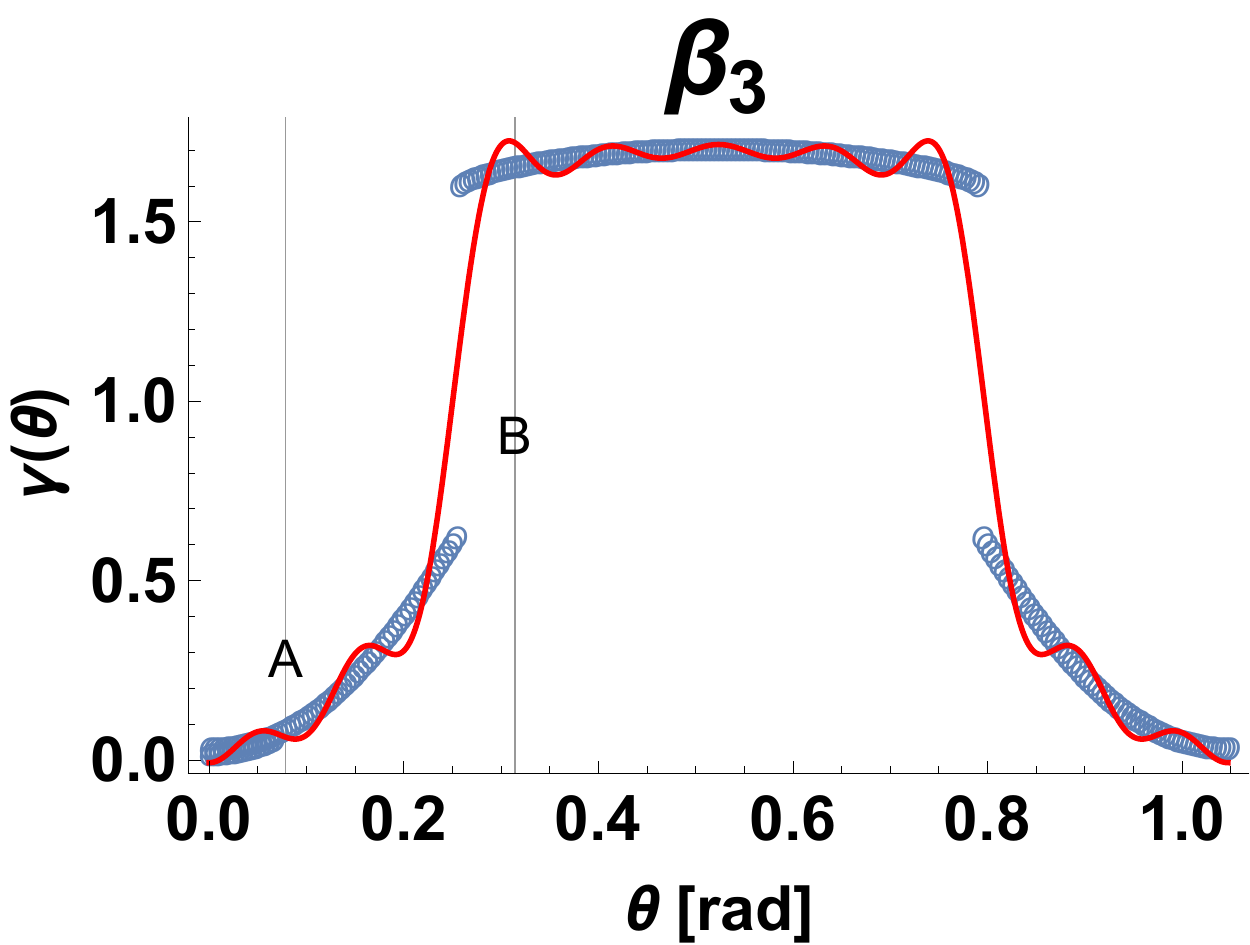}}
\caption{Surface energy $\beta_3$}
\end{subfigure}
\begin{subfigure}[b]{.5\textwidth}
 \includegraphics[width=1.5in]{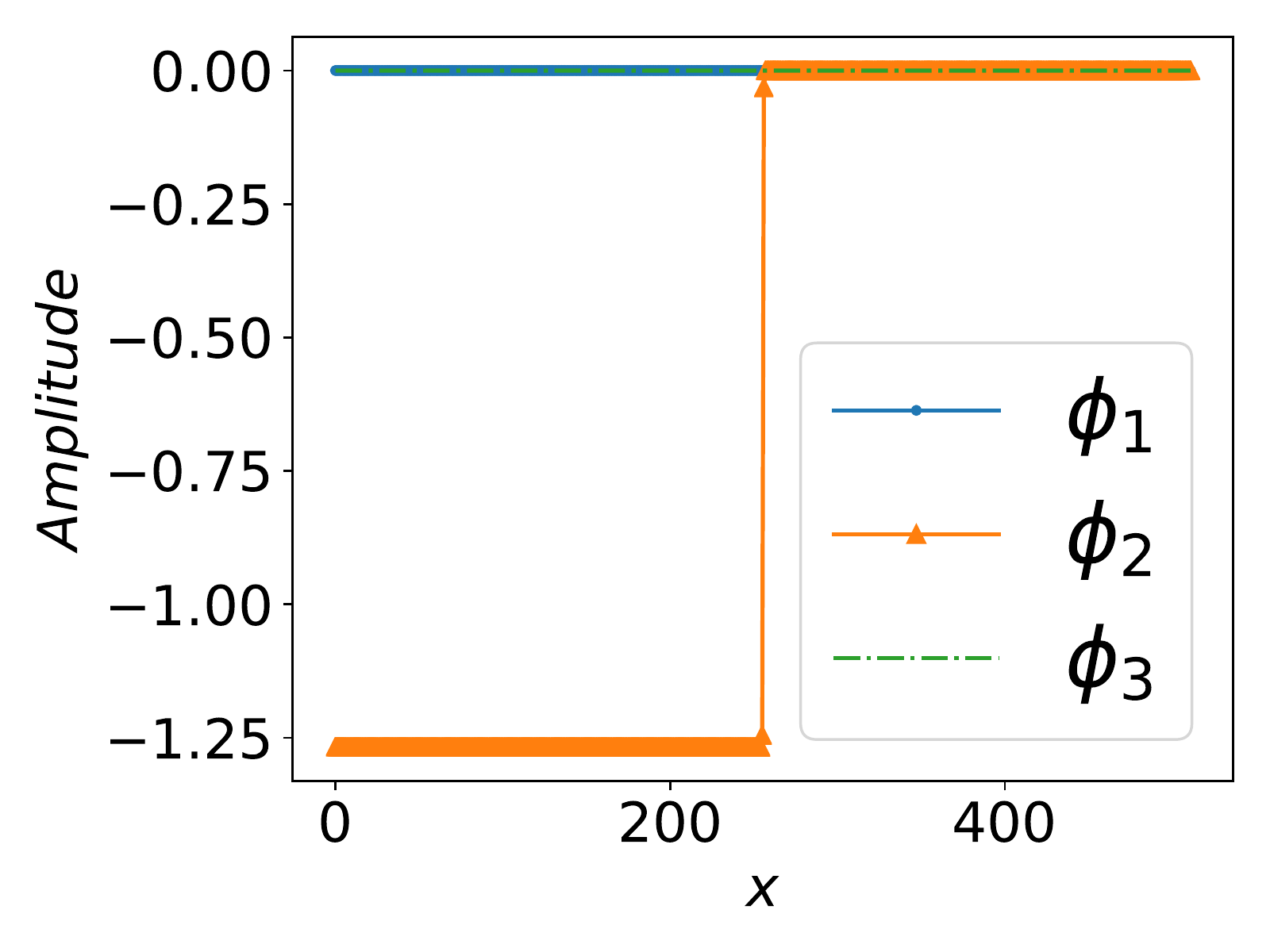} \hspace{0ex}  \includegraphics[width=1.5in]{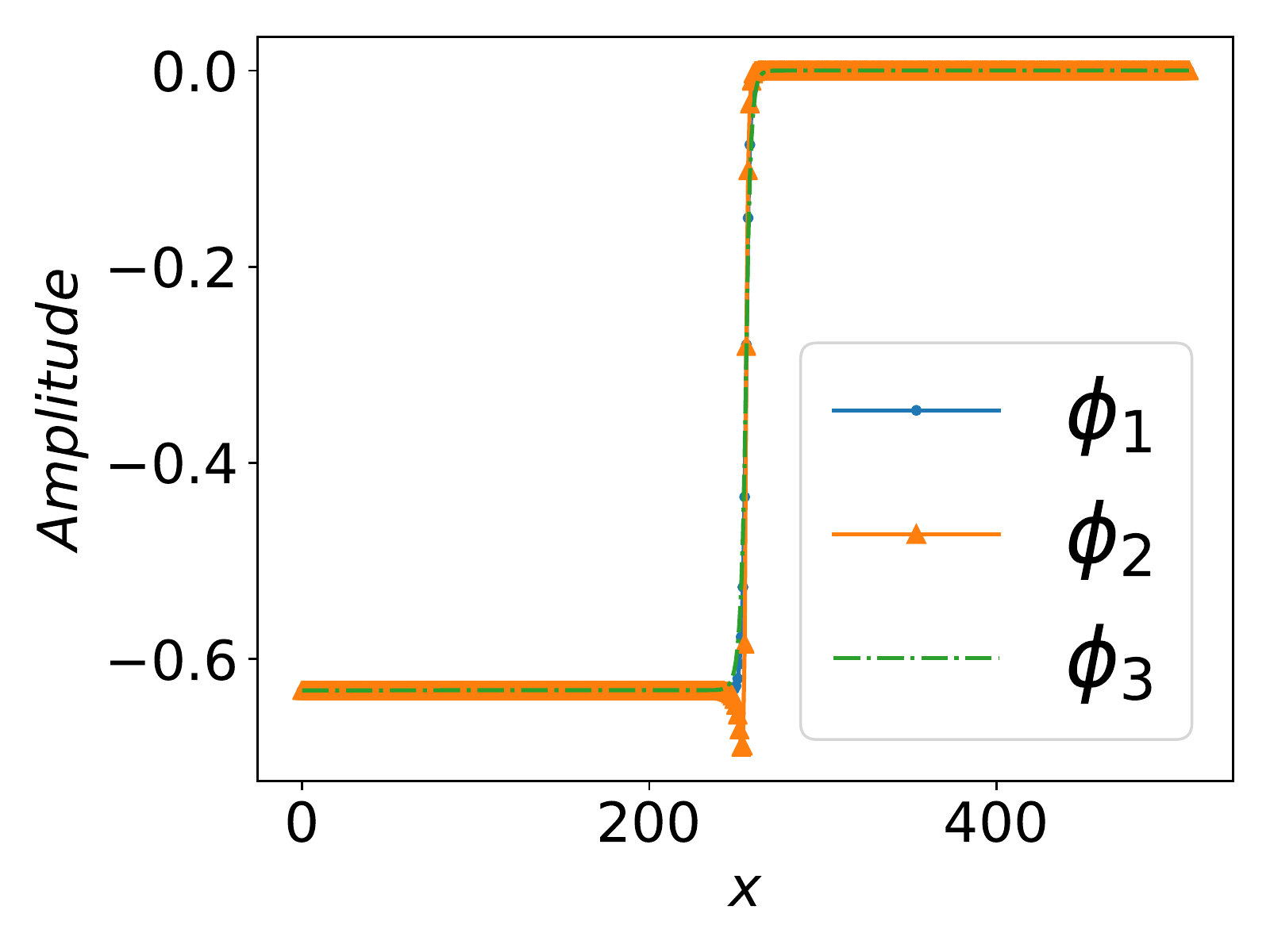}
\vspace{0ex}
\includegraphics[width=1.5in]{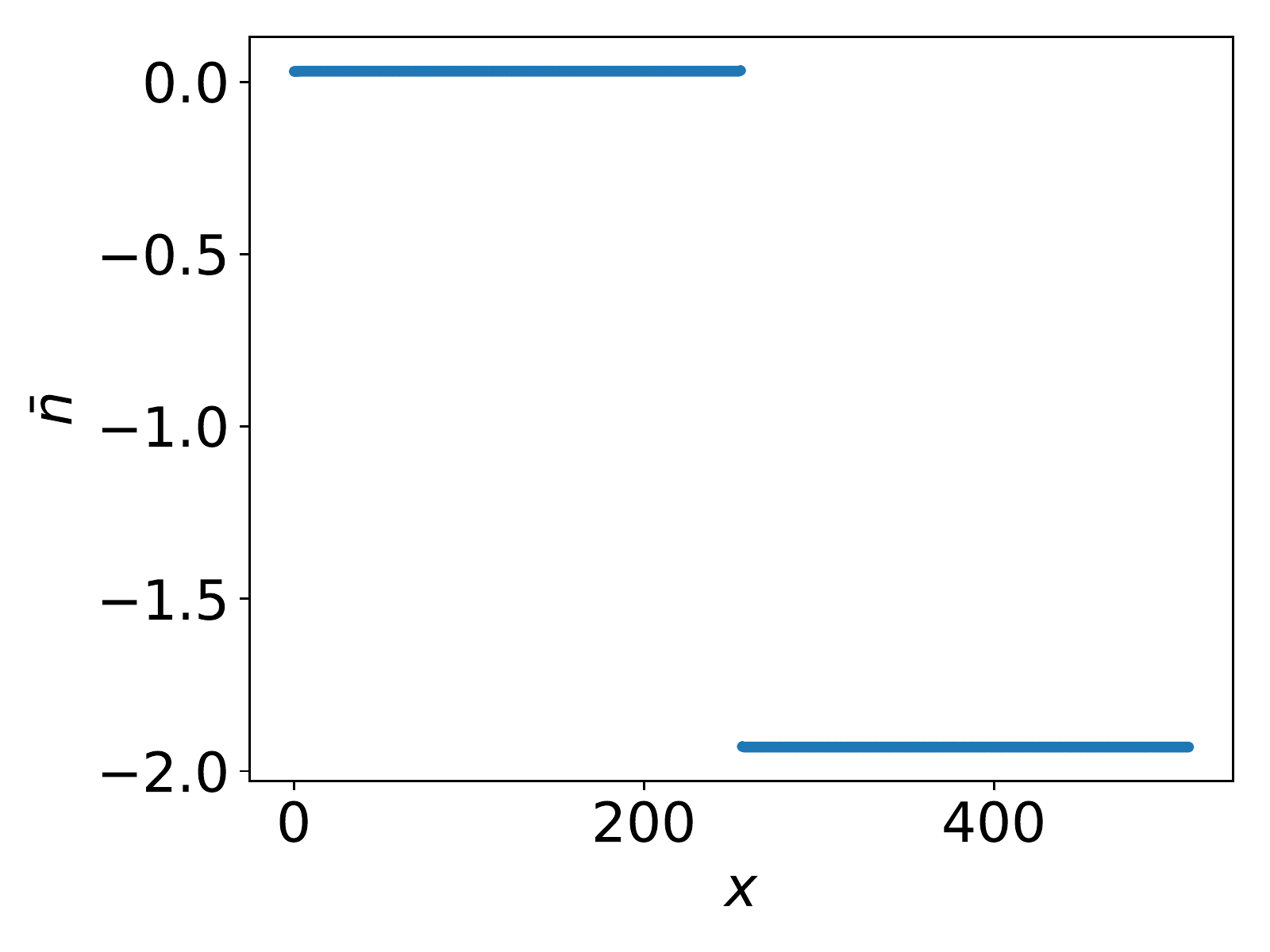} \hspace{0ex}  \includegraphics[width=1.5in]{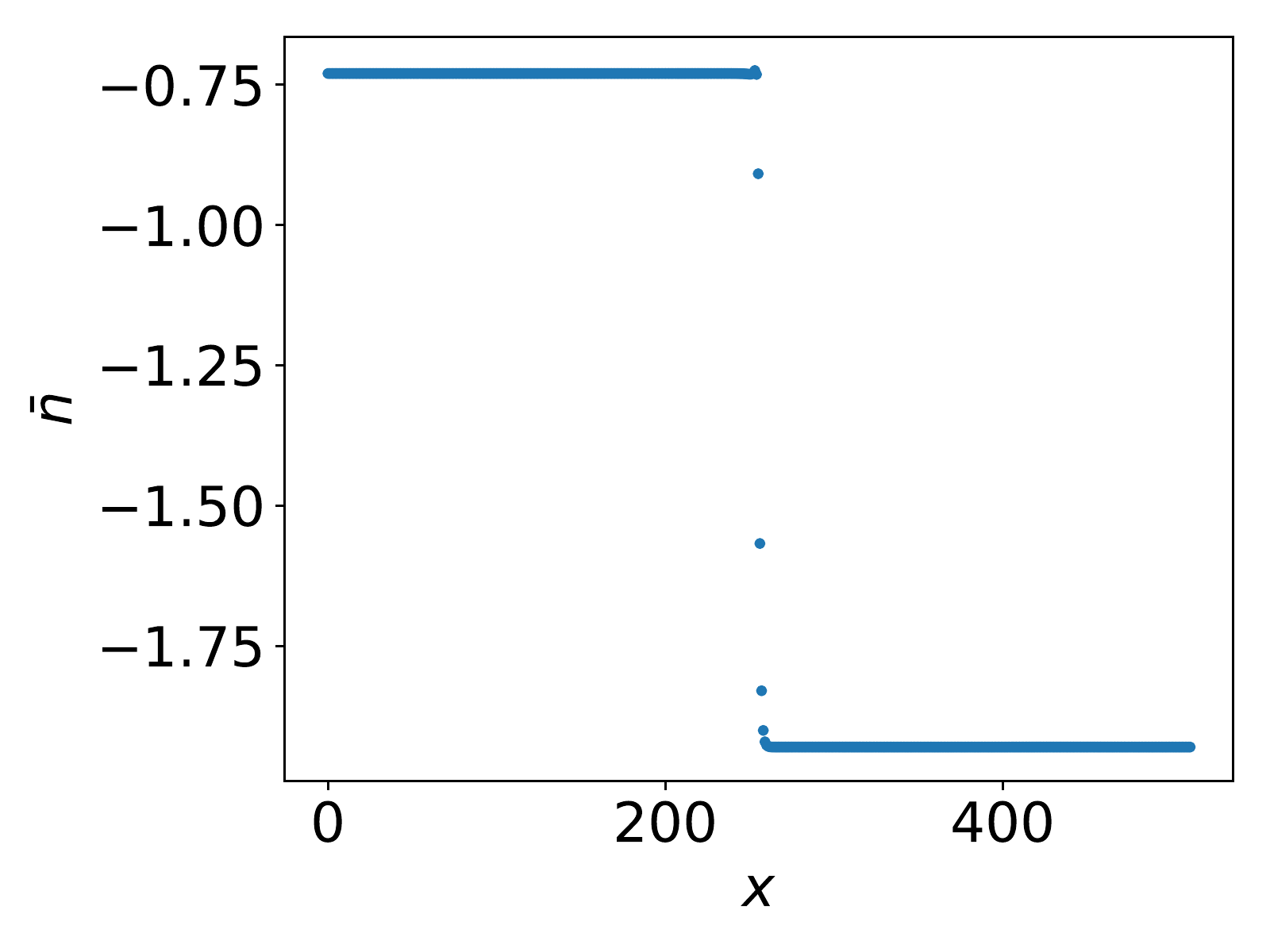}
\caption{Order parameter profiles $\beta_3$}
\end{subfigure}
\begin{subfigure}[b]{.4\textwidth}
\vspace{5ex}
\resizebox{2.75in}{!}{\includegraphics{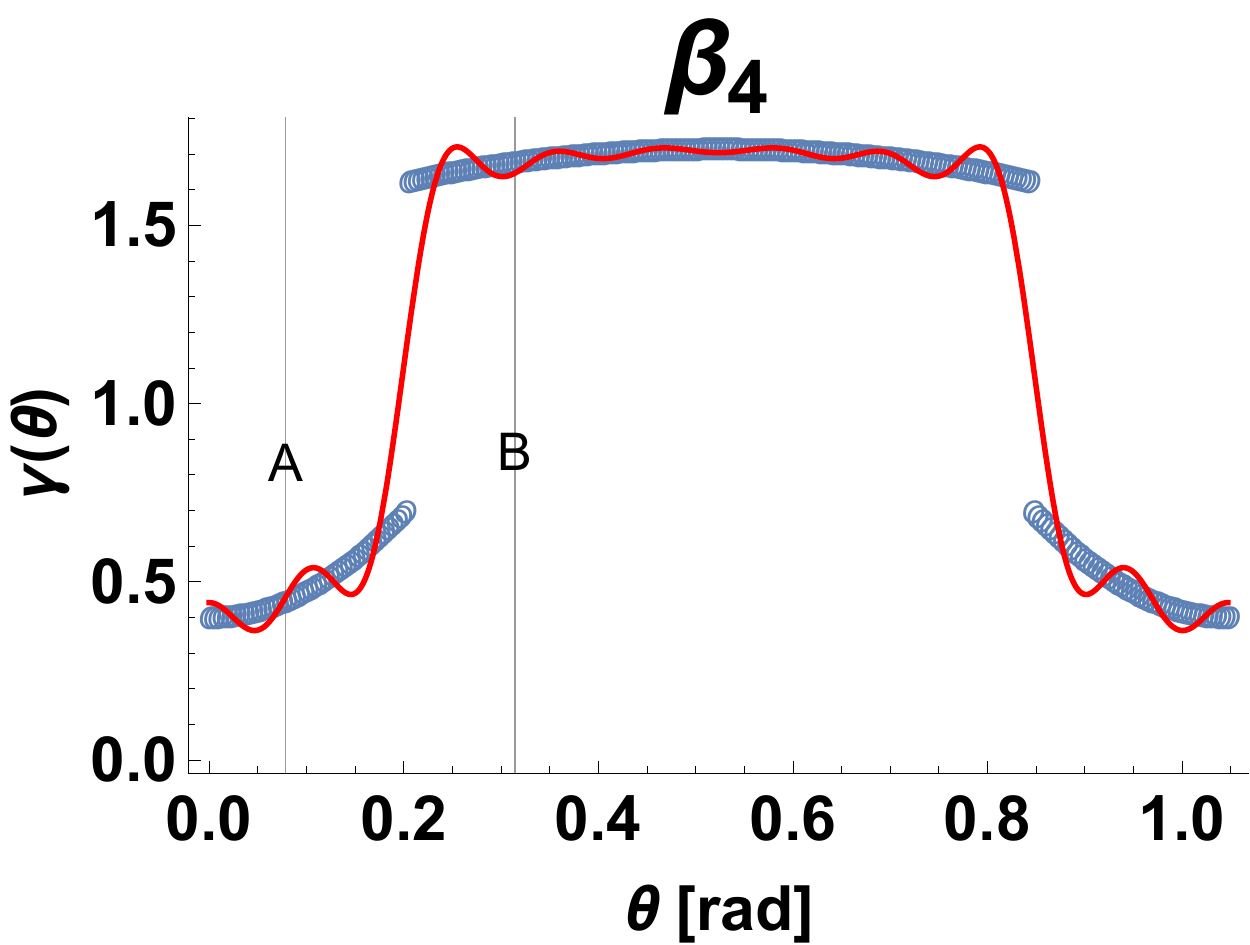}}
\caption{Surface energy $\beta_4$}
\end{subfigure}
\begin{subfigure}[b]{.5\textwidth}
 \includegraphics[width=1.5in]{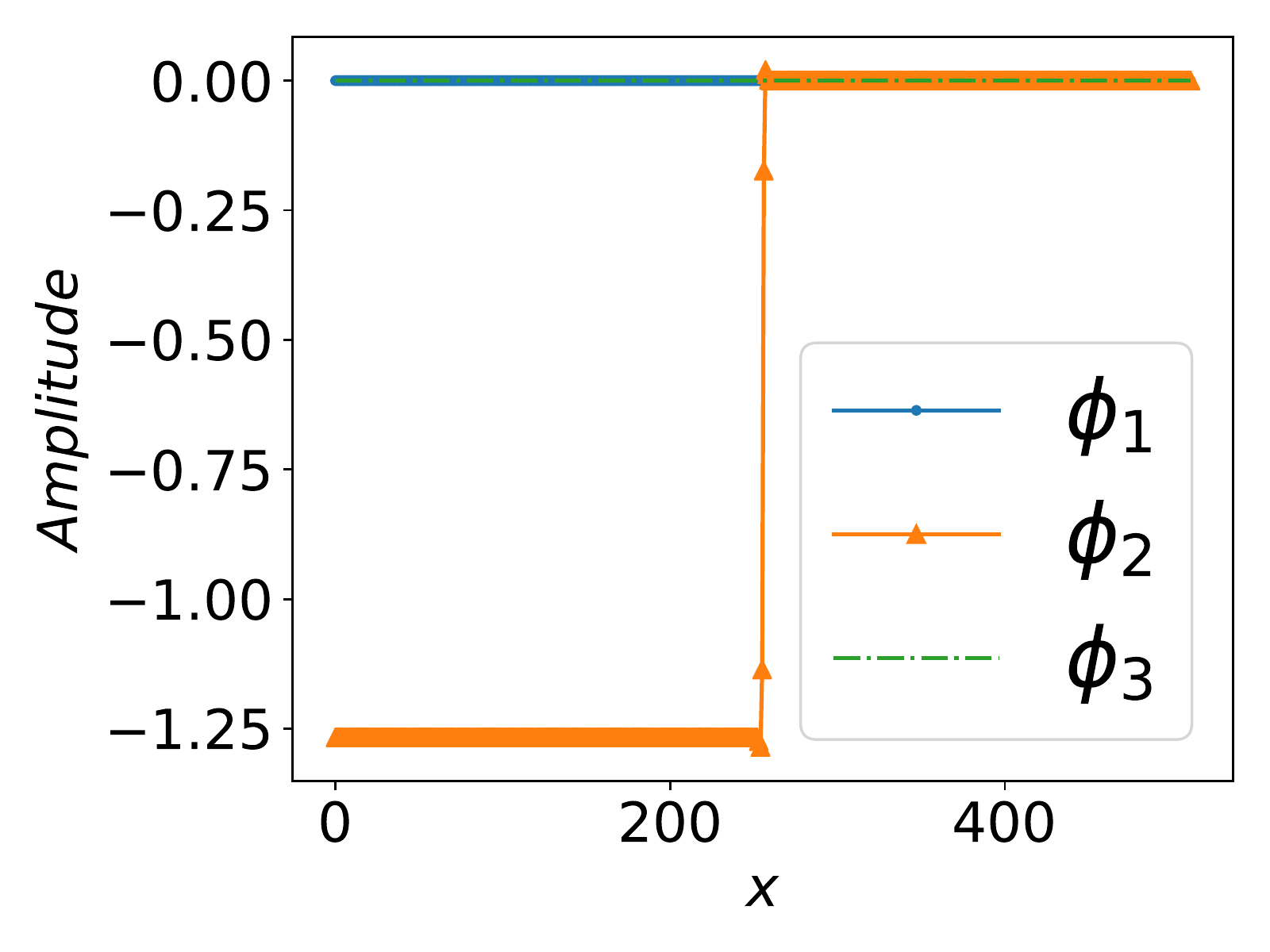} \hspace{0ex}  \includegraphics[width=1.5in]{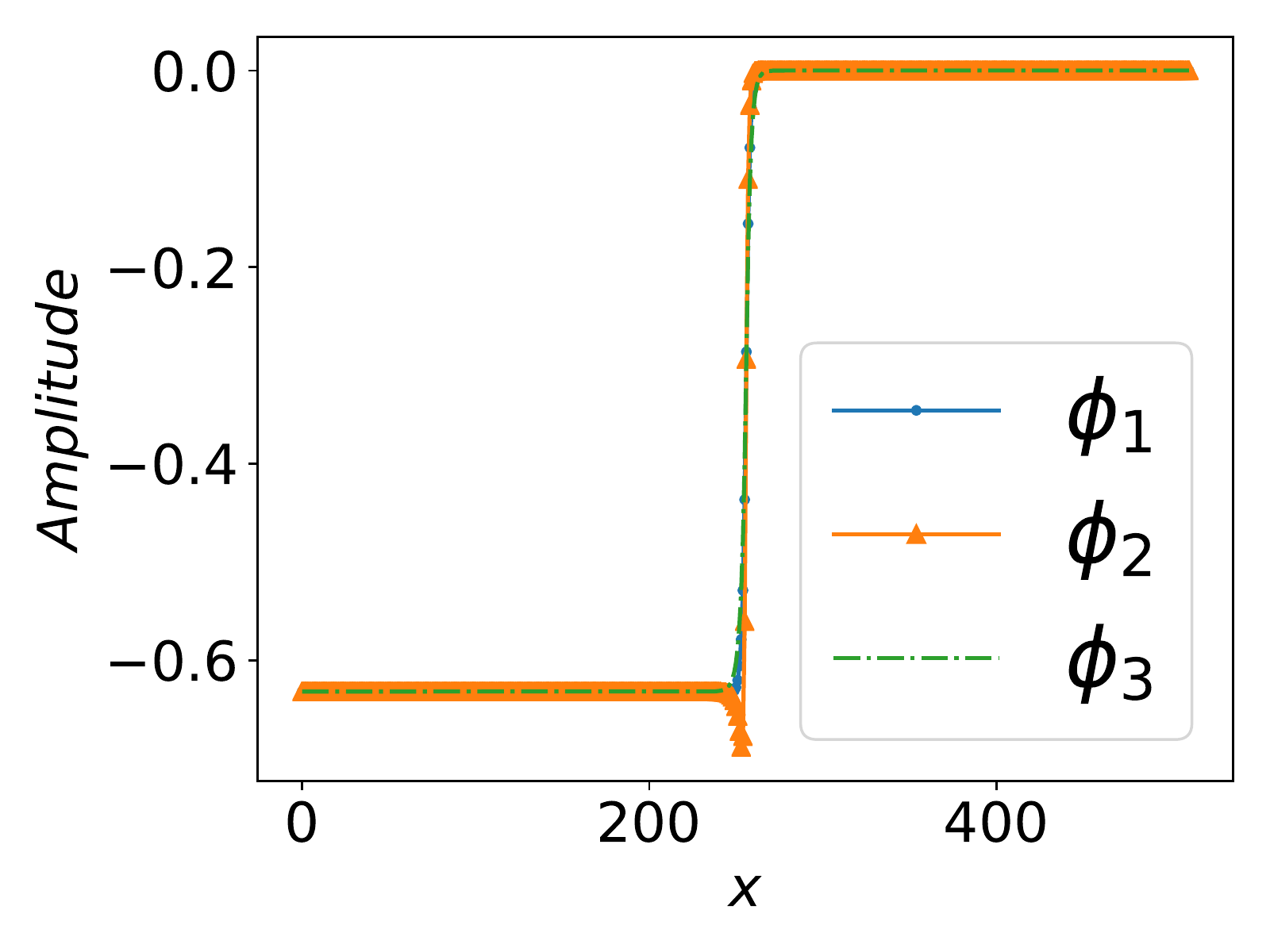}
\vspace{0ex}
\includegraphics[width=1.5in]{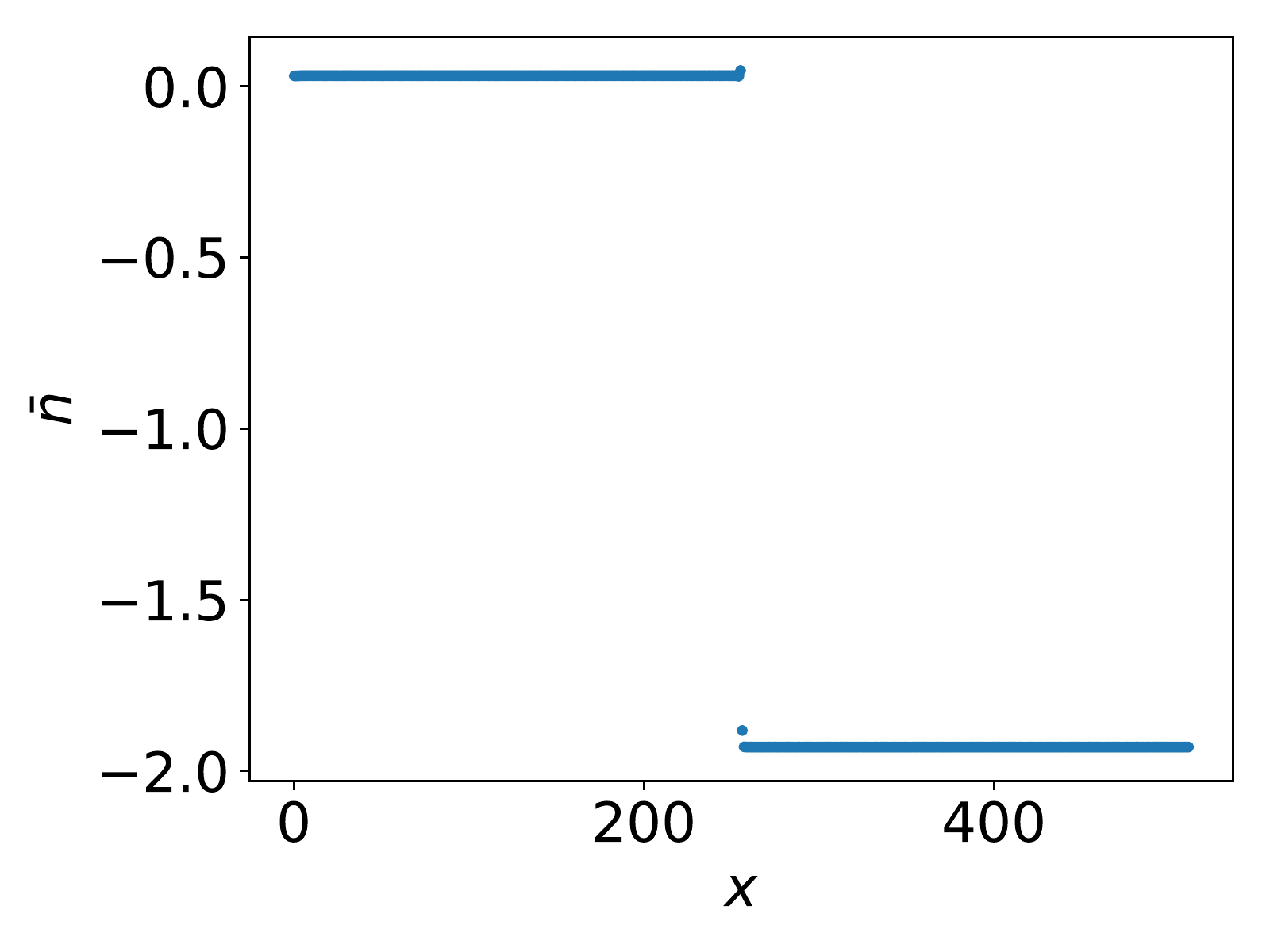} \hspace{0ex}  \includegraphics[width=1.5in]{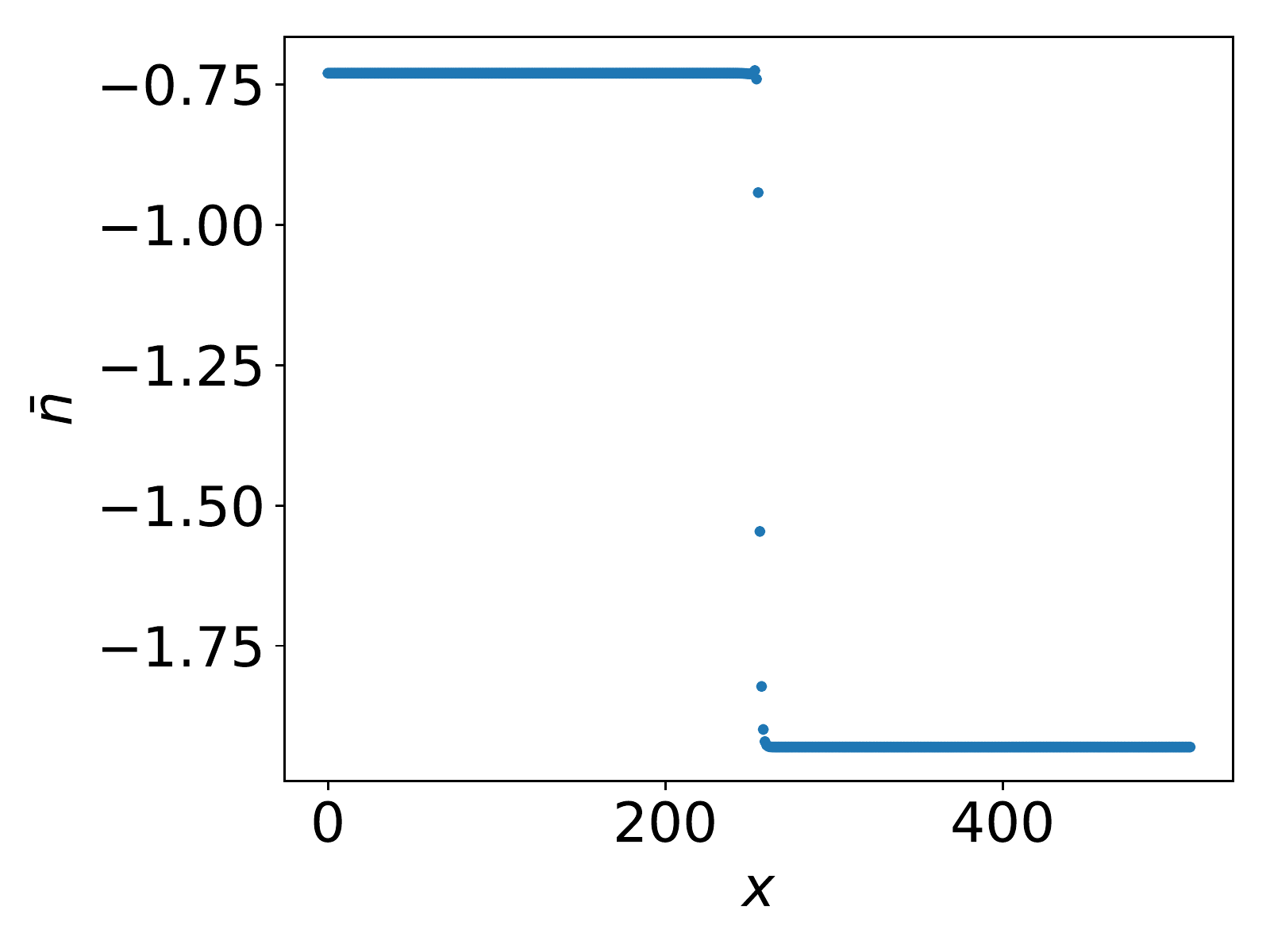}
\caption{Order parameter profiles $\beta_4$}
\end{subfigure} 
\caption{(Color online) Profiles and surface energy plot for $\Delta x_1$ and $\epsilon=-5$. (a) and (b)  $\beta = 1\times 10^{-4}$ while for (c) and (d) $\beta = 1\times 10^{-2}$. Surface energy plots, (a) and (c) have raw simulation data in blue, while the red curve is a fit to \cref{eq:se-expan}. For the order parameter profiles, (c) and (d), amplitudes are shown (top row) and the density (bottom row) at the respective orientation $\theta = 0.025\pi$ and $0.1\pi$ marked by the lines A (left column) and B (right column) respectively.}
\label{fig:se-dx1-T1}
\end{figure}

\Cref{fig:se-dx1-T2} shows the surface energy plots for $\beta_3$ (\cref{fig:se-dx1-T2}(a)) and $\beta_4$ (\cref{fig:se-dx1-T2}(c)) at $\epsilon = -3.08$. In the plots, the simulation is represented by the blue data points, while the red line is the fit using \cref{eq:se-expan}. The lines A and B delineate the orientations $\theta = 0.025\pi$ and $0.1\pi$ respectively, for which we have plotted the order parameter profiles in \cref{fig:se-dx1-T2}(b) and (d) respectively. In the profiles of the order parameters, the amplitudes are displayed in the top row, while the density in the bottom row, with the left column corresponding to line A and the right to line B. The same convention is used in \cref{fig:se-dx1-T1}. In \cref{fig:se-dx1-T2}(a) the surface energy exhibits a discontinuous, two branched behavior. The upper branch represents the allowed, stable orientations on the ECS. While the lower branch, representing those orientations closest to the high symmetry orientations of the crystal, depict missing/unstable orientations. The surface energy for the latter set of orientations were calculated by relaxing the equilibrium constraint on \cref{eq:surf-ene}; see \cref{derive-surfene}. The unstable character of these orientations is clearly visible in the plots of the order parameter fields presented. Choosing the orientations marked by lines A $(\theta = 0.025\pi)$ and B $(\theta = 0.1\pi)$, we see that for the former, only one of the amplitudes is nonzero ($\phi_2 \sim -1$), and its value deviates from the equilibrium value, $\phi_{eq}=-0.48167$, from \cref{tab:eqlm-prop}. Similar with the density $\bar{n}_s = -.1$ when compared with $\bar{n}_{sol}=-0.61819$. Conversely, the order parameters for $\theta = 0.1\pi$, have all converged to their equilibrium values. This is indicative that for the unstable orientations, we have lost integrity of the triangular crystal structure represented by our amplitude expansion and therefore a triangular solid described by this system of equations is unstable. Moreover, our subsequent attempts to fit the full range of surface energy (red line with $N=15$), below the transition, is the likely cause behind the oscillations exhibited in \cref{fig:sec-derv}. The fact that the oscillations mimic a sinusoidal-like behavior is also presumably caused by the functional form of the fitting function we have used here. For $\beta_4$, \cref{fig:se-dx1-T2}(b), we see a well behaved, continuous surface energy plot with no missing orientations. The order parameter profiles further reinforce this for the representative orientations chosen.

At $\epsilon = -5$, displayed in \cref{fig:se-dx1-T1}, we witness two interesting features. First, for \cref{fig:se-dx1-T1}(a), we notice that the number of unstable orientations has increased, a trend towards complete faceting. The order parameter profiles coincide with our previous discussion. Once again for the unstable orientations, we have no solid integrity as $\phi_1=\phi_3=0,\phi_2\sim-1.2$ and $\bar{n}_s = 0$, showing significant deviations from their expected equilibrium values (\cref{tab:eqlm-prop}). Secondly, we see that for $\beta_4$, \cref{fig:se-dx1-T1}(c), a transition has occurred where now unstable orientations are present. The number of unstable orientations however, is smaller, specifically due to the larger corner energy coefficient regularizing the interface. In this case, when we compare the two plots, it becomes apparent that like temperature, the corner energy coefficient in this description, can also influence the number of and transition to unstable orientations.

We should mention here that a discontinuous surface energy is strictly speaking unphysical, even at the limit of metastability. Here, we only observe this behavior after relaxing the equilibrium constraint placed on \cref{eq:surf-ene}, which in doing so allows us to calculate a general relative excess quantity; see \cref{derive-surfene}. A quantity that in the regime of stable orientations, i.e., the upper branch solution, gives exactly the equilibrium constrained surface energy equation. Note that if the constraint of equilibrium was imposed for all orientations, the resulting surface energy values for the unstable orientations discussed above would be negative and therefore prohibited. Applying the equilibrium conditions, we performed simulations that confirmed that indeed by not allowing the negative surface energy values, only the upper branch, i.e., stable orientations, are selected. We further note that the orientations where the surface energy is negative, the interface can varifold and effectively undergo a decomposition process that eliminates these surfaces. We have tested this in 2D simulations of arbitrary initial sinusoidal interface shapes.

\subsection{2D Crystal Growth}
\label{results-2D}
\begin{figure*}[h]
\centering
\begin{tabular}{ccccc}
	$\epsilon = -0.1$ & $\epsilon = -0.5$ & $\epsilon = -1$ & $\epsilon = -2$ & $\epsilon = -3$ \\
	 \hline\hline\\
	 \resizebox{1.2in}{!}{\includegraphics{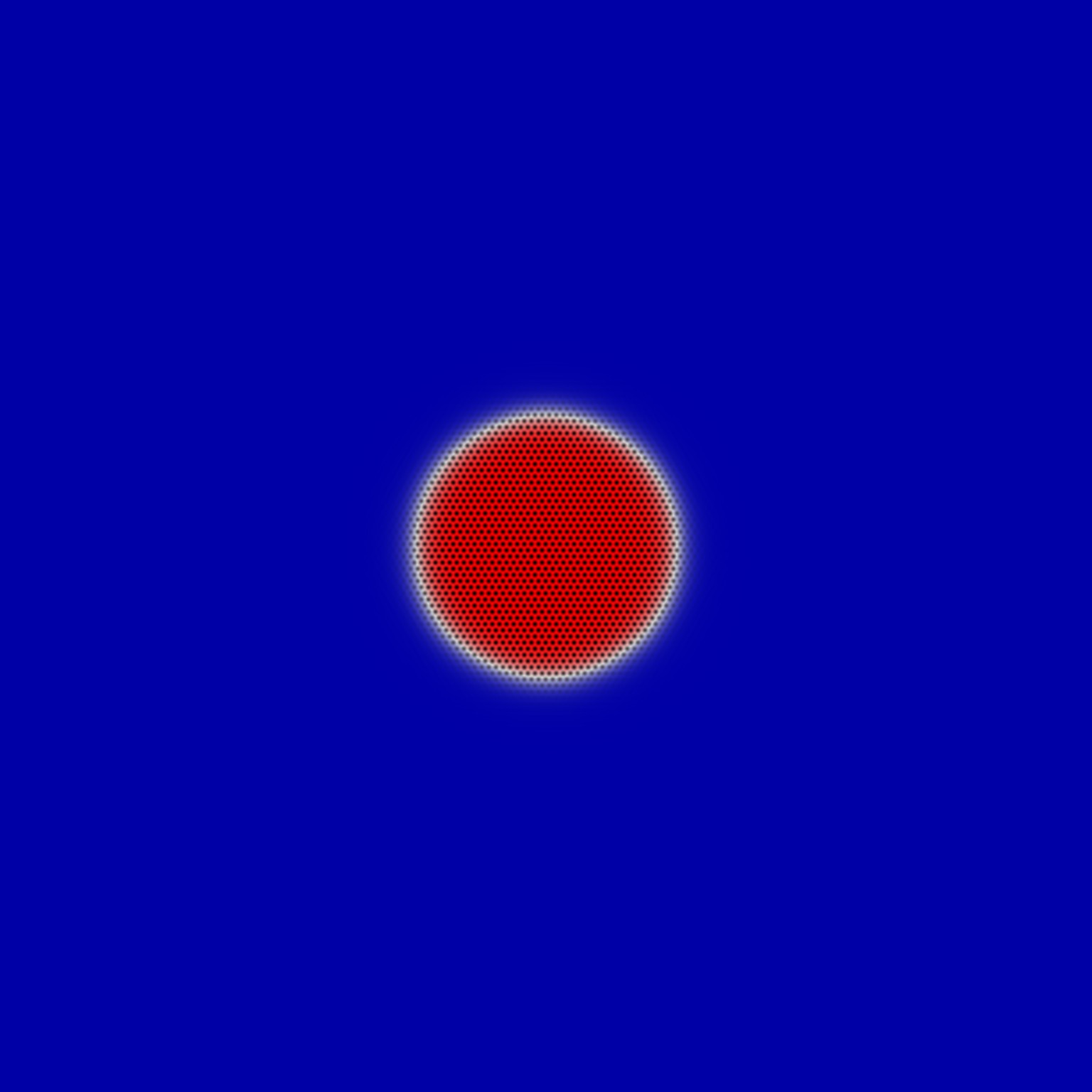}} & \resizebox{1.2in}{!}{\includegraphics{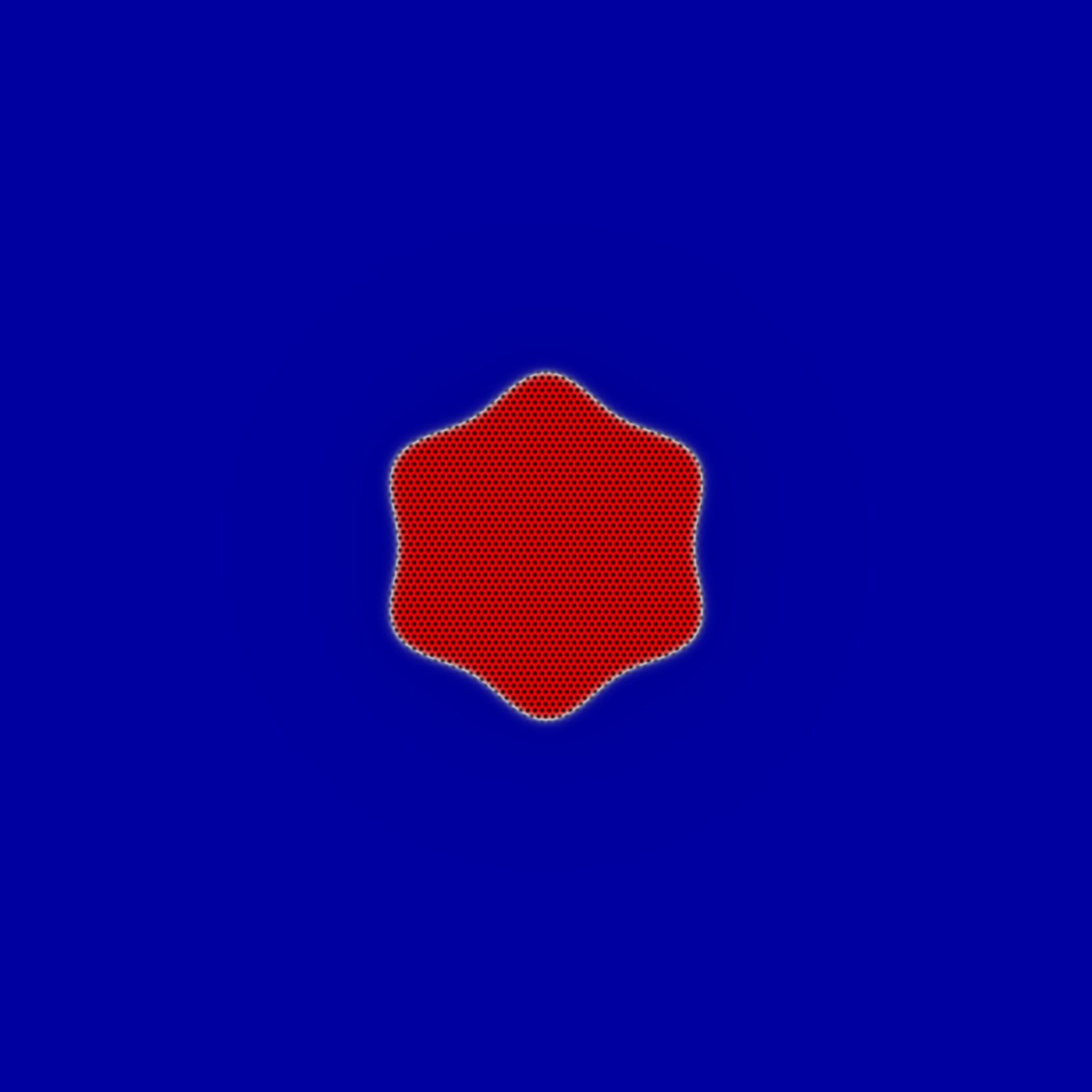}} & \resizebox{1.2in}{!}{\includegraphics{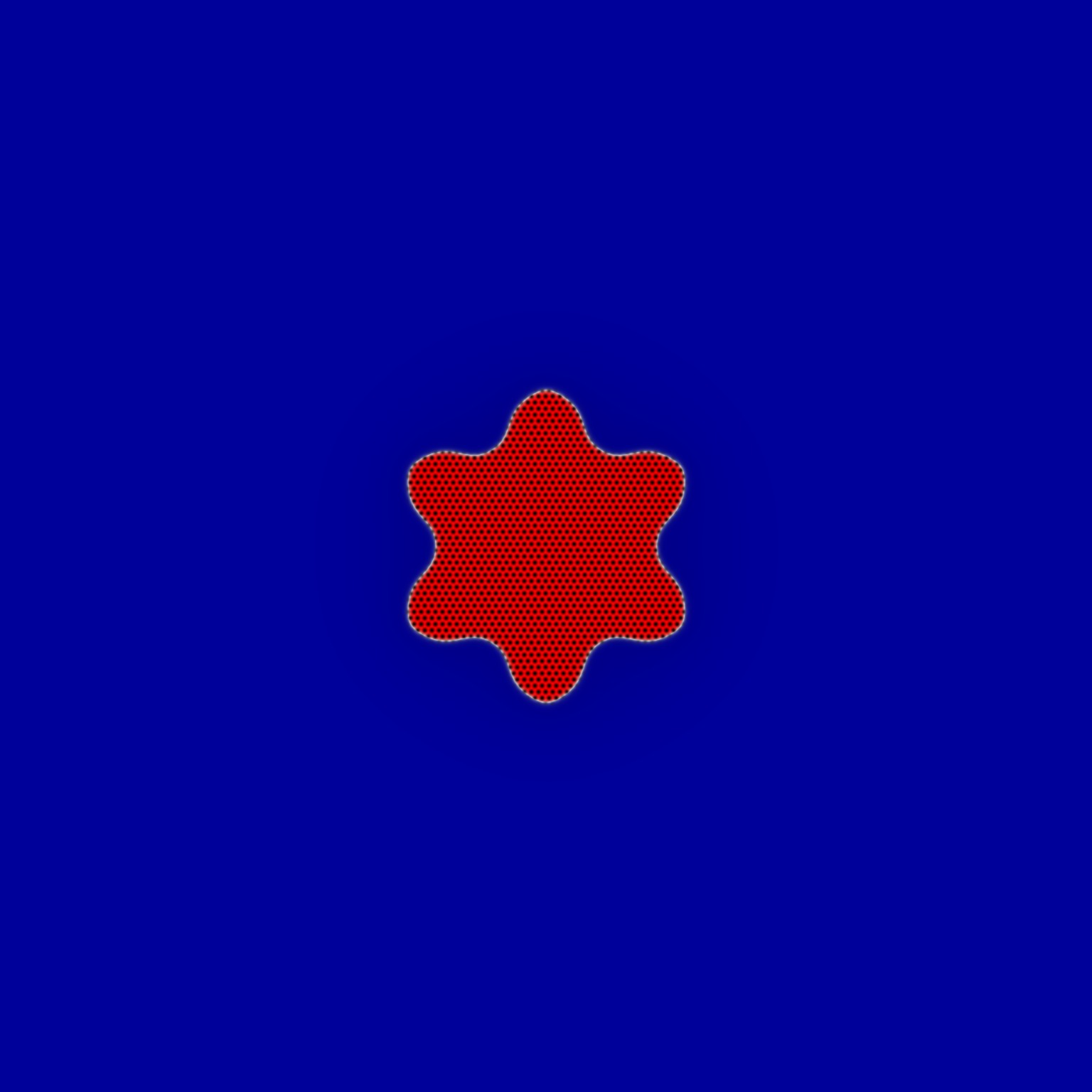}} & \resizebox{1.2in}{!}{\includegraphics{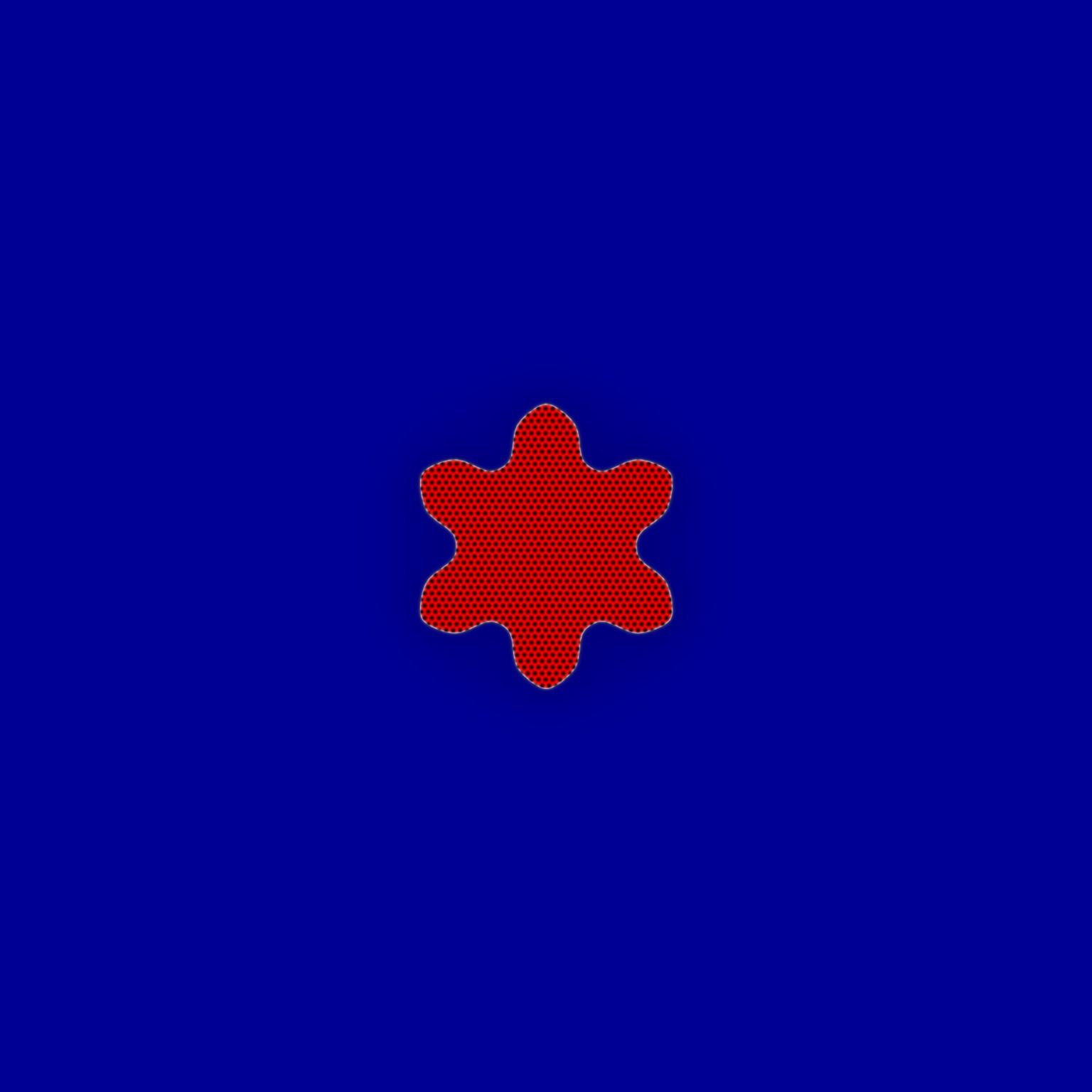}} & \resizebox{1.2in}{!}{\includegraphics{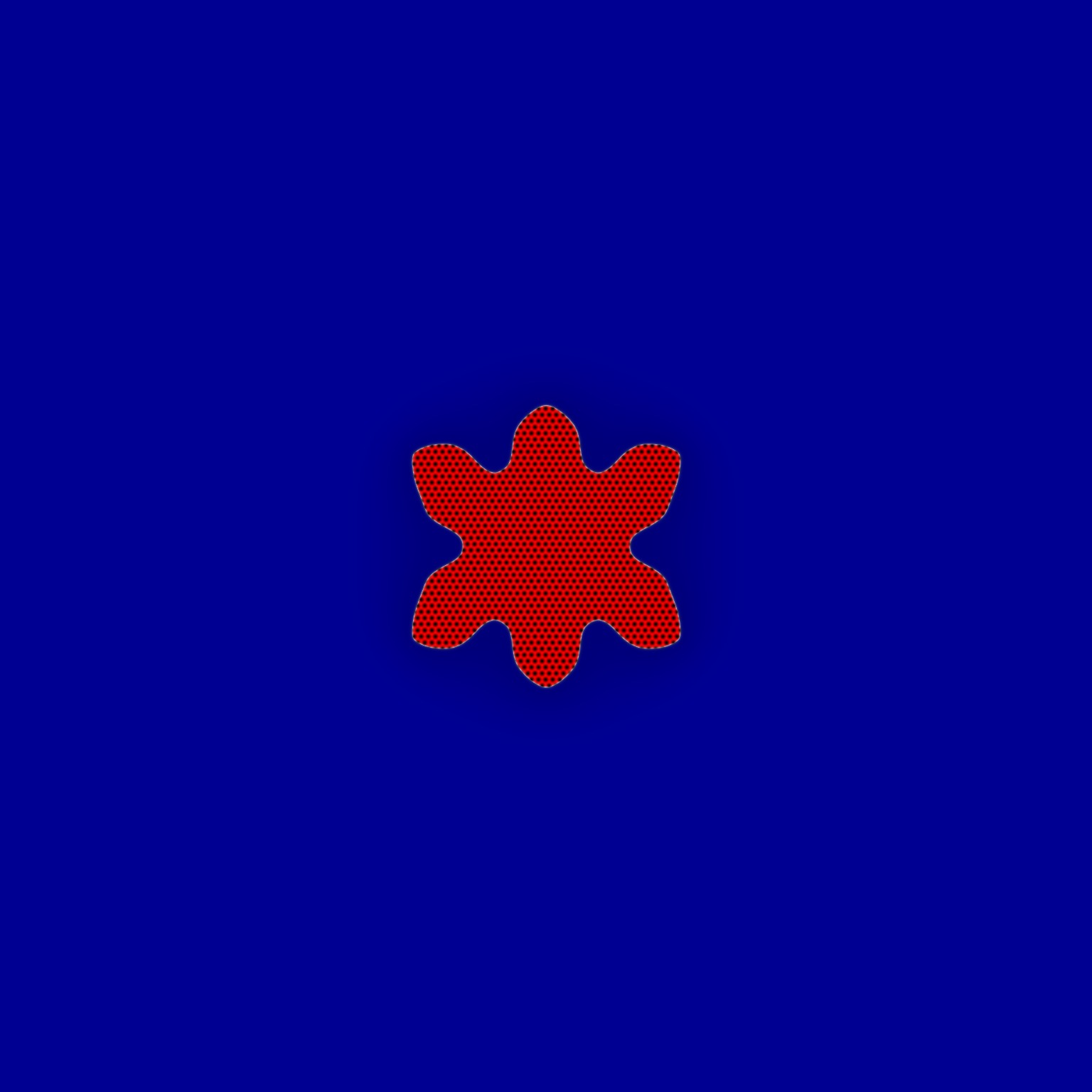}} \\
	$t = 61\tau$ & $t = 57\tau$ & $t = 30\tau$ & $t = 15\tau$ & $t = 12\tau$ \\ \\
	\resizebox{1.2in}{!}{\includegraphics{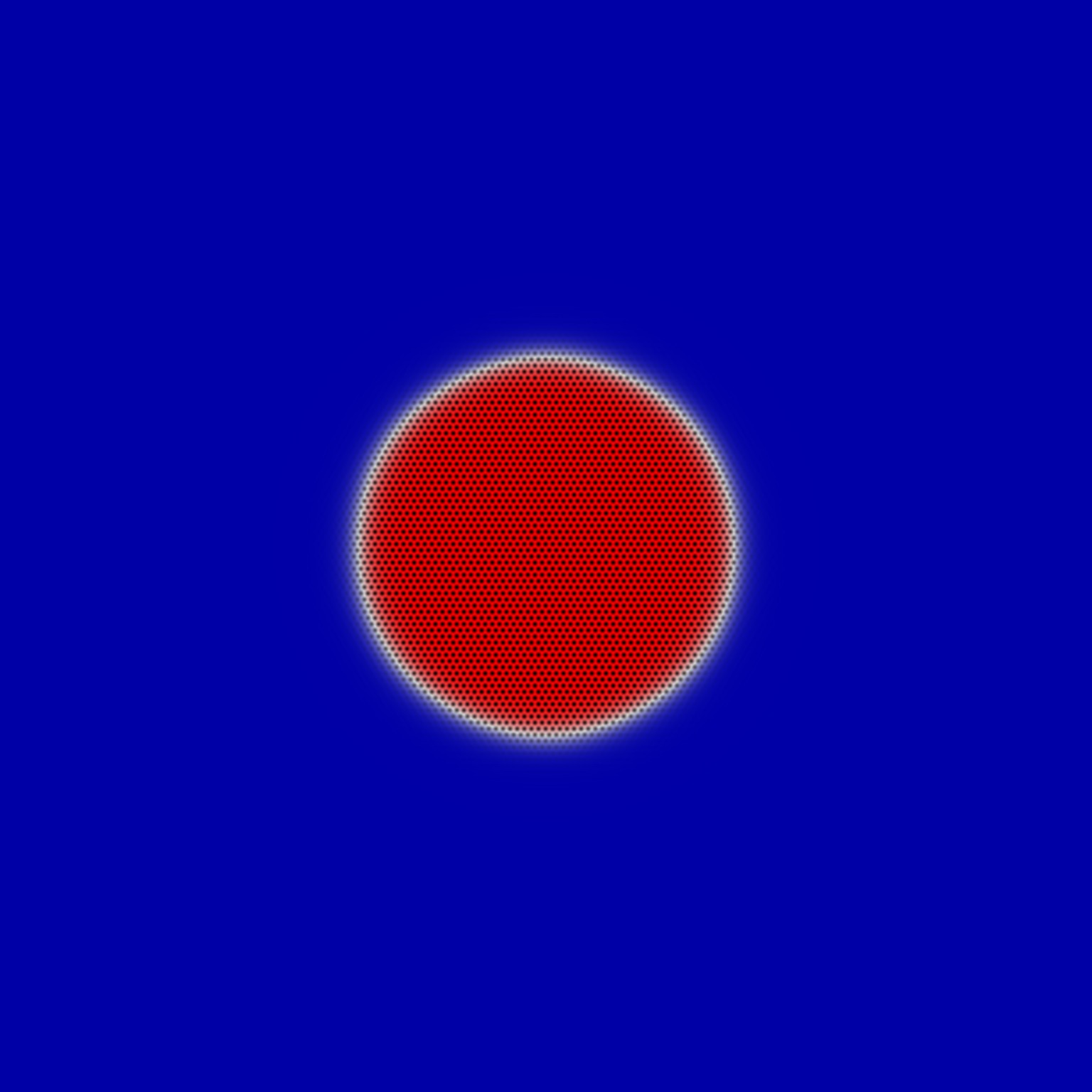}} & \resizebox{1.2in}{!}{\includegraphics{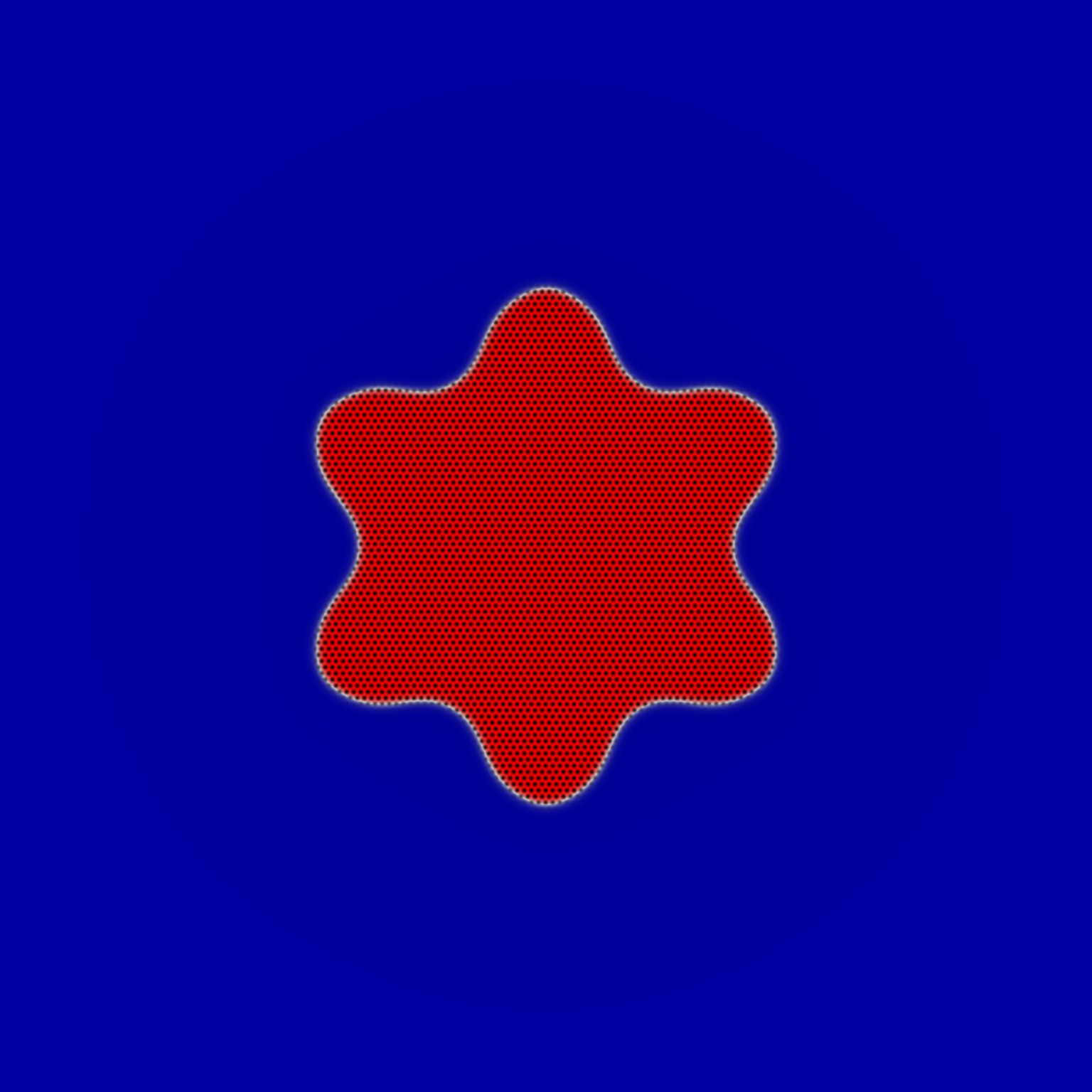}} & \resizebox{1.2in}{!}{\includegraphics{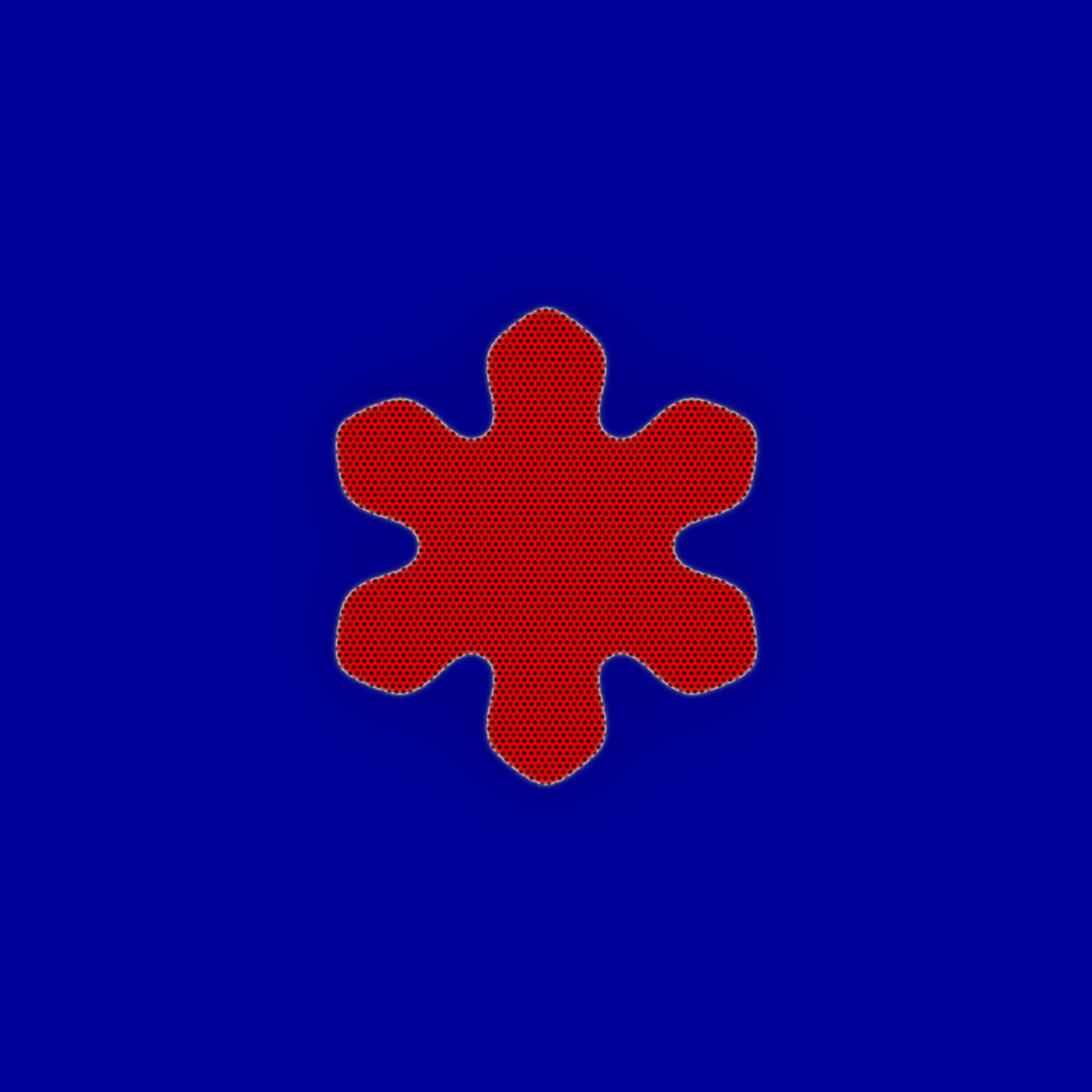}} & \resizebox{1.2in}{!}{\includegraphics{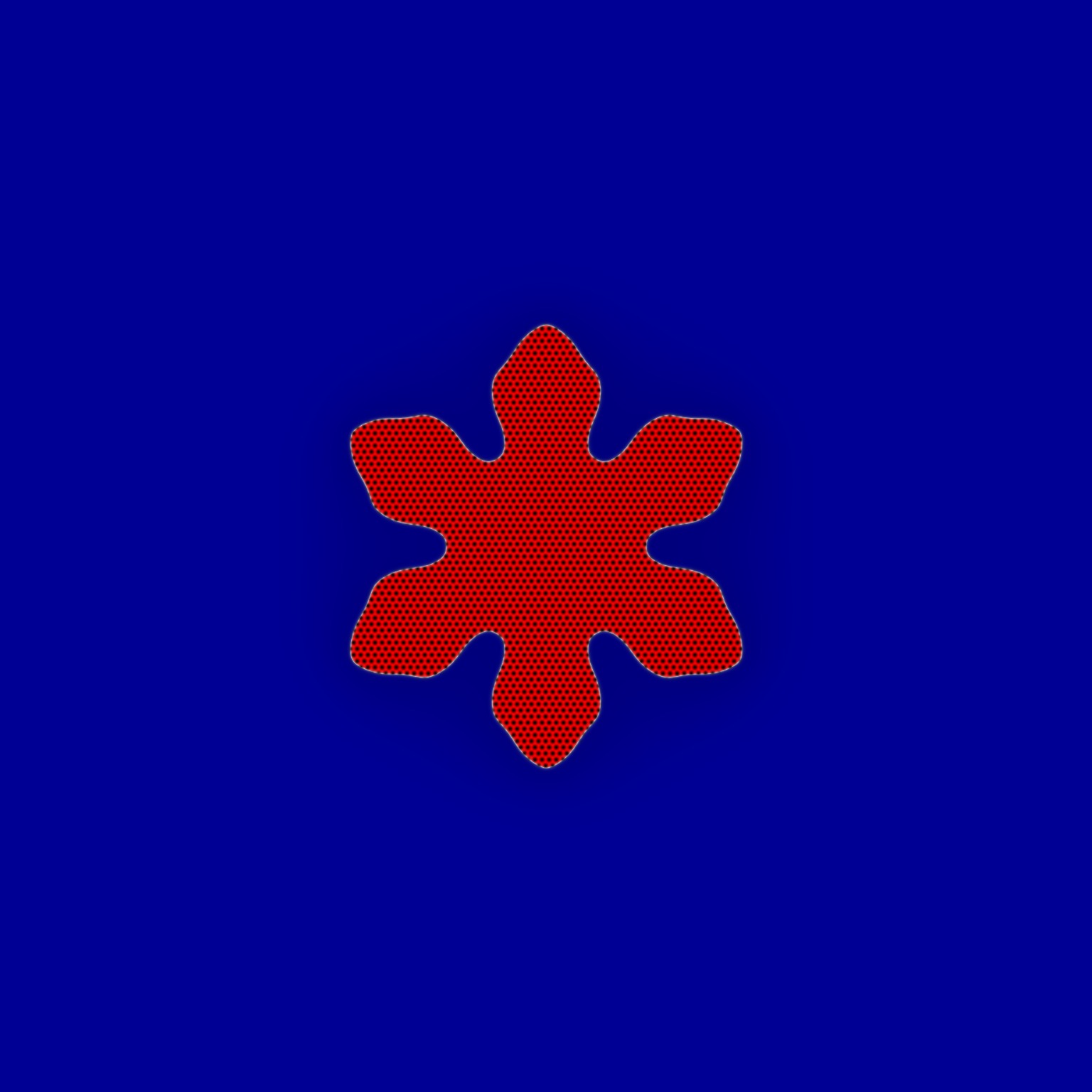}} & \resizebox{1.2in}{!}{\includegraphics{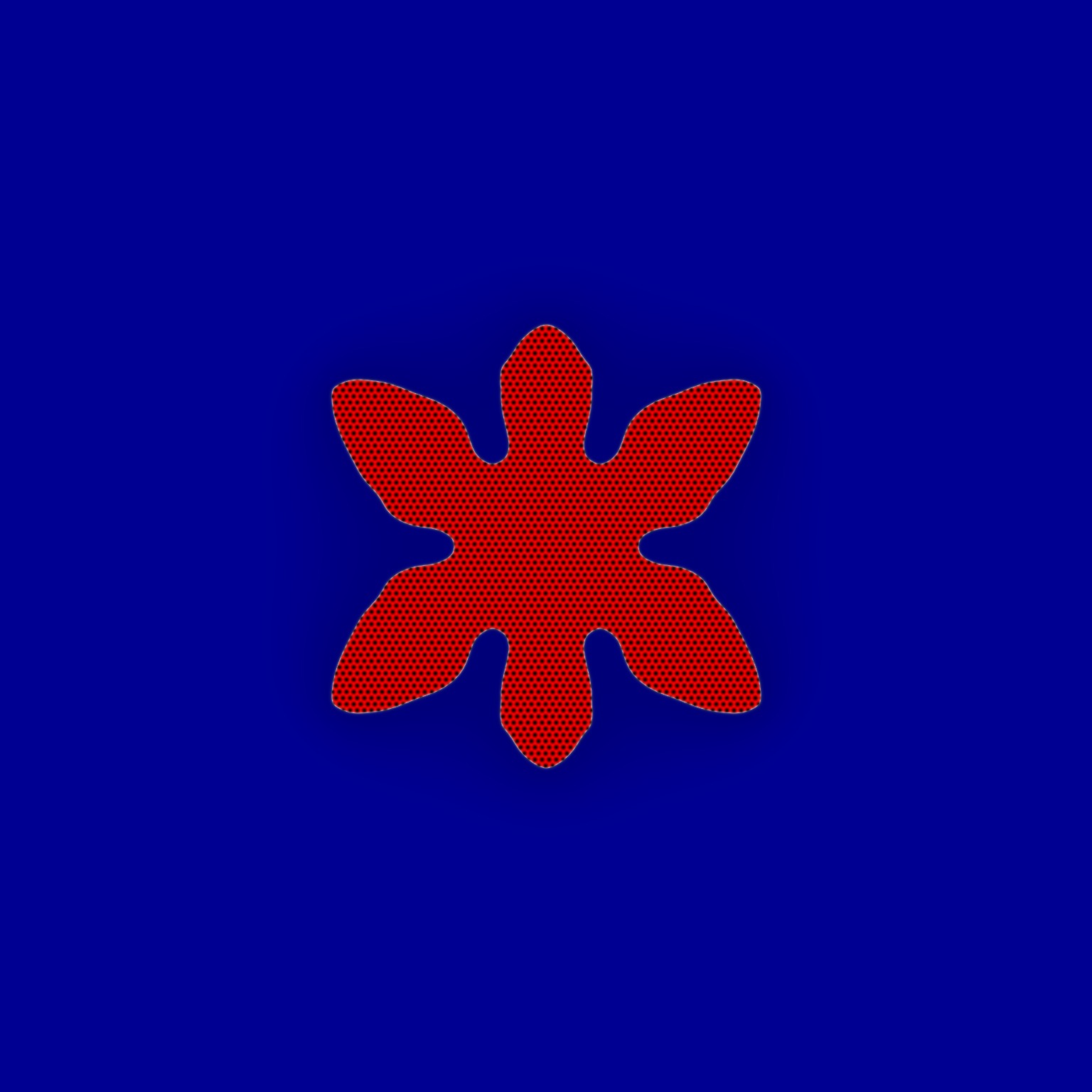}} \\
	$t = 122\tau$ & $t = 114\tau$ & $t = 60\tau$ & $t = 30\tau$ & $t = 24\tau$ \\ \\
	\resizebox{1.2in}{!}{\includegraphics{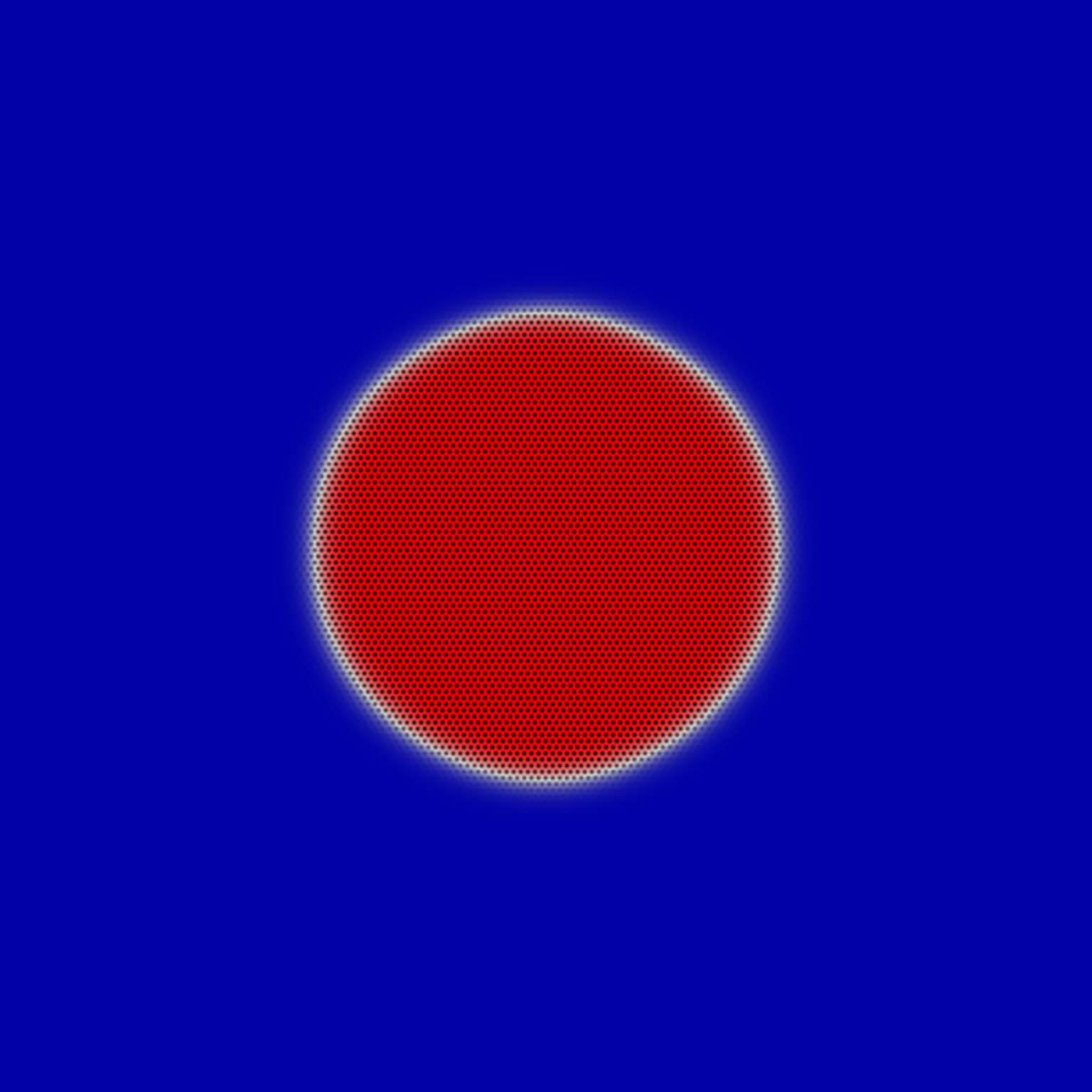}} & \resizebox{1.2in}{!}{\includegraphics{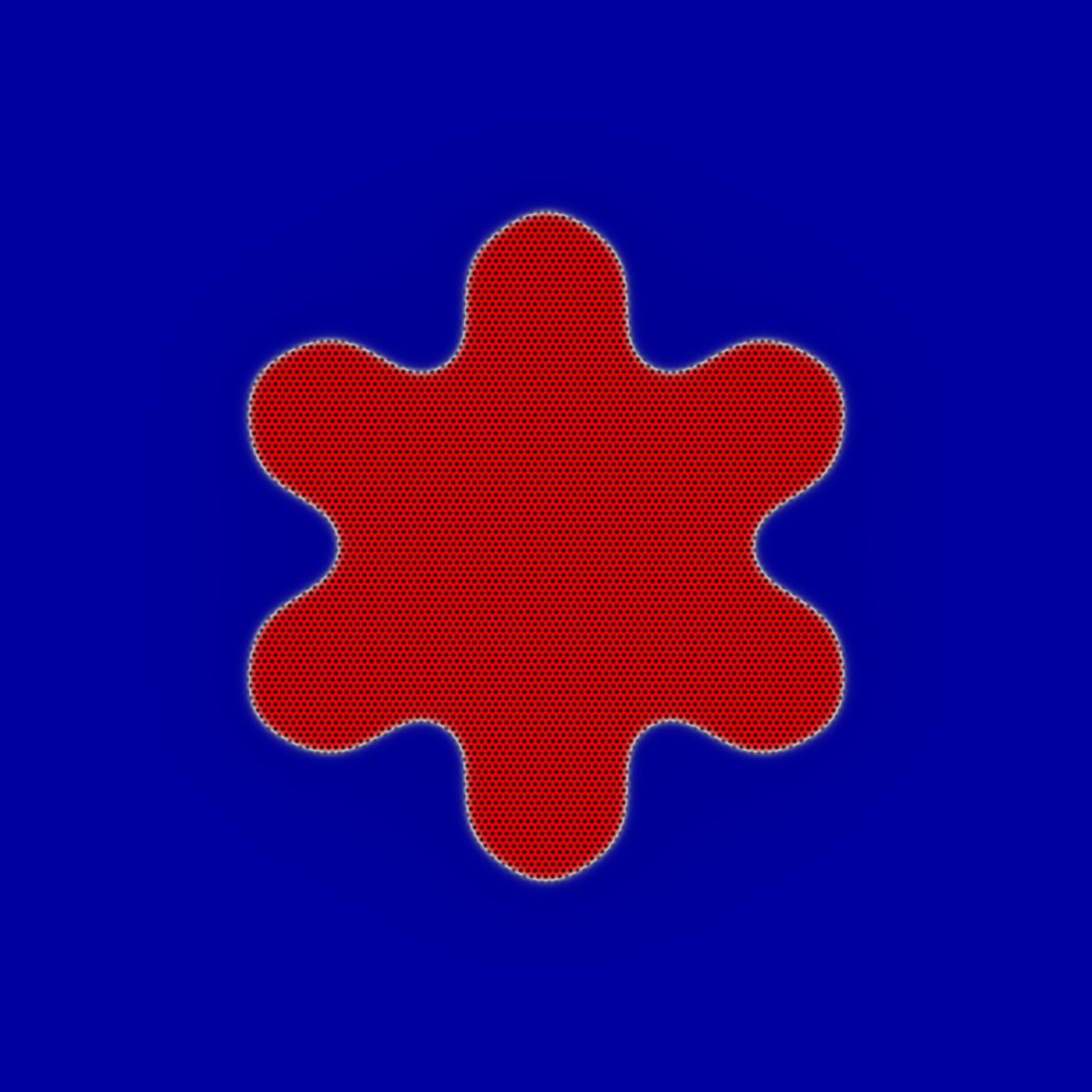}} & \resizebox{1.2in}{!}{\includegraphics{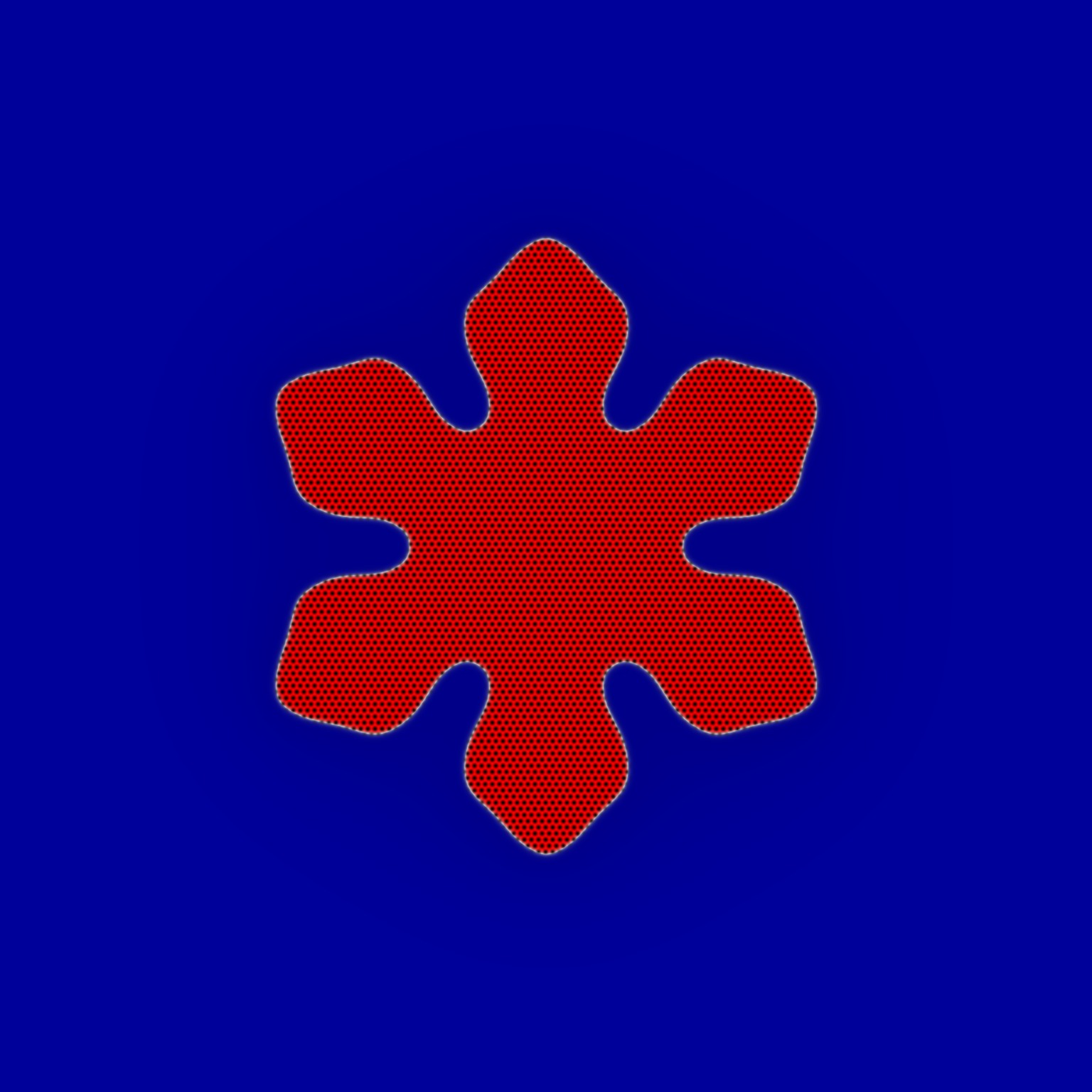}} & \resizebox{1.2in}{!}{\includegraphics{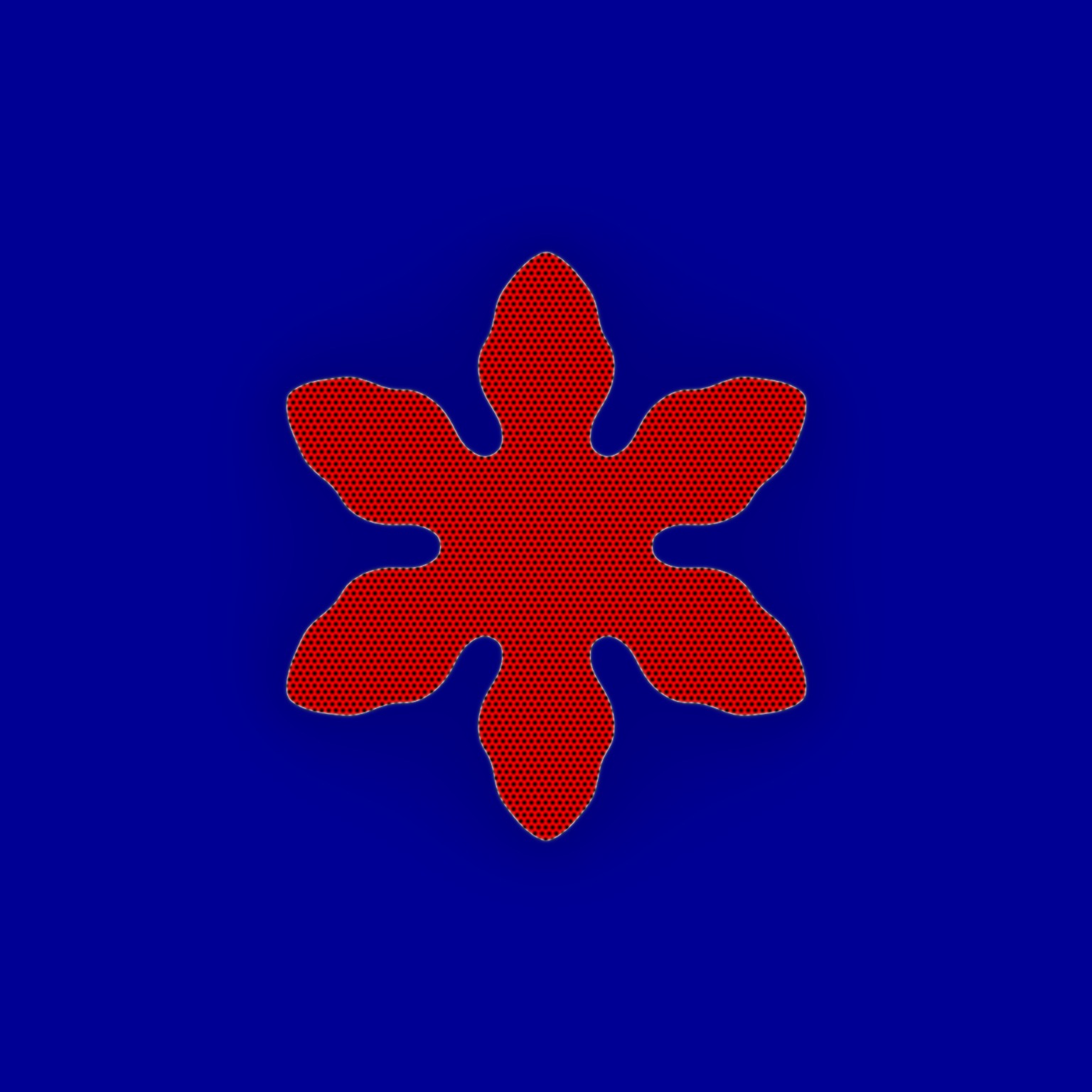}} & \resizebox{1.2in}{!}{\includegraphics{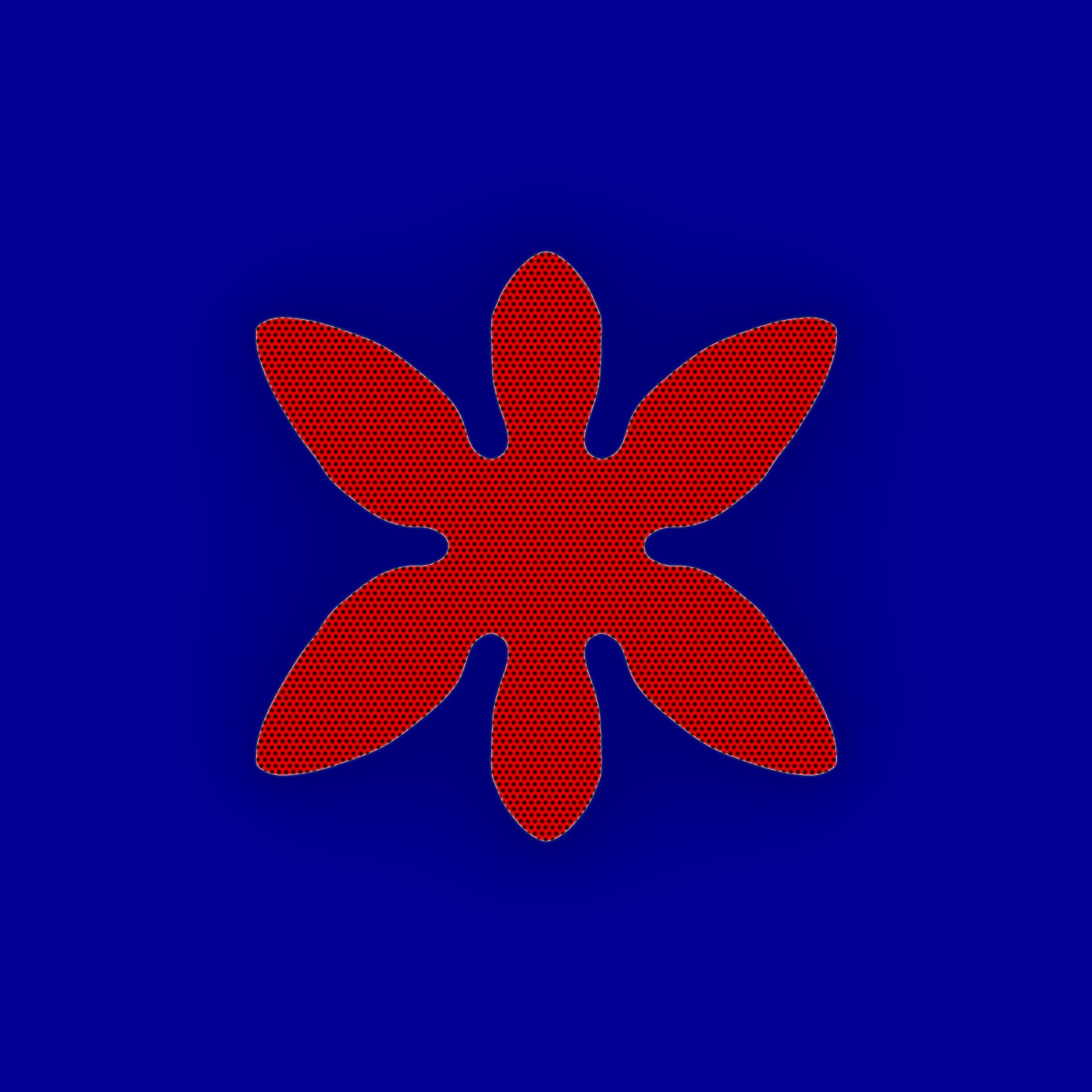}} \\
	$t = 183\tau$ & $t = 171\tau$ & $t = 90\tau$ & $t = 45\tau$ & $t = 36\tau$ \\ \\
	\resizebox{1.2in}{!}{\includegraphics{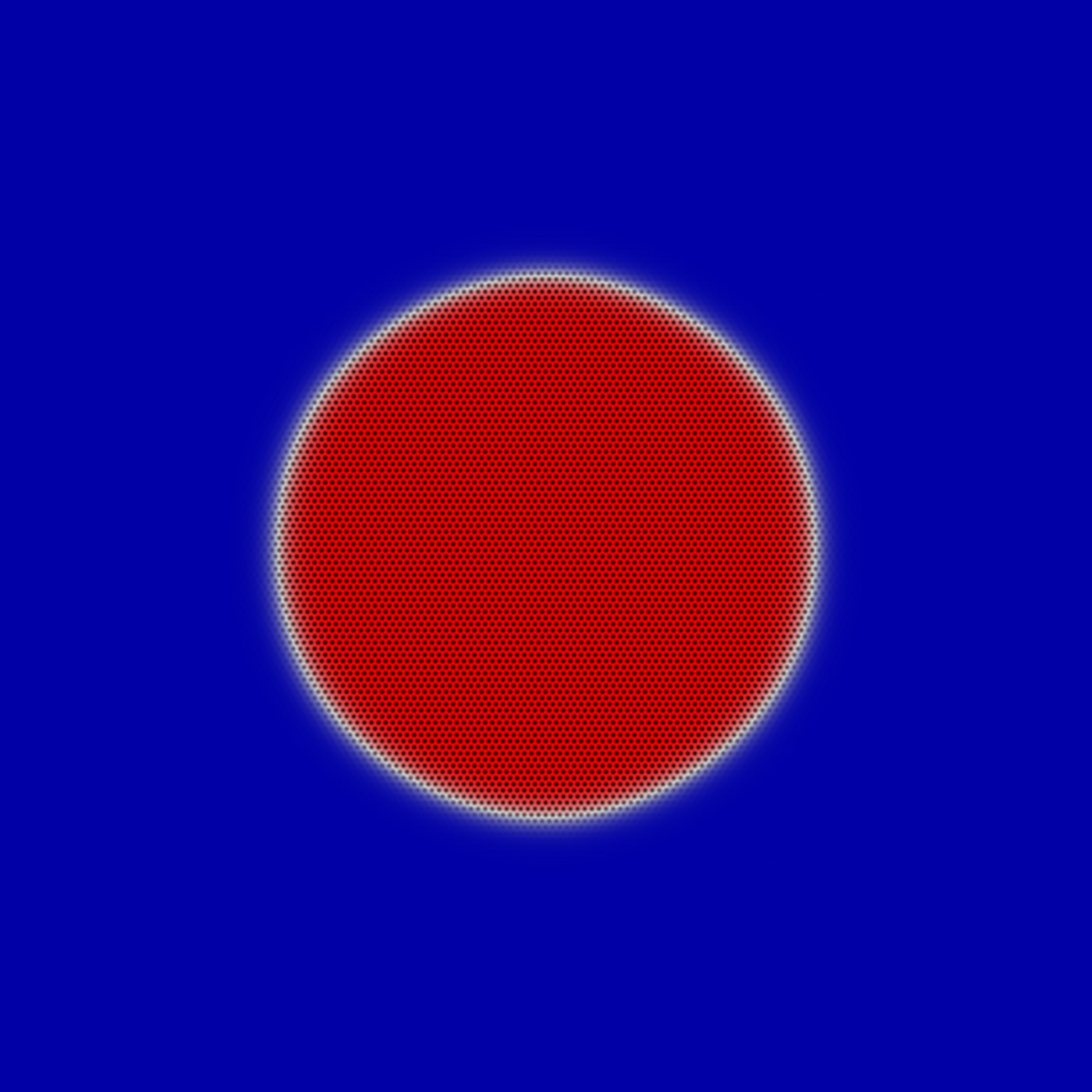}} & \resizebox{1.2in}{!}{\includegraphics{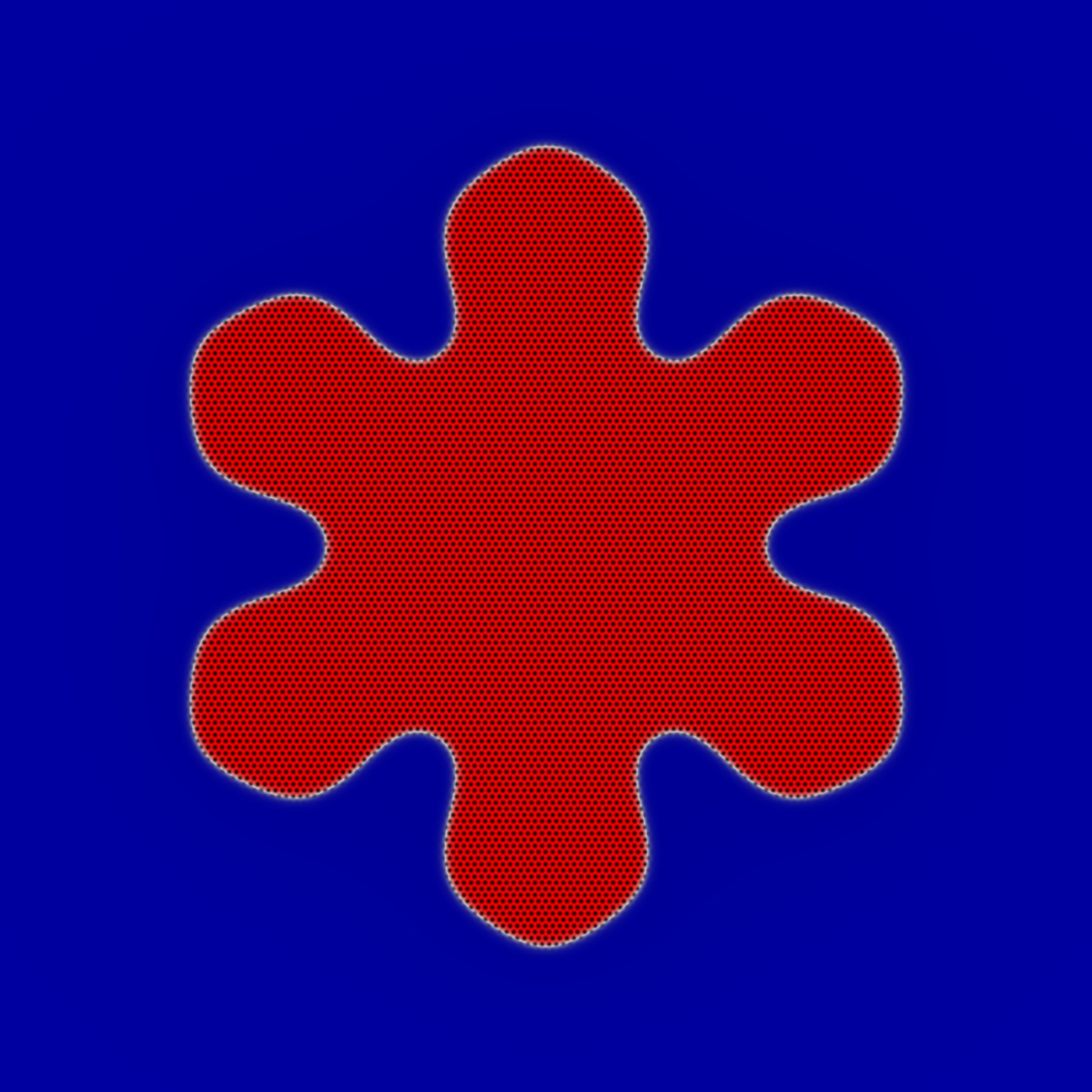}} & \resizebox{1.2in}{!}{\includegraphics{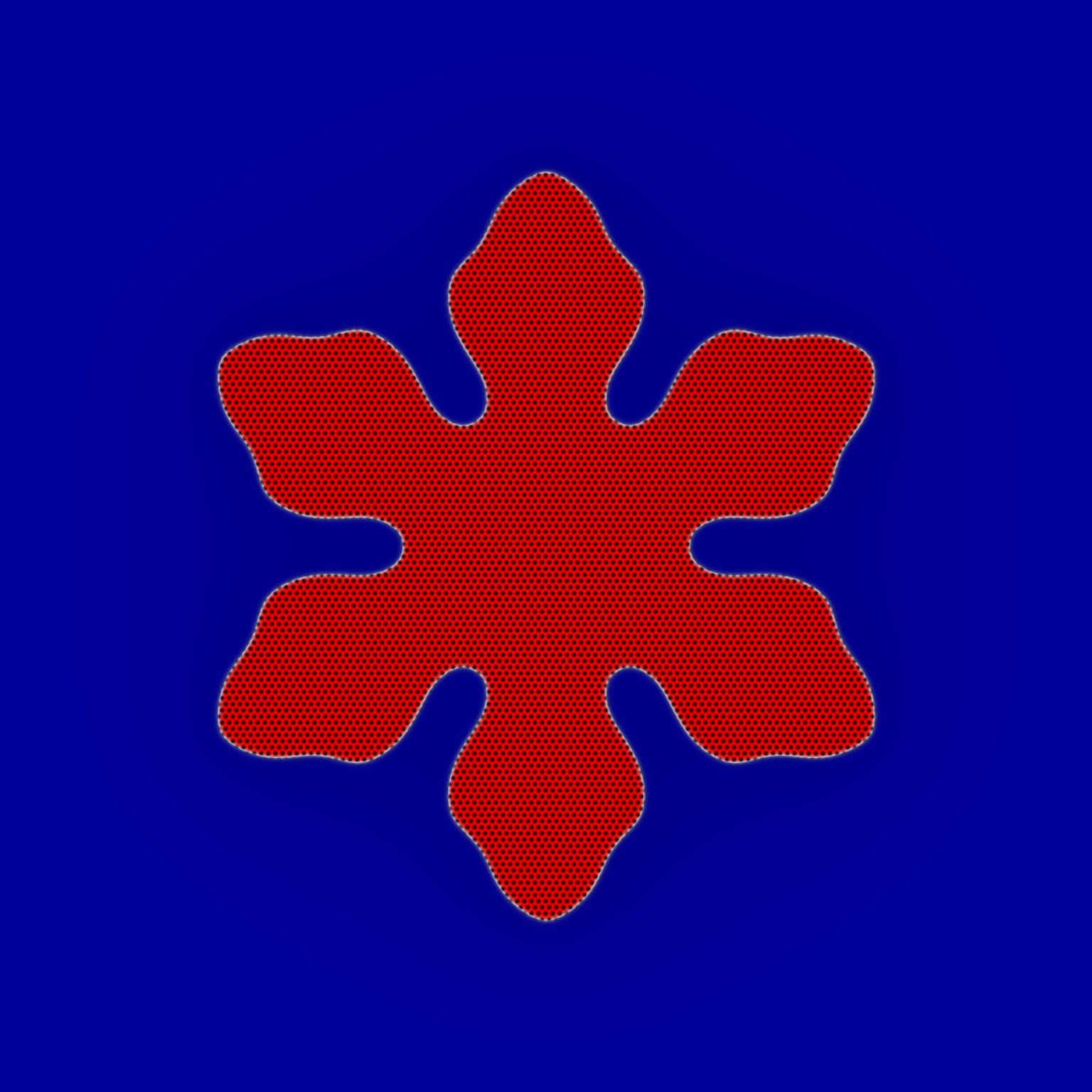}} & \resizebox{1.2in}{!}{\includegraphics{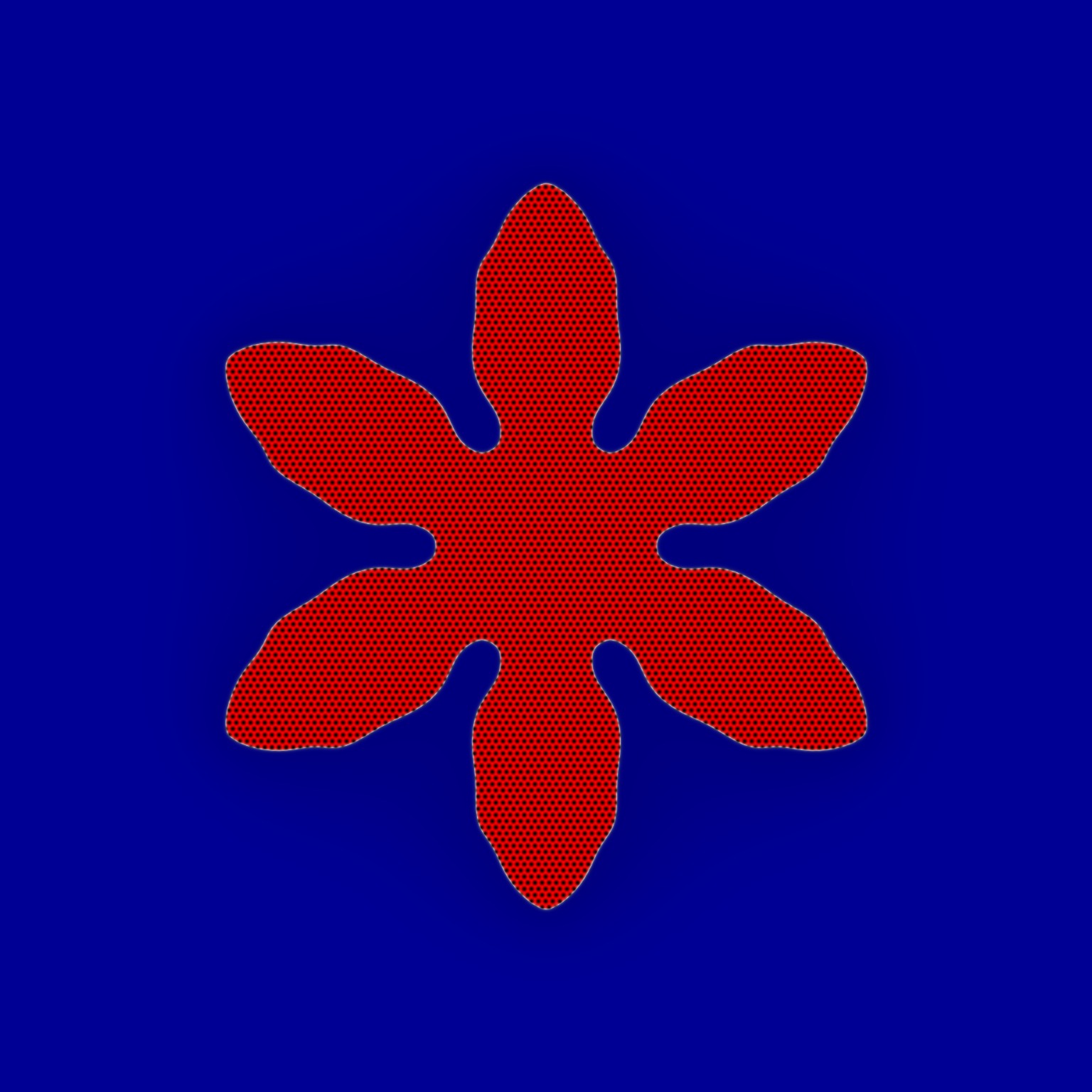}} & \resizebox{1.2in}{!}{\includegraphics{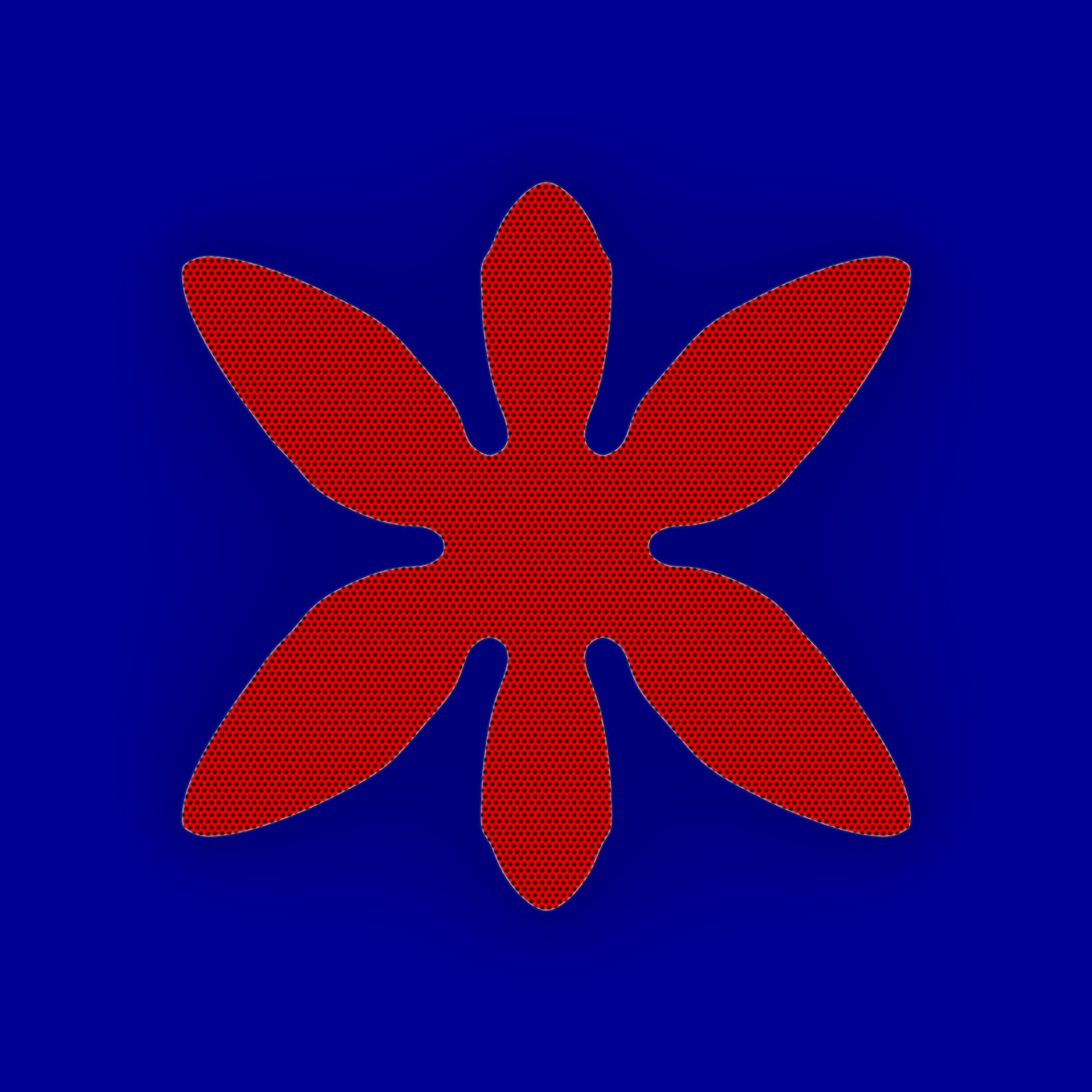}} \\
	$t = 244\tau$ & $t = 228\tau$ & $t = 120\tau$ & $t = 60\tau$ & $t = 48\tau$ \\ \\
\end{tabular}     
\caption{(Color online) Dendritic solidification of the effective phase field model demonstrating effects of highly anisotropic surfaces as function of temperature. Each column of images represents a different temperature, which decreases from left to right. Time increases from top to bottom, with $\tau = 10^4 \Delta t$, where we have attempted to match time according to similar solid fractions, terminating at an approximate solid fraction of $f_s \approx .12$. The sum of the modulus of amplitudes, i.e., $\sum_{j}^{3}|\phi_j|$, is superimposed on the atomic density as reconstructed from \cref{eq:dens}, with red being large amplitude for the solid phase, while blue is zero representing the liquid phase and black the density peaks.}
\label{fig:dend-growth}
\end{figure*}

Two dimensional simulations of dendritic growth were performed to support the calculations in the previous section. Primarily, we want to demonstrate the increasing of the surface energy anisotropy as a function of decreasing temperature and the consequences it has on dendritic behavior. We chose temperatures, $\epsilon = -0.1, -0.5, -1, -2,$ and $-3$, corner energy coefficient $\beta_4$ and grid resolution $\Delta x = a/10$, where $a$ is the equilibrium lattice spacing. All simulations were performed with a time resolution of $\Delta t = 1\times10^{-3}$.  We start with a circular seed of radius $r = 40\Delta x$, where the solid is initialized with the equilibrium amplitude and density, $\bar{n}_{\rm{sol}}$, as specified by the phase diagram. The liquid density is set to satisfy the supersaturation, $\omega = \lt( \bar{n}_{\rm{liq}} - \bar{n}\rt)/\lt(\bar{n}_{\rm{liq}} - \bar{n}_{\rm{sol}} \rt)$, where $\bar{n}$ is the average density of the system. The supersaturation was chosen such that in equilibrium the system will have attained a solid fraction $f_s=0.55$. The results are displayed in \cref{fig:dend-growth}, where we show time slices throughout the simulation for the various temperatures. Temperature decreases from left to right and time increases from top to bottom. Though the simulations were performed in a domain of $3072\Delta x \times 3072\Delta x$, only a portion of the domain, $1536\Delta x \times 1536\Delta x$, centered around the growing crystal is shown. In the images, we plot the sum of the modulus of amplitudes, $\sum_{j}^{3}|\phi_j|$, overlaid on the reconstructed density  described by  \cref{eq:dens}. At high temperature, $\epsilon = -0.1$, the crystal grows isotropically, with only slight variations. However, with decreasing temperature, dendrites emerge, with the dendritic instability occurring at successively earlier times for decreasing temperature, characteristic of increasing anisotropy. In addition to the early onset of the dendritic morphology for decreasing temperature, we also observe that due to increasing anisotropy the  form and shape of the dendrites also change. Specifically, we note that the primary arms become finer -- almost needle like -- with decreasing temperature, and the interfaces become smoother eliminating fluctuations that would eventually cause the emergence of side branching.

\section{Summary and Conclusions}
\label{summ}
%
%
This paper reports the development of an effective phase-field model, derived from the diffusive atomistic PFC model through a coarse grained formalism, for the modeling of highly anisotropic interfaces. In the derived model, the gradient energy responsible for surface energy was shown is related to atomistic level information through the reciprocal lattice vectors of the underlying lattice symmetry of choice. This relationship was anticipated by  Caginalp and Fife~\cite{Caginalp1986,Caginalp1986a,Caginalp1987}, who envisioned anisotropic Ginzburg-Landau models not directly through assumed coefficients of gradients but through general non-local interactions based on lattice symmetry. Further, the model self-consistently gives rise to a biharmonic contribution known to act as a regularizing term in other diffuse interface theories, giving rise to explicit corner energy contributions~\cite{Wise2005,Wise2007,Wheeler2006}. Using 1D calculations, we perform an analysis of the interfacial energy and consequently, the Wulff shape properties of the model as a function of the surface orientation. The role of corner energy is explored, as well as the appropriate resolution with which to perform simulations to adequately account for the multiple length scales involved in the physics of highly anisotropic surfaces. We find that a resolution similar to the full PFC model needs to be considered. The phase-field model captures a transition to missing orientations for specified corner energy coefficients. Dendritic growth simulations show the clear role of strongly anisotropic surface energy on the morphology of growing dendrites. We find that decreasing temperature, and hence increasing anisotropy, enhanced the onset of dendritic instability while also leading to finer primary arms with smooth surfaces. A deeper investigation of dendritic growth in this regime would be instructive in detailing other features of our model as they compare with standard theories. Finally, the phase-field model does not yield faceted interfaces or a roughening transition. This is a likely consequence of the coarse-grained smoothing of the model, but it is a matter worth further investigation.

This is an initial step in determining the feasibility of PF-type models of this form in adequately describing highly anisotropic properties. As such, several things are not considered that are worth exploration. For example, neither the role of elastic and stress effects are not explicitly considered, nor the kinetic factors that have an effect on the behavior of interfaces. Nonetheless, the phase-field model presented here, capable of describing the presence of cusps in the surface energy, dictated by capillary effects, is nontrivial. Also the model we derived, though capable of describing surface orientation, cannot describe multiple grain orientations, a feature crucial to the modeling and understanding of polycrystalline systems that exhibit highly anisotropic interfaces and boundaries. To this end, a natural extension of this work is to examine the complex amplitude formalism, derived in \cref{eq:compAmp}, which naturally includes elasto-plastic effects and the ability to model multiple crystal orientations. For the kinetic aspects of anisotropic interface growth and behavior, one can consider a capillary fluctuation approach \cite{Saidi2017}, on the level of the complex amplitudes or the effective phase-field model. Also an asymptotic analysis of the governing equations might shed light on the role of kinetics to first order. Finally, there are emergent PFC formulations that are based on descriptions of the two-point direct correlation function that closely mimic the more fundamental classical density functional theory~\cite{Greenwood2010,Greenwood2011,Seymour2016}. It would be instructive to use the effective models derived from these models to examine anisotropic surface properties using the same approach outlined here. Doing so would naturally give a temperature dependence on parameters such as the corner energy coefficient, $\beta$, and would also allow us to relate these properties to a fundamental measure theory. This will be considered elsewhere.

\acknowledgements{We acknowledge the following financial assistance award 70NANB14H012 from the U.S. Department of Commerce, National Institute of Standards and Technology as part of the Center for Hierarchical Materials Design (CHiMaD). This research was supported in part through the computational resources and staff contributions provided for the Quest high performance computing facility at Northwestern University which is jointly supported by the Office of the Provost, the Office for Research, and Northwestern University Information Technology, and the resources at the NIST Center for Theoretical and Computational Materials Science (CTCMS).}

\appendix

\section{Equations for Surface Energy}
\label{derive-surfene}
Starting from \cref{eq:surf-omega}, there are numerous ways that one can calculate the surface energy. The simplest is to choose a reference grand potential, notably the equilibrium from one of the bulks, and subtract that from the full grand potential. This reads as

\begin{align}
\label{eq:surf-omega-1}
\gamma &=  \frac{\Omega^{ph}\lt[\{\phi_j\},\bar{n}\rt]  - \Omega^{ph}_{\ell}}{A}\nline
 &=  \frac{\Omega^{ph}\lt[\{\phi_j\},\bar{n}\rt]  - \omega^{ph}_{\ell}(\bar{n}_{\ell})V}{A} \nline
  &=  \frac{1}{A}\int_V d\mbfr \lt( \omega^{ph} - \omega^{ph}_{\ell}\rt), 
\end{align}
where $\omega^{ph}$ represents the grand potential density of our effective phase-field model. In the second line we have used the definition that for a bulk phase $\Omega^{ph} = -pV$ (where $-p \equiv \omega^{ph}$) and have chosen the liquid as our reference. This definition can be further expanded by using the explicit definition of the grand potential, and noting that the equilibrium chemical potential is defined by $\mu = \lt(f^{ph}_{s}-f^{ph}_{\ell} \rt)/ \lt( \bar{n}_{s}-\bar{n}_{\ell}\rt)$. We then have
\begin{align}
\label{eq:surf-omega-2}
 \gamma  &= \frac{\Omega^{ph}\lt[\{\phi_j\},\bar{n}\rt]  - \Omega^{ph}_{\ell}}{A}\nline
 & = \frac{1}{A}\int_V d\mbfr \lt( f^{ph} - \frac{f^{ph}_s - f^{ph}_{\ell}}{\bar{n}_{s}-\bar{n}_{\ell}}n(\mbfr) - f^{ph}_{\ell} + \frac{f^{ph}_{s} - f^{ph}_{\ell}}{\bar{n}_{s}-\bar{n}_{\ell}}\bar{n}_{\ell} \rt)\nline
  & = \frac{1}{A}\int_V d\mbfr \lt( f^{ph} - f^{ph}_s \frac{n(\mbfr) - \bar{n}_{\ell}}{\bar{n}_{s}-\bar{n}_{\ell}} + f^{ph}_{\ell} \frac{n(\mbfr) - \bar{n}_{s}}{\bar{n}_{s}-\bar{n}_{\ell}} \rt),
\end{align}
where in the second line, we have used the definition of the grand potential, and this is the form of the surface energy we have reported in the text. $f^{ph}_{\nu}$ and $\bar{n}_{\nu}$ represents the equilibrium values for the free energy and density for the liquid and solid respectively. In equilibrium, \cref{eq:surf-omega-1,eq:surf-omega-2} are identical. However, for out of equilibrium or for unstable configurations that still have constant bulk contributions, we can define the following $f^{ph}_{\nu} \rightarrow f^{ph*}_{\nu} $ and $\bar{n}_{\nu} \rightarrow \bar{n}_{\nu}^*$ (where `*' denotes evaluation in the bulk) and refer to this as a {\it relative excess} quantity~\cite{Godreche1991}.

Further to the definitions above, we can also examine the surface energy contributions using an alternate method. In 1D for simplicity, we begin with the Euler-Lagrange equations (ELE) of \cref{eq:phAmp2}. This method explicitly reveals some of the underlying properties of our derived effective phase-field model. The ELE read
\begin{align}
\label{eq:phAmp2-ele}
&\fxfy{\freem^{ph}}{\bar{n}} = \pxpy{f^{ph}}{\bar{n}}  \equiv \mu \\
&\fxfy{\freem^{ph}}{\phi_j} = \pxpy{f^{ph}}{\phi_j} - \partial_x\pxpy{f^{ph}}{(\partial_x \phi_j)} + \partial_x^2\pxpy{f^{ph}}{(\partial_x^2 \phi_j)} \equiv 0 \quad \forall~  j\nonumber.
\end{align}
We multiply both equations by their respective gradients to get,
\begin{align}
\label{eq:phAm2-ele-proj}
&\bar{n}_{x}\pxpy{f^{ph}}{\bar{n}}  = \bar{n}_{x}\mu \\
&\sum_{j}^{3} \phi_{j,x}\pxpy{f^{ph}}{\phi_j} - \phi_{j,x}\partial_x\pxpy{f^{ph}}{\phi_{j,x}} + \phi_{j,x}\partial_x^2\pxpy{f^{ph}}{\phi_{j,xx}} = 0 \nonumber,
\end{align}
where we have used shorthand notation for the gradient derivatives of the fields. The general gradient of the energy reads
\beq
\label{eq:phAm2-grad}
\pxpy{f^{ph}}{x} =  \pxpy{f^{ph}}{\bar{n}}\bar{n}_{x} + \sum_{j}^{3} \pxpy{f^{ph}}{\phi_j}\phi_{j,x} + \pxpy{f^{ph}}{\phi_{j,x}}\phi_{j,xx}  + \pxpy{f^{ph}}{\phi_{j,xx}}\phi_{j,xxx}.
\eeq
We substitute \cref{eq:phAm2-grad} into the addition of the ELE, \cref{eq:phAm2-ele-proj}, and arrive at the following
\begin{align}
\label{eq:phAmp2-firstint}
\pxpy{f^{ph}}{x} - \bar{n}_{ x}\mu =&
- \sum_{j}^{3} \Biggl( \pxpy{f^{ph}}{\phi_{j,x}}\phi_{j,xx} + \pxpy{f^{ph}}{\phi_{j,x}}\phi_{j,xxx} + \phi_{j,x}\partial_x\pxpy{f^{ph}}{\phi_{j,x}} - \phi_{j,x}\partial_x^2\pxpy{f^{ph}}{\phi_{j,xx}}  \Biggr).
\end{align}
Next, we perform, under suitable boundary conditions, partial integration by parts on the right hand side of \cref{eq:phAmp2-firstint} to yield,
\begin{align}
\pxpy{f^{ph}}{x} - \bar{n}_{ x}\mu = \sum_{j}^{3} \Biggl( \phi_{j,x}\pxpy{f^{ph}}{\phi_{j,xx}} - \phi_{j,x}\pxpy{f^{ph}}{\phi_{j,x}} + 2\phi_{j,xx}\pxpy{f^{ph}}{\phi_{j,xx}} \Biggr).
\end{align}
Lastly, we integrate both sides across the 1D domain, i.e., $-\infty$ to $\infty$. Note that the left hand side is nothing but the integral of the grand potential, where in equilibrium we expect the bulk phases to yield equivalent contributions, this integral is nothing but the surface energy. That is
\begin{align}
\int_{-\infty}^{\infty} \lt( \pxpy{f^{ph}}{x} - \bar{n}_{ x}\mu \rt) & = \int_{-\infty}^{\infty} \pxpy{\omega^{ph}}{x} dx \nline
& \equiv \gamma .
\end{align}
For the right hand side, we write the derivatives explicitly to attain
\begin{align}
\sum_{j}^{3} &\lt( 8(\mbfk_j \cdot\mathbf{n} )^2\lt(\pxpy{\phi_j}{x} \rt)^2   +  \beta \lt(\pxpy{\phi_j}{x}\rt)  \lt(\pxpy{^2\phi_j}{x^2}\rt)  +  4 \beta \lt(\pxpy{^2\phi_j}{x^2}\rt)^2 \rt).\nonumber
\end{align}
The normal vector is defined as $\mathbf{n} = (\cos{\theta},\sin{\theta})$. The equilibrium profiles of the amplitudes, $\{\phi_j\}$, to first order can be represented by hyperbolic tangents, an integral over the domain then yields
\begin{align}
\label{eq:surf-ene-firstint}
\gamma(\theta) = \int_{-\infty}^{\infty} \sum_{j}^{3}  \Biggl\{ &8\lt(\mbfk_j^x\cos{\theta}+ \mbfk_j^y\sin{\theta}\rt)^2\lt(\pxpy{\phi_j}{x} \rt)^2 + 4\beta\kappa_j^2\lt(\pxpy{\phi_j}{x} \rt)^4   \Biggr\},
\end{align}
where $\mbfk_j^x$ and $\mbfk_j^y$ are the $x$- and $y$-components of the reciprocal lattice vectors and we have used the fact that the curvature can defined by $\kappa_j = \partial_x^2\phi_j/(\partial_x\phi_j)^2$. We explicitly see the anisotropic nature of the surface energy through the surface orientation, $\theta$. In rewriting the biharmonic contribution in terms of the curvature, we immediately see the role of the regularizing term in the rounding of corners. The above calculation can also be performed more rigorously through perturbation expansions involving matched asymptotic analysis.

We have seen in \cref{eq:surf-ene-firstint}, the contributions from corners to the total excess energy of the interface, which we called the surface energy. If we define the actual corner energy, $\sigma$, to simply be the difference of the the total excess energy with and without the higher order contribution, then it is trivial to obtain that the corner energy is defined by
\begin{align}
\label{eq:corner-ene}
\sigma = \int_{-\infty}^{\infty} 4\beta\,\sum_{j}^{3} \kappa_j^2\lt(\pxpy{\phi_j}{x} \rt)^4.
\end{align}
For the singular case where $\beta$ might be zero, the corner energy based on \cref{eq:corner-ene} is necessarily zero. This is pathological and for a more exact treatment of cases such as these, an asymptotic expansion is necessary, where one considers $\beta \rightarrow 0$ as was done in Ref. \cite{Wheeler2006}.

\section{Transition of anisotropy}
\label{anis-transition}
\begin{figure*}[h]
\centering
\begin{tabular}{ccccc}
	 \resizebox{1.1in}{!}{\includegraphics{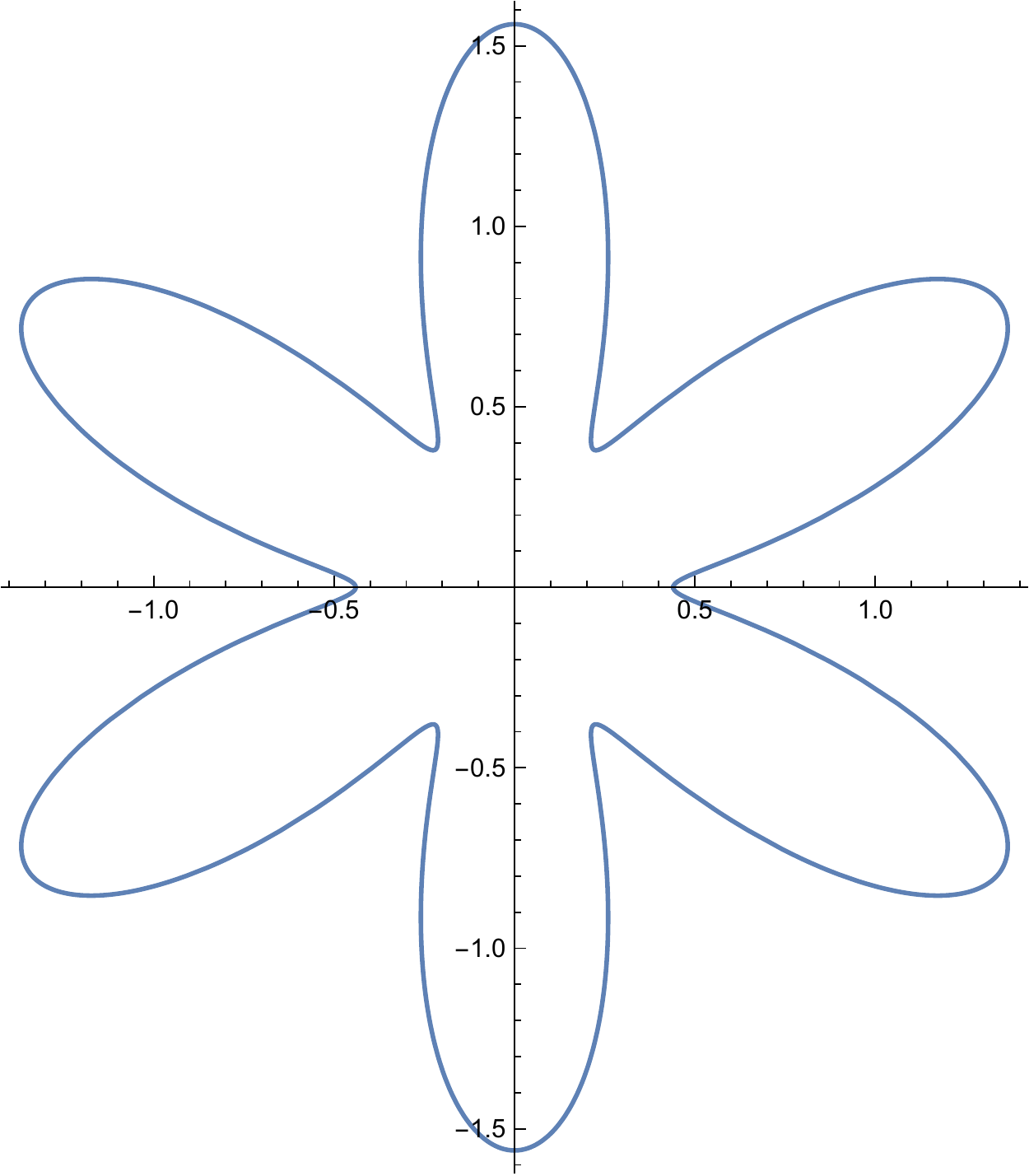}} &  \resizebox{1.1in}{!}{\includegraphics{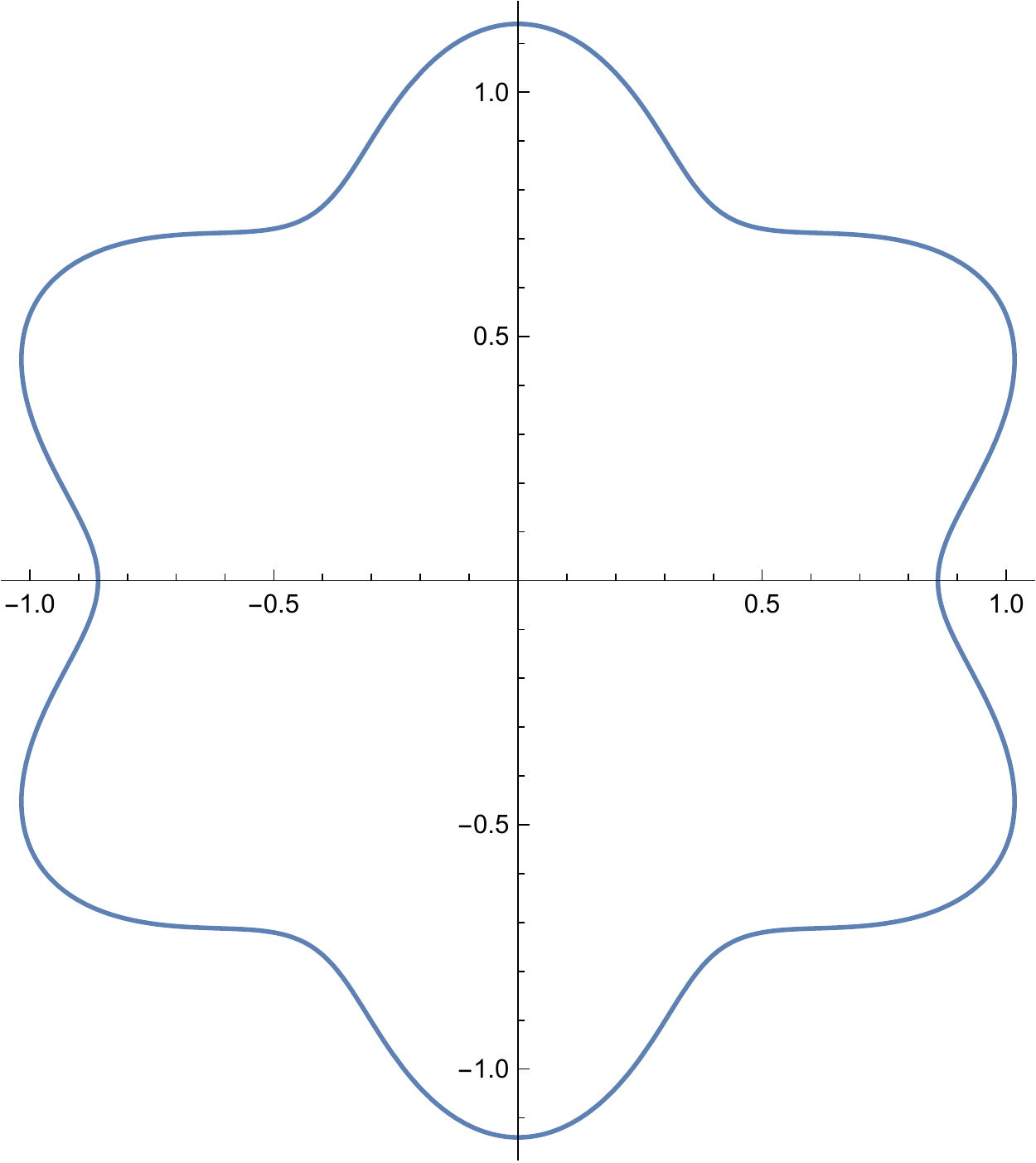}}  &  \resizebox{1.225in}{!}{\includegraphics{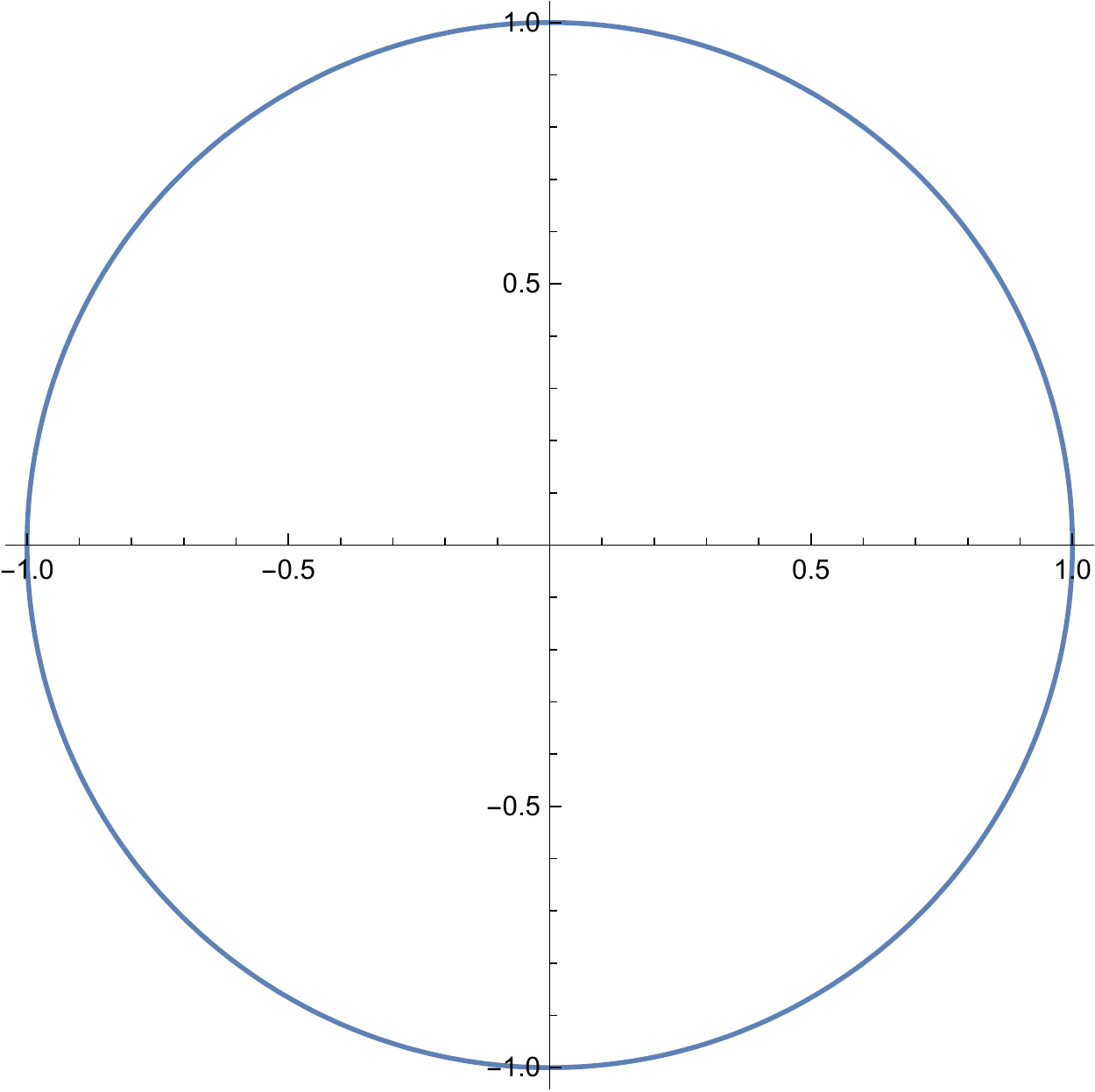}}  & \resizebox{1.35in}{!}{\includegraphics{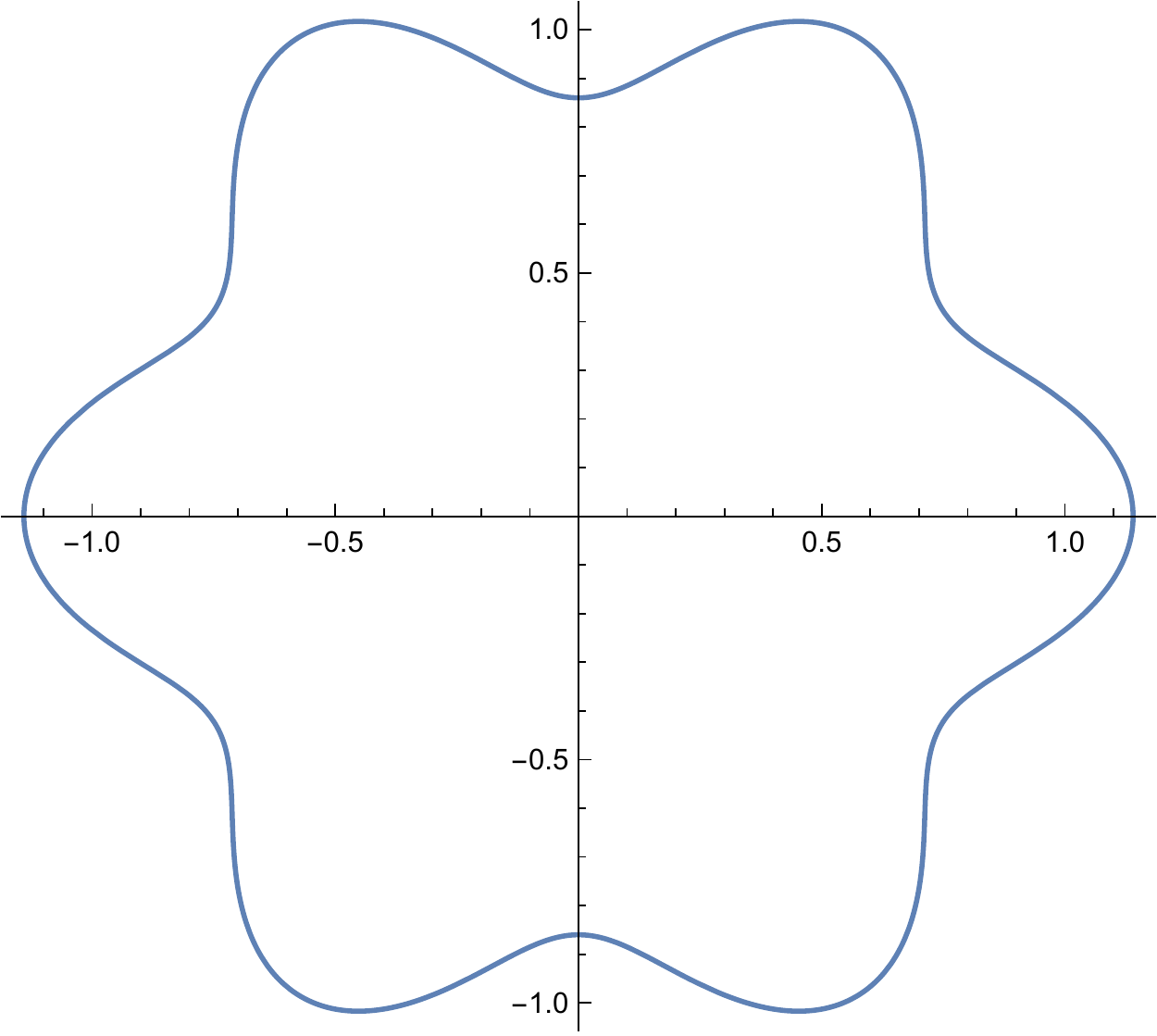}}  &  \resizebox{1.375in}{!}{\includegraphics{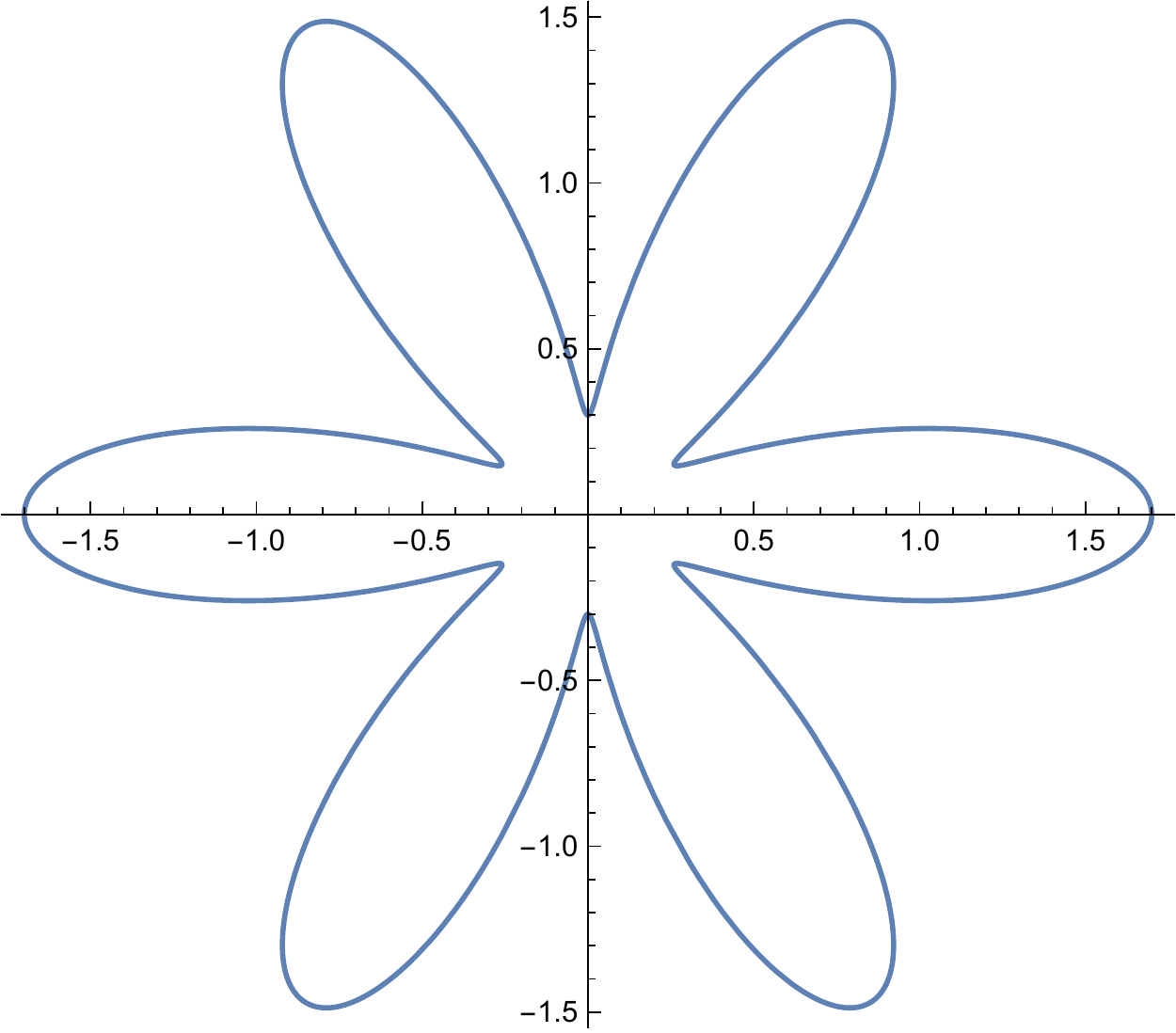}}  \\
\end{tabular}
\caption{Stiffness plots, from the anisotropy function $\eta(\theta) = 1+\delta\cos{6\theta}$, showing the change of the growth directions as a function of the anisotropy strength $\delta$, which decreases from left to right. The middle middle panel represents $\delta = 0$.}
\label{fig:anis-change}
\end{figure*}

In \cref{results-1D}, when we considered the surface energy and Wulff shapes, we found that for select parameters of the corner energy coefficient, $\beta$, and the resolution, $\Delta x$, that the system underwent an anisotropy transition. This is the first for 2D structures that we are aware of. However, this mechanism has been reported in the literature for both experiments and computations \cite{Haxhimali2006,Friedli2013,Dantzig2013} for 3D structures. For pure systems, specifically for fcc structures, this was a result of a negative anisotropy coefficient for the second harmonic used to describe the surface energy. In alloys, this resulted from solute additions changing the anisotropy coefficient. Although, the transition reported here is the result of an under resolved system,  we have found this is caused by a sign change in the anisotropy coefficients as a function of temperature, which can also be understood in the light of solute additions changing the temperature behavior in alloys. Here we demonstrate this for a simple case for an anisotropy function of the form, $\eta(\theta) = 1+\delta\cos{6\theta}$.  In \cref{fig:anis-change}, we show the stiffness plot of the aforementioned anisotropy function for various anisotropy strengths, $\delta$, which decreases from left to right. For this to exhibit the hyper-branching resulting from an anisotropy change, as described in the models and experiments of Refs. \cite{Haxhimali2006,Friedli2013,Dantzig2013}, higher order terms would need to be considered in our six-fold function.

We note that conventionally, anisotropy coefficients do not change as function of temperature or alloy content in the myriad of phase-field models present in the literature. In models such as the one presented here, derived from atomistics, because {\it a priori} there is no predetermined form of the anisotropy function, the system is allowed to explore various effects. Of course assuming that the resolution is sufficient.

\bibliography{facetBib}
\end{document}